\documentclass[useAMS,usenatbib]{mn2e}
\usepackage{graphicx,amsmath,multirow,amssymb}
\usepackage{subfig}
\usepackage{natbib}
\newcommand{\comment}[1]{}

\def\simgt{\lower.5ex\hbox{$\; \buildrel > \over \sim \;$}}
\def\simlt{\lower.5ex\hbox{$\; \buildrel < \over \sim \;$}}

\title{AGB stars in the LMC: evolution of dust in circumstellar envelopes}
\author[Dell'Agli et al.]{F. Dell'Agli$^{1,2}$, P. Ventura$^2$, R. Schneider$^2$, 
M. Di Criscienzo$^{2}$, 
\newauthor
D. A. Garc\'{\i}a--Hern\'andez$^{3,4}$, C. Rossi$^{1}$, E. Brocato$^2$   \\
$^1$Dipartimento di Fisica, Universit\`a di Roma ``La Sapienza'', P.le Aldo Moro 5, 00143, Roma, Italy \\
$^2$INAF -- Osservatorio Astronomico di Roma, Via Frascati 33, 00040, Monte Porzio Catone (RM), Italy \\
$^{3}$Instituto de Astrof\'{\i}sica de Canarias, E-38200 La Laguna, Tenerife, Spain \\
$^{4}$Departamento de Astrof\'{\i}sica, Universidad de La Laguna (ULL), E-38206 La Laguna, Tenerife, Spain\\
}

\begin{document}

\date{Accepted, Received; in original form }

\pagerange{\pageref{firstpage}--\pageref{lastpage}} \pubyear{2012}

\maketitle

\label{firstpage}

\begin{abstract}
We calculated theoretical evolutionary sequences of asymptotic giant branch (AGB) stars,
including formation and evolution of dust grains in their circumstellar envelope. By
considering stellar populations of the Large Magellanic Cloud (LMC), we calculate synthetic 
colour--colour and colour--magnitude diagrams, which are compared with those
obtained by the Spitzer Space Telescope.

The comparison between observations and theoretical predictions outlines that
extremely obscured carbon--stars and oxygen--rich sources experiencing hot bottom burning (HBB)
occupy well defined, distinct regions in the colour--colour ($[3.6]-[4.5]$, $[5.8]-[8.0]$) 
diagram. The C--rich stars are distributed along a diagonal 
strip that we interpret as an evolutionary sequence, becoming progressively more obscured 
as the stellar surface layers enrich in carbon. Their circumstellar envelopes host solid carbon dust
grains with size in the range $0.05 < a <  0.2 \mu m$. The presence of SiC particles is 
expected only in the more metal--rich stars. The reddest sources, with $[3.6]-[4.5] > 2$, 
are the descendants of 
stars with initial mass $M_{in} \sim 2.5 - 3 M_{\odot}$ in the very latest phases of the 
AGB life. The oxygen--rich stars with the reddest colours ($[5.8]-[8.0] > 0.6$) are those 
experiencing HBB, the descendants of $\sim 5 M_{\odot}$ objects formed 
$10^{8}$ yr ago; alumina and silicates dust start forming at different distances from the central star. 
The overall dust production rate in 
the LMC is $\sim 4.5 \times 10^{-5} M_{\odot}/yr$, the relative percentages due to 
C-- and M-- star being respectively 85$\%$ and 15 $\%$.
\end{abstract}


\section{Introduction}
The stellar sources most relevant for dust production are low-- and intermediate--
mass stars during the asymptotic giant branch (AGB) phase and supernovae (SNe).
A full understanding of the amount of dust produced by these stellar sources is 
essential for accounting the presence of large quantity of dust in galaxies.
The analysis of the spectral energy distribution (SED) of high--redshift quasars 
shows that large reservoirs of dust are observed up to redshift $z \sim 6.4$ 
\citep{bertoldi03, priddey03, robson04, beelen06, wang08, wang13}. 
Several investigations have addressed the dust evolution in galaxies of the
Local Group \citep{dwek98,calura08,zhukovska08,boyer13,raffa14,matteo} and in
high--redshift galaxies \citep{calura08, morgan03}. These works were aimed
not only at providing an exhaustive explanation of the observational scenario, but also
to explain the presence of dust at early epochs \citep{valiante09,valiante11,mattson11,
dwek11,pipino11}. Concerning the high redshifts, a lively debate is still open on the 
dominant source of dust. Early investigations suggested the
dominant role of SNe \citep{maiolino04}, a conclusion challenged by subsequent studies, 
that stressed the importance of dust destruction by SNe reverse shock's'
\citep{nozawa06,bianchi07, silvia10, silvia12}. Recent studies 
showed that the dust present at high redshift is produced by all the stars more massive 
than $\sim 3M_{\odot}$, which points in favour of a non negligible contribution from AGBs 
\citep{valiante09}.

The study of the dust produced by AGB stars has been proved extremely useful in several astrophysical contexts, given the central role played by these stars on dust production and evolution in galaxies.
The pioneering investigations by the
Heidelberg group \citep{gs85, gs99, fg06, zhukovska08} were the first attempts to 
track down numerically the dust condensation sequences in AGB stars:
their scheme is based on a basic description of the stellar 
wind, assumed to expand isotropically from the stellar surface, and accelerated by radiation 
pressure acting on the dust grains. This approach represents an extremely simplified 
schematisation of a more complex situation, where formation and growth of dust particles 
is likely favoured by periodic shocks, triggered by large amplitude pulsations experienced 
by AGBs \citep{wood79, Bertschinger85, bowen88, fleischer92}. 
However, the hydrostatic atmosphere approximation is currently the only description that 
can be easily interfaced with stellar evolution codes, and was adopted more recently by 
different groups involved in this research \citep{paperI, paperII, paperIII, paperIV, 
nanni13a, nanni13b, nanni14}. The results are still far from being completely reliable,  
owing to the uncertainties affecting the AGB evolution (mainly the description of mass loss
and the treatment of the convective instability) and the dust formation process (e.g. 
sticking coefficients of some species, extinction properties, etc.). The differences among 
the results presented by the various groups outline the need for a further refinement of 
the models \citep{nanni13b, paperIV}.

The comparison with the observational scenario is therefore mandatory to
confirm or disregard the theoretical models so far produced. The analysis of dusty
AGBs in the Galaxy is not straightforward, owing to the obscuration from its own
interstellar medium, and the unknown distances, that render uncertain the luminosity
of the sources observed. The Large Magellanic Cloud (LMC) is a much more favourable
target as it is relatively close \citep[$50$kpc,][]{feast99} and 
with a low average reddening \citep[$E(B-V) \sim 0.075$,][]{schlegel98}. A growing
body of observational data, based on dedicated photometric surveys, has been made 
available to the community: the Magellanic Clouds Photometric Survey 
\citep[MCPS,][]{zaritsky04}, the Two Micron All Sky Survey \citep[2MASS,][]{skrutskie06},
the Deep Near Infrared Survey of the Southern Sky \citep[DENIS,][]{epchtein94}, 
Surveying the Agents of a Galaxy's Evolution Survey \citep[SAGE--LMC with the 
{\it Spitzer} Space Telescope,][]{meixner06}, and {\it HERschel} Inventory of the Agents of
Galaxy Evolution \citep[HERITAGE,][]{meixner10, meixner13}. 

Additional data allowed to reconstruct the Star Formation History (SFH) of the LMC
\citep{harris09, weisz13}, and the age--metallicity relation \citep[AMR,][]{carrera08, piatti13}. 
These results, in combination with the models of dust production by AGBs currently
available, were used to determine the dust production rates by AGB stars, to be 
compared with the observations of evolved stars in the LMC \citep{zhukovska13, raffa14}.

The many data available, particularly in the infrared bands, where most of the emission
from dust--enshrouded stars occurs, stimulated a series of investigations, with the scope
of interpreting the observed colour--colour and color--magnitude diagrams,
based on the reprocessing of the radiation emitted from the central star by dust particles
present in the circumstellar envelope \citep{srinivasan09, srinivasan11, boyer11, sargent11}. 
These works are based on a wide exploration of the various quantities
relevant for the determination of the spectrum of a single object (effective temperature, 
surface gravity, luminosity, optical depth, size of the individual dust species formed),
in the attempt of selecting the combinations of parameters allowing the best agreement
with the observations, and to further refine the suggested classification of dust obscured
stars in the LMC \citep{cioni06, blum06}.

In the present work we tackle this problem from a different point of view. Our goal is to provide 
a full and exhaustive interpretation of the observed distribution of dust obscured AGB stars 
in the LMC, in the various colour--colour and colour--magnitude diagrams (CCD and CMD, respectively)
obtained with the Spitzer Space Telescope bands. In particular, we use the bands of the InfraRed 
Array Camera (IRAC: 3.6, 4.5, 5.8 and 8.0 $\mu$m) and the $24\mu$m band of the Multiband 
Imaging Photometer (MIPS). This choice allows a detailed analysis of the properties of the
individual star+dusty envelope systems, whose Spectral Energy Distribution (SED) peaks in
the infrared.

Our theoretical description is based on a complete modelling of the AGB phase, that also 
account for dust formation in the wind, published in \citet{paperI, paperII, paperIV}.

This work represents an important step forward compared to the previously mentioned investigations,
for the following reasons:  i) the choice of the parameters
relevant for the determination of the synthetic spectra is not free, rather it is 
based on the results coming from AGB modelling;
ii) the mass distribution of the stars is calculated 
based on the LMC SFH by \citet{harris09}, according to the
evolutionary times of the individual stars;
iii) we do not use a single metallicity, rather we account for the distribution among
different chemistries given by \citet{harris09}, considering models with
$Z=10^{-3}$, $Z=4\times 10^{-3}$, $Z=8\times 10^{-3}$; iv) points (ii) + (iii) allow not
only a qualitative comparison between the observed colours of the individual stars
and the expectations from the evolutionary tracks, but also
a quantitative, statistical analysis, based on the relative number of stars
populating different zones of the CCD and CMD considered.

A first step in this direction was recently made by \citet{flavia14}, who analysed the group 
of dust obscured stars in the LMC classified as "extreme" \citep{blum06}; following
\citet{srinivasan11}, they interpreted their distribution in the colour--colour 
($[3.6]-[4.5]$, $[5.8]-[8.0]$) diagram using $Z=8\times 10^{-3}$ models of AGB evolution 
+ dust production in the winds. Here we extend this work further, 
by considering a much larger sample of over 6000 AGB candidates, 
and adopting the appropriate metallicity distribution suggested by \citet{harris09}.

The present investigation proves important not only to interpret the properties of
AGB stars in the LMC; it constitutes also an important test for the AGB modelling and for the
description of the dust formation process in the winds of AGBs.

The paper is organized as follows: section 2 gives the numerical and physical input adopted 
to build the evolutionary sequences and to describe dust formation in the winds; an 
overview on the main evolutionary properties of AGBs is given in section 3; we propose a
classification of AGBs based on their position in the colour--colour and colour--magnitude 
diagrams obtained with the Spitzer filters in section 4. Section 5 presents 
an overview of the observations of dust obscured AGBs in the LMC and the synthetic
modelling used for the population synthesis. The interpretation of the data is addressed 
in section 6, while in section 7 we suggest the evolutionary status of the stars in 
pre--existing classifications. Section 8 is focused on the expected dust production 
rate by AGBs. The conclusions are given in section 9.

\section{Numerical and physical inputs}
\label{inputs}
To produce the synthetic diagrams to be compared with the observed distribution of
LMC AGB stars in the different colour--colour and colour--magnitude diagrams 
in the Spitzer bands we used stellar evolution models with a detailed description of the 
AGB phase. The formation and growth of dust grains was
described following the isotropic expansion of the stellar wind, from the stellar surface.
The magnitudes in the 5 IRAC and MIPS bands were calculated by convolution of the synthetic spectra
with the instrumental transmission curves of the different filters.
In the following of this section we give more details of the steps required.

\subsection{Stellar evolution modelling}
\label{agbmodel}
The stellar evolution models were calculated by means of the ATON code for stellar
evolution. The interested reader may find in \citet{ventura98} a detailed description of
the numerical and physical input of the code; the most recent updates, concerning in
particular the adopted cross--sections for the reactions included in the nuclear
network, can be found, e.g., in \citet{ventura09}.
The evolutionary sequences were calculated from the pre--main sequence phase until 
the almost complete loss of the star's hydrogen envelope (the end of the AGB phase).
To follow the different populations present in the LMC we considered three
different sets of models, with metallicity of $Z=10^{-3}$, $Z=4\times 10^{-3}$ and 
$Z=8\times 10^{-3}$. In the first case we adopted an $\alpha-$enhancement $[\alpha/Fe]=+0.4$,
whereas for the two more metal--rich models we used $[\alpha/Fe]=+0.2$. The initial 
abundances of elements other than $\alpha$-elements and carbon are scaled from the 
Solar abundance \citep{gs98}, accordingly to the Fe abundance.

The physical properties of the three sets of models, and the variation of their surface
elemental abundance as they evolve on the AGB, are extensively discussed in the previous papers
published by our group \citep{ventura13, ventura14}. In Table 1 we report the 
evolutionary timescales of the various stellar models considered, in terms of the duration of the
core hydrogen--burning phase ($\tau_{ev}$) and of the whole AGB phase ($\tau_{AGB}$). The
latter timescale was determined by considering the time elapsed from the phase following the exhaustion
of central helium to the total loss of the external mantle. 

We briefly recall the main physical and chemical input, most relevant for the 
description of the AGB phase.

\begin{itemize}

\item{The temperature gradient in regions unstable to convective motions was found by
means of the full spectrum of turbulence (FST) model developed by \citet{cm91}. As extensively 
discussed by \citet{vd05}, use of the FST scheme leads to strong HBB
in models with initial mass above $\sim 3M_{\odot}$; the ignition of HBB favours a severe
destruction of the surface carbon \citep{vd05}, that prevents the star from reaching the 
C--star stage.}

\item{In convective regions, mixing of chemicals and nuclear burning are self--consistently
coupled via a diffusive approach \citep{cloutmann}. Overshoot from convective borders is
simulated by an exponential decay of convective velocities from the neutrality layer,
fixed via the Schwartzschild criterion \citep{freytag}. During the core H- and He--burning 
phases we assumed an e--folding distance for the exponential decay of velocities from the 
border of the core of $0.02H_p$, where $H_p$ is the pressure scale height; this was
obtained by comparison of the theoretical isochrones with open clusters colour--magnitude
diagram by \citet{ventura98}. During the AGB phase some overshoot is assumed from 
the borders of the convective shell that develops during the thermal pulse and from the
base of the convective envelope; in this case the e--folding distance is $0.002H_p$,
following the calibration aimed at reproducing the luminosity function of carbon 
stars in the LMC, given in \citet{paperIV}.
}

\item{Mass loss was described following the description by \citet{blocker95} in all 
the AGB phases before the C--star phase begins. This treatment, based on hydrodynamic 
simulations by \citet{bowen88}, assumes a steep dependence of mass--loss on luminosity 
($\dot M \sim L^{4.7}$). Consequently, stars experiencing HBB are expected to loose mass 
at large rates. For carbon stars we used the formula giving the mass 
loss rate as a function of luminosity and effective temperature for the 
LMC, from the Berlin group \citep{wachter02, wachter08}. This treatment is based on pulsating
hydrodynamical models, in which mass loss is driven by radiation pressure on dust grains.}

\item{The molecular opacities in the stellar surface layers (temperatures below 10,000K) were
calculated by means of the AESOPUS tool \citep{marigo09}, which is necessary because of
the increase in the opacity of the external regions when the mixture becomes enriched 
in carbon; this is mainly due to the formation of CN molecules, occurring when the C/O ratio 
becomes larger than unity \citep{marigo02}. The tables generated with the AESOPUS code
are available in the range of
temperatures $3.2 \leq \log T \leq 4.5$. The reference tables assume the same initial 
composition of the models used in the present work. For each combination of 
metallicity and $\alpha-$enhancement, additional tables are generated, in which the reference
mixture is altered by varying the abundances of C, N and O. This step is done by introducing 
the independent variables $f_C$, $f_N$, $f_{CO}$, that correspond to the enhancement (in comparison with the initial stellar chemistry) of carbon, nitrogen and of the C/O ratio, respectively. In each stellar layer the opacity is found via
interpolation on the basis of the value of temperature and density and of the mass
fractions of the CNO elements. \citet{vm09, vm10} showed that once the star
enters the C--star phase the surface layers expand, which accelerates loss of the 
remaining hydrogen envelope, owing to the increase in the rate at which mass loss occurs.}

\end{itemize}

\subsection{Dust formation in the winds of AGBs}
\label{dustmodel}
The winds of AGBs are rather cool, given the small effective temperatures
($T_{eff}$ below $\sim 4000$K); in these cool winds dust particles can form more easily
in regions sufficiently close to the surface of the star (typically at distances in the 
range $1-10R_*$, where $R_*$ is the stellar radius), where the densities are large enough 
to allow atoms and molecules to condense into dust grains. Towards the end of their 
evolution, the effective temperature of AGBs decreases considerably, as a consequence of the
loss of their external envelope; this effect is stronger in carbon stars,
owing to the gradual enrichment in carbon of their surface layers, that leads to a general
cooling of the whole zone close to the surface \citep{marigo02}. Therefore, in the final phases of their 
life AGBs suffer very large mass loss rates, which also contributes to form large quantities 
of dust, because more gas particles are available for condensation. In a self--consistent approach, mass--loss should be 
calculated based on the dust formed, and the interaction of dust particles with radiation 
pressure. However, in the simple schematization adopted here, mass loss is assumed a priori, 
as a boundary condition. An interested reader may find in \citet{paperIV} an exhaustive 
discussion on this limitation.

In the present work the growth of dust particles is calculated with a simple
model for the stellar wind. The outflow is assumed spherically symmetric, and
to have an initial expanding velocity from the surface of the star of $v=10^5 cm/sec$.
The description is done via a set of equations describing the rate of growth of each 
individual dust species (starting from the innermost point where it forms). These
relations are completed by mass and momentum conservation. Mass conservation allows to determine
the radial density of the gas \citep[see e.g. Eq.2 in][]{paperIV}.
The equation of momentum conservation considers the acceleration to the gas particles, 
due to effects of the radiation pressure on the already formed grains; it allows the description
of the increase of the velocity of the wind after dust formation begins, until an
asymptotic value is reached \citep[typically from 5 to 30 $km/s$ as observed in AGB stars;][]{knapp85}. 
The necessary input for
the description of the wind is the evolutionary status of the central object: 
the effective temperature, luminosity, mass--loss rate, and surface chemical
composition, particularly the C/O ratio. An interested reader may find in 
\citet{paperI, paperII, paperIII, paperIV} all the relevant equations,
and the discussion of the various uncertainties affecting the robustness of this approach,
which is based on the pioneering explorations by the Heidelberg group 
\citep{gs85, gs99, fg01, fg02, fg06, zhukovska08}, and was also adopted in more recent
investigations by \citet{nanni13a, nanni13b, nanni14}.

A common result from all these investigations is that in the winds of oxygen--rich stars
formation of silicates occurs, at a distance that, depending on the effective 
temperature, is in the range $d \sim 5-10R_*$ from the stellar surface; also small quantities of 
solid iron are present. \citet{flavia14b} find that the most stable
dust species (alumina dust, $Al_2O_3$) in oxygen--rich stars form at a typical distance of $d \sim 1-3$ stellar radii
from the stellar surface. The quantity of alumina dust formed is extremely sensitive to the
amount of aluminium in the circumstellar envelope.
Silicon is more abundant than aluminium at the surface of AGBs. Therefore, silicates, 
though less stable, form in larger quantity compared to alumina grains and represent the 
dominant compound produced by oxygen--rich stars; in addition, alumina is highly transparent to 
radiation, thus the wind is only scarcely accelerated by radiation pressure on $Al_2O_3$
grains. 

In carbon--rich environments we find a similar situation \citep{flavia}, with a stable and extremely 
transparent dust species, silicon carbide (SiC), forming close to the surface of the 
star ($d \sim 1-2R_*$), surrounded by a more external region ($d \sim 5R_*$), where solid 
carbon grains form and grow \citep[see Fig. 4.2 in][]{flavia}. The latter species form in much 
greater quantities than SiC, owing to the much larger availability of carbon atoms in 
comparison to silicon. Also in this case some solid iron can form. 

To calculate the extinction coefficients entering the momentum equation and the synthetic spectra, 
as described in next section, we used the following
optical constants: for silicates, we used results from \citet{ossenkopf92}; for 
corundum, we used \citet{koike95}; for iron, we adopted the optical constants by
\citet{ordal88}; for SiC, we based on the compilation by \citet{pegourie88}; finally,
results from \citet{jaeger94} were used for solid carbon.

Formation of silicates (oxygen--rich stars) and carbon (C--stars) particles
favours a strong acceleration of the wind, whose velocity reaches rapidly an asymptotic
behaviour. On the mathematical side, this allows to limit the computations to a region
sufficiently far from the surface of the star ($d \sim 100R_*$), beyond which the velocity 
of the wind and the size of the dust species considered undergo only minor changes.

\subsection{Synthetic spectra}
\label{spectramodel}
The modelling of the AGB phase, discussed in section \ref{agbmodel}, allows to determine
the variation with time of the main physical quantities of the central star, namely the
luminosity, effective temperature, surface gravity, rate at which mass--loss occurs. At the
same time we follow the variation of the surface elemental abundances, for the species
included in our network (from hydrogen to silicon). Because the nucleosynthesis
in AGBs is limited to species lighter than silicon, we assume that the iron content remains
unchanged during the whole evolution.
We use all these ingredients, as also the results of dust formation models,
to calculate synthetic spectra, for some selected points along
the evolutionary tracks; this step allows to determine the magnitudes in the various bands,
via convolution with the corresponding transmission curves.

\begin{figure*}
\begin{minipage}{0.47\textwidth}
\resizebox{1.\hsize}{!}{\includegraphics{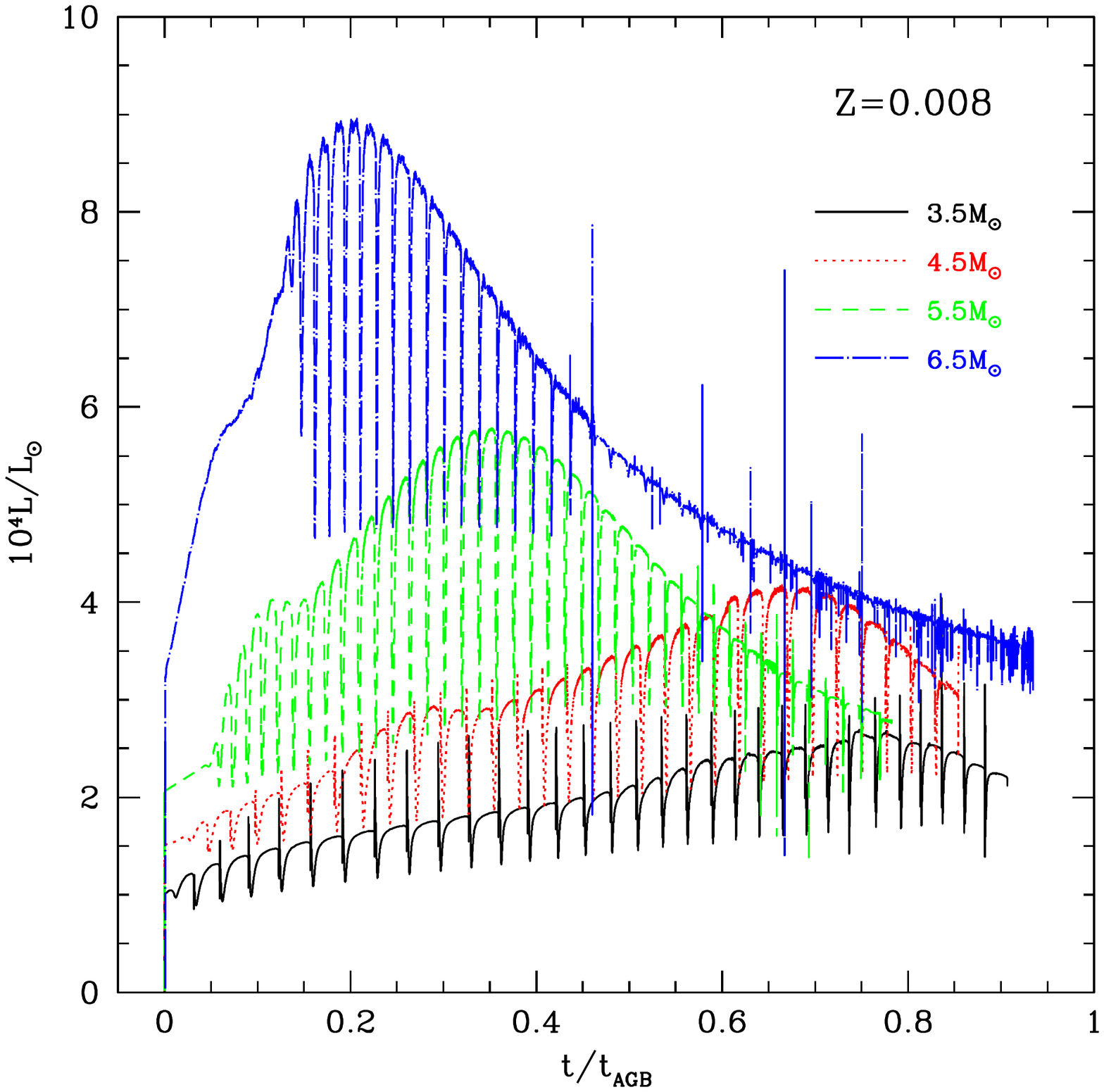}}
\end{minipage}
\begin{minipage}{0.47\textwidth}
\resizebox{1.\hsize}{!}{\includegraphics{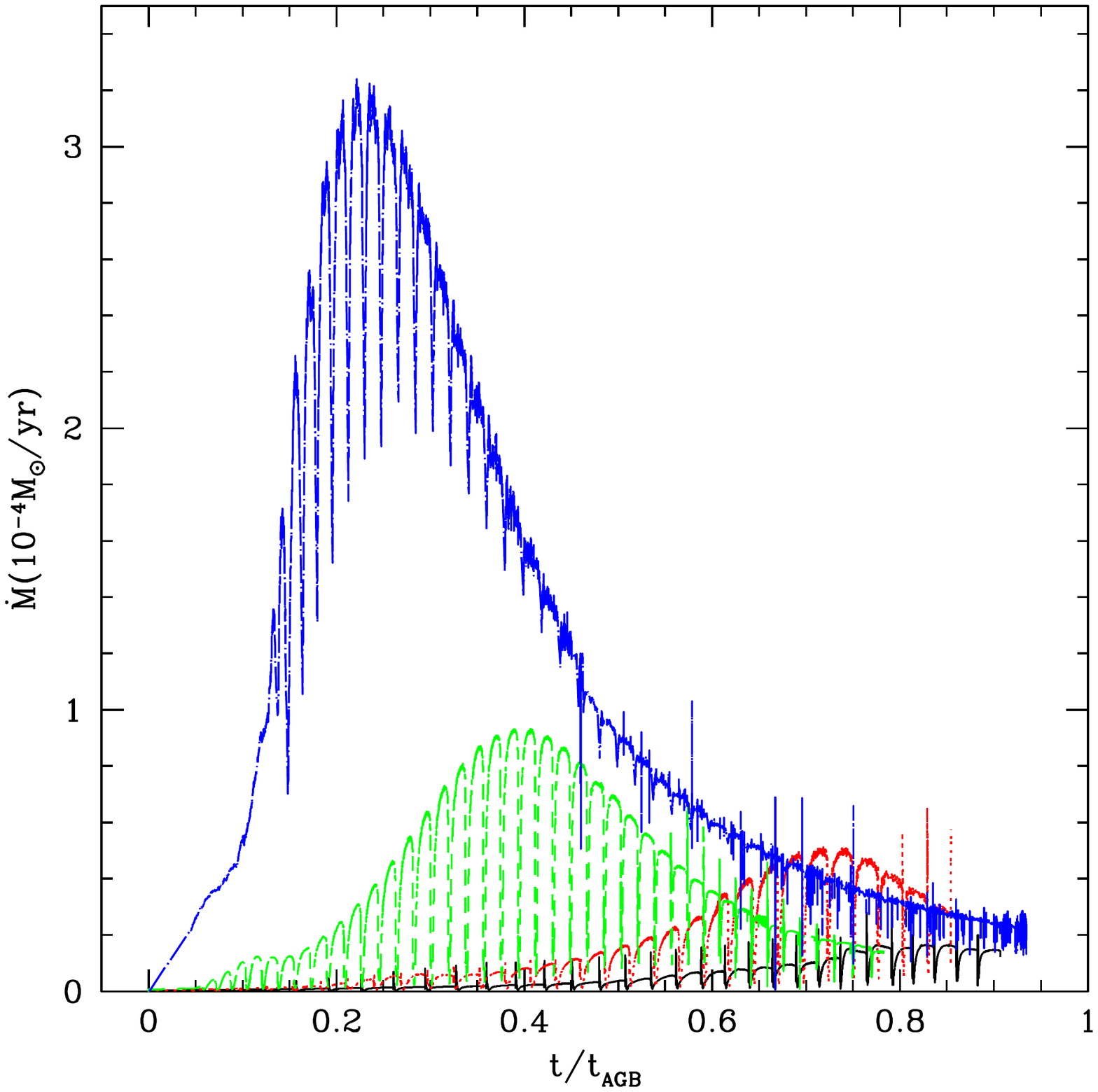}}
\end{minipage}
\vskip-50pt
\caption{The variation of the luminosity (left) and of the mass loss rate (right)
of models of different initial mass during the AGB
phase. Times are normalised to the total duration of the AGB phase for each mass,
indicated in Table 1. The various tracks refer to models of metallicity $Z=8\times 10^{-3}$ 
and initial mass $3.5M_{\odot}$ (black, solid track) $4.5M_{\odot}$ (red, dotted), 
$5.5M_{\odot}$ (green, dashed) and $6.5M_{\odot}$ (blue, dotted--dashed).
}
\label{fhbb}
\end{figure*}

The synthetic spectra were calculated in two steps, by means of the code DUSTY 
\citep{dusty}.

\begin{enumerate}

\item{For carbon stars, we first considered the region from the surface  up to the 
beginning of the solid carbon formation zone. In this preliminary step the only dust
species considered is SiC. As input radiation, we used the spectral energy distribution found by
interpolation in surface gravity, effective temperature and C/O ratios among COMARCS atmospheres
\citep{grams} of the appropriate metallicity. An analogous procedure was followed for oxygen--rich stars,
with the difference that the only species considered is alumina, and the input spectrum
is obtained by interpolating among the NEXTGEN atmospheres \citep{nextgen} of the same
metallicity; in this case the computations are extended until the formation of silicates begins.
}

\item{The spectral energy distribution obtained in step (i), emerging from the SiC (alumina) dust layer for 
C--stars (oxygen--rich stars), is used as input for the second layer, where we consider reprocessing of the 
radiation by SiC and solid carbon for C--stars (alumina and silicates for oxygen--rich stars).}

\end{enumerate}

The input necessary to steps (i) and (ii) are not assumed a priori; they
are found via the description of the dust formation in the wind, discussed in
section \ref{dustmodel}, that allows the determination of
the dust grains size and composition, the temperature of the region where the various dust species form,
and the radial distribution of the gas density.
Concerning the latter point, although we know the radial variation of the dimension
of the dust particles, we use a single grain size for each species, corresponding to the 
asymptotic value reached. This choice is motivated by the fact that the asymptotic value
is reached rapidly once the dust begins to form, thus leading to a grain size distribution
strongly peaked towards such asymptotic value. To determine the optical depth, $\tau_{10}$, we
integrate along the radial direction the product of the number density of dust particles 
and the extinction cross section, based on the knowledge of the optical constants and the
grain size.

To compare the fluxes observed with those found via our spectral analysis, we 
adopted a distance to the LMC of 50kpc \citep{feast99}.

\begin{table*}
\begin{center}
\caption{Time scales of stars of intermediate mass of different metallicity in terms of the duration of the
core hydrogen--burning phase ($\tau_{ev}$) and of the whole AGB phase ($\tau_{AGB}$). The
latter timescale was determined by considering the time elapsed from the phase following the exhaustion
of central helium to the total loss of the external mantle.} 
\begin{tabular}{c|c|c|c|c|c|c}
\hline
$M/M_{\odot}$ & $\tau_{ev}$ (Myr) & $\tau_{AGB}$ (Kyr)& $\tau_{ev}$ (Myr)& $\tau_{AGB}$ 
(Kyr) & $\tau_{ev}$ (Myr)& $\tau_{AGB}$ (Kyr)   \\ 
\hline
& $Z=10^{-3}$ & & $Z=4\times 10^{-3}$ & & $Z=8\times 10^{-3}$ &  \\
\hline
1.00  &  5400   &  1600   &  5700   &  1160  &  8020  &  1500   \\ 
1.25  &  2500   &  1520   &  2690   &  1240  &  3270  &  1520   \\ 
1.50  &  1480   &  1520   &  1630   &  1760  &  1890  &  1510   \\ 
2.00  &  700    &  2000   &  768    &  4480  &  888   &  1500   \\ 
2.50  &  410    &  880    &  566    &  1570   &  690   &  3000   \\ 
3.00  &  279    &  450    &  282    &  611   &  413   &  990    \\ 
3.50  &  196    &  360    &  235    &  315   &  268   &  450    \\ 
4.00  &  146    &  280    &  170    &  237   &  190   &  290    \\ 
4.50  &  130    &  200    &  130    &  197   &  141   &  210    \\ 
5.00  &  92     &  140    &  102     &  152   &  111   &  175    \\ 
5.50  &  74     &  80     &  73     &  55    &  89    &  150    \\ 
6.00  &  63     &  65     &  70     &  49    &  74    &  80     \\ 
6.50  &  53e    &  50     &  59     &  51    &  62    &  80     \\ 
7.00  &  47     &  27     &  51     &  47    &  53    &  35     \\ 
7.50  &  41     &  15     &  44     &  43    &  46    &  35     \\ 
\hline
\end{tabular}
\end{center}
\label{tabtimes}
\end{table*}

\section{Evolutionary properties and Spitzer colours of AGB stars}
The spectra of AGBs change during their evolution, owing to variation 
in effective temperatures, luminosities, and the 
composition and mass of dust grains in the circumstellar envelopes. The C/O 
ratio is the relevant quantity to assess whether carbon dust or silicates form.

The surface content of carbon and oxygen changes during the AGB phase under
the effects of the two physical processes that are able to alter the elemental 
abundance of the circumstellar envelope: HBB and third dredge--up 
(TDU) \citep{herwig05}. In the first case we have the activation of an advanced 
proton--capture nucleosynthesis at the bottom of the convective envelope \citep{renzini81, blocker91}. 
The elemental abundances change via enhanced CN--cycle, with 
the increase in the nitrogen content at the expenses of carbon and (for temperatures above
$\sim 50-60$MK) oxygen. The third dredge--up, occurring immediately 
after each TP, consists in the inwards penetration of the convective envelope, down 
to regions of the star previously touched by $3\alpha$ nucleosynthesis: the main effect 
of TDU is the increase in the surface carbon.

We analyze separately the main evolutionary features of AGB models dominated
by either mechanisms.

\begin{figure*}
\begin{minipage}{0.47\textwidth}
\resizebox{1.\hsize}{!}{\includegraphics{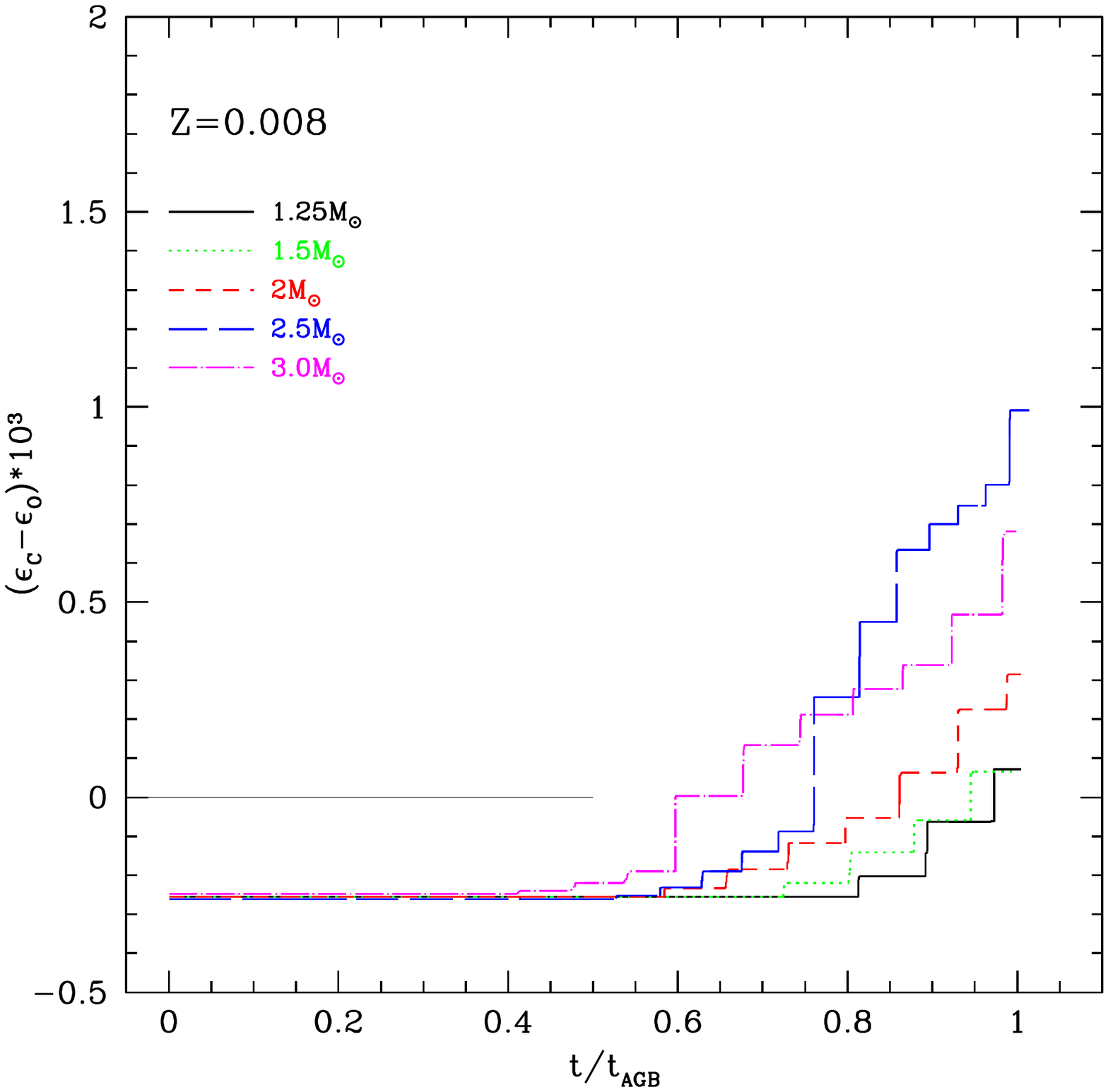}}
\end{minipage}
\begin{minipage}{0.47\textwidth}
\resizebox{1.\hsize}{!}{\includegraphics{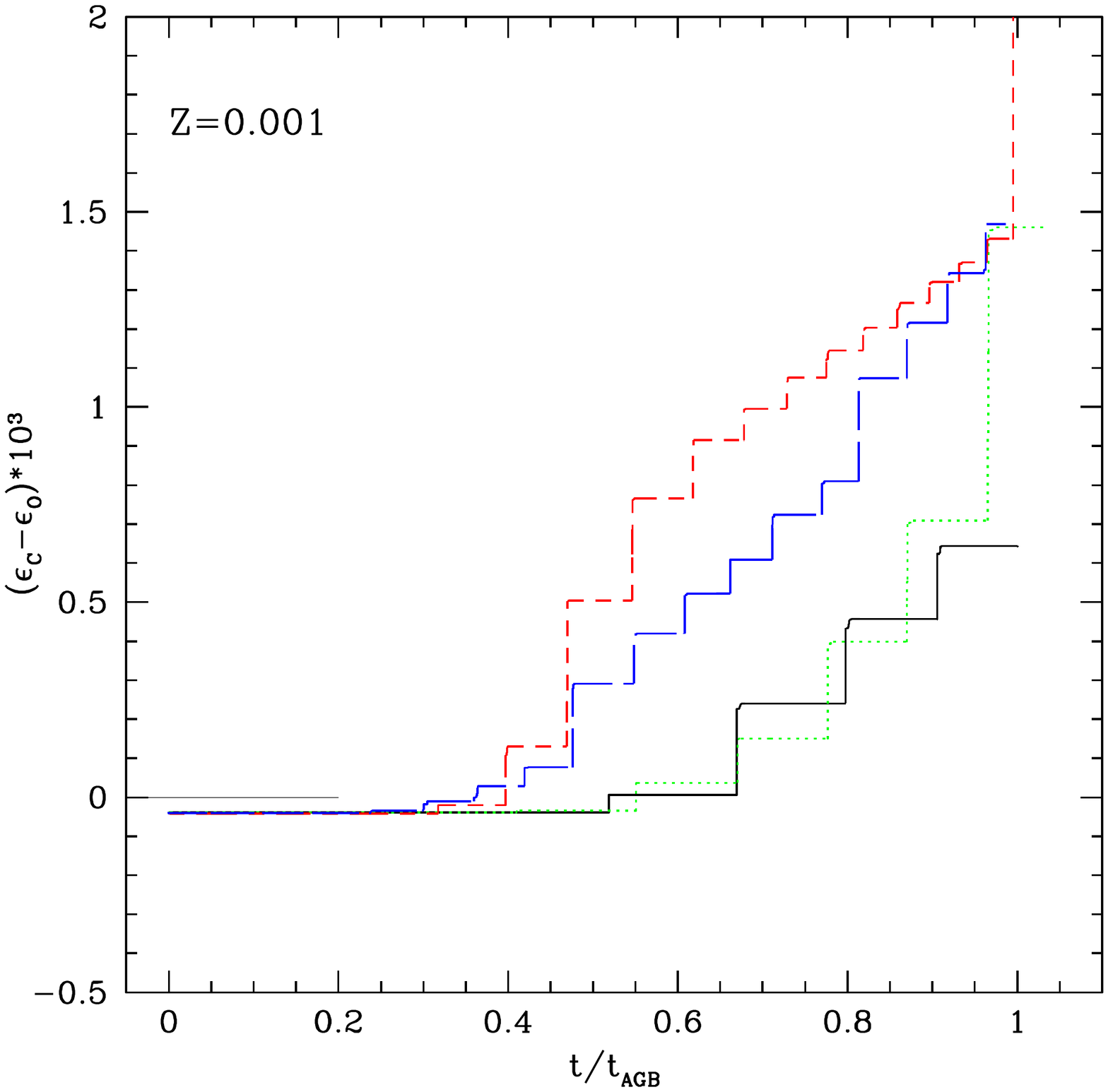}}
\end{minipage}
\vskip-50pt
\caption{The evolution of the relative excess of carbon with respect to oxygen
(see text for definition) during the AGB phase of stars with masses in the range $1.25-3M_{\odot}$ and
metallicity $Z=8\times 10^{-3}$ (left panel) and $Z=10^{-3}$ (right). Times are normalised
to the total duration of the AGB life. Thin, horizontal lines indicate the $C=O$ condition;
tracks above these lines correspond to C--stars. The values on the vertical axis
when the abscissa is unity indicate the carbon excess reached at the very end of the 
AGB evolution, when the envelope is lost.}
\label{fcstar}
\end{figure*}

\subsection{The effects of hot bottom burning}
In stars with initial mass above $\sim 3M_{\odot}$ the occurrence of HBB keeps the C/O ratio
below unity, because carbon is destroyed via proton capture at the bottom of the convective 
envelope. This threshold mass partly depends on the metallicity, being $3M_{\odot}$ for 
$Z=8\times 10^{-3}$ and $2.5M_{\odot}$ for $Z=10^{-3}$ \citep{ventura13}. Stars of mass 
above this limit never become carbon stars, thus the only dust particles formed in their 
winds are alumina and silicates\footnote{This conclusion holds as far as convection is
modelled within the FST framework. When the traditional Mixing Length scheme is adopted,
the temperatures at the bottom of the convective envelope are smaller, the strength of
HBB is consequently reduced, thus the range of masses potentially able to reach the C--star 
stage is larger \citep{ventura09, doherty14}.}. 
The surface abundances of the key--particles for these dust species, i.e. aluminium and 
silicon, remain practically constant during the AGB phase\footnote{This holds strictly for 
stars with $Z \geq 8\times 10^{-3}$, where the temperatures at the bottom
of the surface convection zone do not allow any Mg--Al nucleosynthesis \citep{ventura13}.
For metallicities $Z \leq 10^{-3}$, massive AGBs are expected to increase the surface
aluminium by a factor 10, and to increase the silicon content by $\sim 10-20 \%$
\citep{ventura11}. However, at these low metallicities only a modest production of
dust is expected, owing to the small abundances of the key--elements required to form
dust.}, thus dust formation is mainly driven by the strength of HBB. A strong HBB favours 
a considerable increase in the luminosity and mass loss rate, that, in turn, increases 
the density of gas particles available to dust formation in the wind \citep{paperI}.

The evolution of the luminosity and mass loss rate of models with initial mass $M>3M_{\odot}$ is shown in
Fig. \ref{fhbb}. For clarity reasons, we only show the $3.5M_{\odot}$, $4.5M_{\odot}$, 
$5.5M_{\odot}$ and $6.5M_{\odot}$ cases. We focus on models with
metallicity $Z=8\times 10^{-3}$, because i) stars within this range of mass have ages
younger than $\sim 2\times 10^8$ yr, and were born in an epoch when most of the stars 
formed in the LMC have metallicity $Z \geq 4\times 10^{-3}$ (see Fig. \ref{fsfr},
showing the SFR that we use in the present work);
ii) in models of lower metallicity, owing to scarcity of silicon, dust is produced with smaller
rates in comparison to more metal--rich stars: the highest rate found in the
$Z=10^{-3}$ case is ${\dot M}_d \sim 3\times 10^{-8}M_{\odot}/yr$, whereas in the
$Z=8\times 10^{-3}$ models typical values are above 
${\dot M}_d \sim 2\times 10^{-7}M_{\odot}/yr$ \citep{paperII}.

We see in Fig. \ref{fhbb} (left panel) that models of higher mass evolve at larger luminosities; 
this is because the higher the initial mass of the star, the heavier the core mass
becomes, hence, the higher luminosity a star can reach \citep{paczynski}. 
Also, they experience a stronger HBB. The luminosity reaches a maximum during the AGB 
evolution, then decreases, because of the gradual loss of the external mantle. The higher 
the initial mass is, the quicker the star reaches the maximum luminosity.

Within the scheme we apply to describe dust formation, where the mass loss rate is adopted
as a boundary condition, we find that the amount of dust produced increases with initial 
stellar mass, because more massive models suffer higher mass loss rates (see right panel of Fig. 1).
This is a straight consequence of the mass conservation law, on the basis of which 
higher mass loss rates favour higher densities, thus more gas molecules available for
condensation.
Therefore, having more dust formed at the inner region, 
stars with higher initial masses tend to emit higher fluxes at mid--infrared wavelengths. These oxygen--rich stars 
emit the largest mid--infrared flux during the phase of maximum luminosity, when mass is 
lost at the highest rates. However, as shown in \citet{flavia14} (see their Figure 1) 
the optical depth of these models will not be
significantly decreased during the following AGB phases, because the decrease in the
mass loss rate is partly counterbalanced by the smaller effective temperatures towards the
final AGB stages, that favour the formation and growth of dust grains.

\begin{figure}
\resizebox{1.\hsize}{!}{\includegraphics{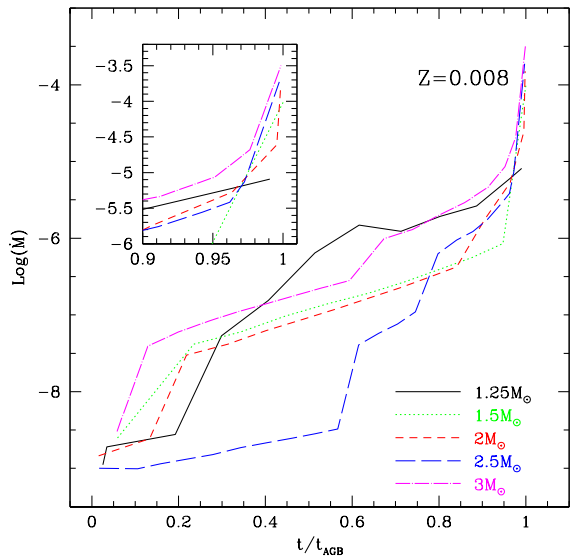}}
\vskip-50pt
\caption{The variation of the mass loss rate experienced by stars reaching
the C--star stage during the AGB evolution. The models, of metallicity
$Z=8\times 10^{-3}$, are the same shown in the left panel of Fig. \ref{fcstar}. In the inset we show a zoom of the mass loss rate at the end of the AGB phase.}
\label{fmloss8m3}
\end{figure}

\begin{figure}
\resizebox{1.\hsize}{!}{\includegraphics{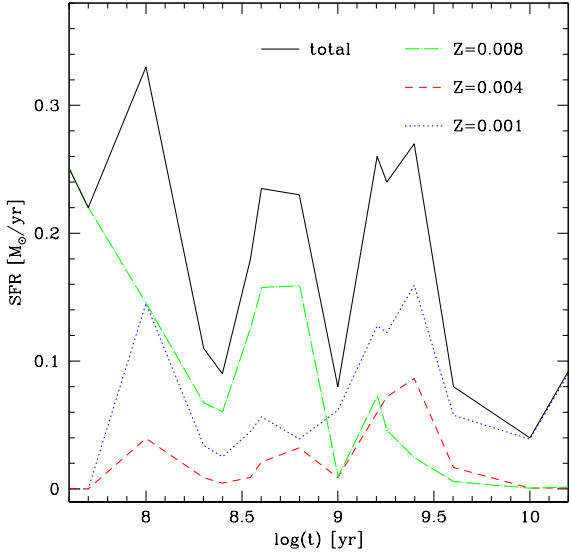}}
\vskip-60pt
\caption{The time variation of the Star Formation Rate in the LMC according to 
\citet{harris09}. The different lines give the total star formation rate
(black, solid track), and the fractional contribution of the metallicities
$Z=8\times 10^{-3}$ (green, dotted--dashed), $Z=4\times 10^{-3}$ (red, dashed), and $Z=10^{-3}$ 
(blue dotted line), as given in \citet{harris09}. The latter component here represents the cumulative 
contribution of the two $Z=10^{-3}$ and $Z=2.5\times 10^{-3}$
tracks of the original \citet{harris09} SFH.}
\label{fsfr}
\end{figure}

\subsection{The AGB evolution towards the C--star regime}
\label{cstarmod}
In stars with $M < 3M_{\odot}$ repeated episodes of Third Dredge--Up favour a gradual
increase in the surface carbon, eventually leading to the formation of carbon stars.
The possibility of reaching the C--star stage depends not only on whether HBB is active, but 
also it requires that at the photosphere, the abundance of carbon exceeds that of oxygen, 
before the hydrogen envelope is completely lost. 
The range of mass that become carbon star is $1.25M_{\odot} \leq M < 3M_{\odot}$ \footnote{Indeed 
in the $Z=10^{-3}$ case the 3$M_{\odot}$ model ignites HBB, thus restricting the range of masses 
becoming carbon stars to $M\leq2.5M_{\odot}$ }.

Stars of $M\leq 1 M_{\odot}$, never become carbon stars. They evolve as oxygen--rich stars through out 
their AGB phase. Their surface chemistry is  changed only by the first Dredge Up, occurring while ascending 
the Red Giant Branch.

In the two panels of Fig. \ref{fcstar} we show the difference between 
the number density of carbon and oxygen nuclei in AGB models with 
metallicity $Z=8\times 10^{-3}$ (left) and $Z=10^{-3}$ (right). The difference is
normalized to the density of hydrogen atoms, i.e. $\epsilon_{C,O}=n_{C,O}/n_H$.
This quantity indicates the efficiency with which the surface envelope 
is enriched in carbon, and is strongly related to the amount of solid carbon formed. 
Unlike their more massive counterparts, here we show both the highest and  the smallest 
metallicities of LMC stars. 
Indeed these low mass stars evolve slowly, with a timescale of 0.3-15 Gyrs. Within
this timescale, the LMC has formed stars with range of metallicities (see Table 1 and 
Fig. \ref{fsfr}), which we represent with three metallicity grids from $Z=10^{-3}$ to 
$8 \times 10^{-3}$. Note that \citet{harris09} also identified the $Z=2.5 \times 10^{-3}$ 
component, but we have binned this component together with the $Z=10^{-3}$ one.
All the models with initial mass $1.25M_{\odot} < M < 3M_{\odot}$
evolve initially as oxygen--rich stars, with $C/O < 1$ 
(i.e. $\epsilon_C-\epsilon_O < 0$). In the last fraction of the AGB evolution (ranging from 
$\sim 30\%$ to $70\%$, depending on the values of M and Z) they evolve as 
carbon stars. Due to the smaller initial oxygen abundance, $\epsilon_C - \epsilon_O$ 
is on the average larger for the $Z=10^{-3}$ population than for the high--metallicity
counterpart. This causes the star to reach the carbon-rich phase earlier in the evolution.
Models of higher mass ($M \geq 2M_{\odot}$) are more enriched in carbon, because they 
experience more TDU events; this trend with mass is reversed close to the limit for HBB 
ignition, because the models experience TDU episodes of smaller efficiency 
\citep{paperIV}. 

The increase in the surface carbon has an important feedback on the AGB evolution: the 
consequent increase in the molecular opacities favours the expansion of the surface 
layers, with the decrease in the surface temperatures, and the increase in the rate at
which mass loss occurs \citep{vm09, vm10}. This effect can be clearly seen in 
Fig. \ref{fmloss8m3}, showing the mass loss rate experienced by stars of metallicity
$Z=8\times 10^{-3}$, with initial mass $1.25\leq M \leq 3M_{\odot}$. We note the fast 
increase in $\dot M$ in the very latest evolutionary phases, associated with the increase 
in the surface carbon (see left panel of Fig. \ref{fcstar}).
The cooling of the external regions and the increase in the mass loss rate 
concur in forming larger quantities of carbon grains, leading to a progressive obscuration 
of the radiation from the star. At odds with the stars experiencing HBB, here the colours 
become redder and redder as the stars loose the external mantle.

\begin{figure*}
\begin{minipage}{0.33\textwidth}
\resizebox{1.\hsize}{!}{\includegraphics{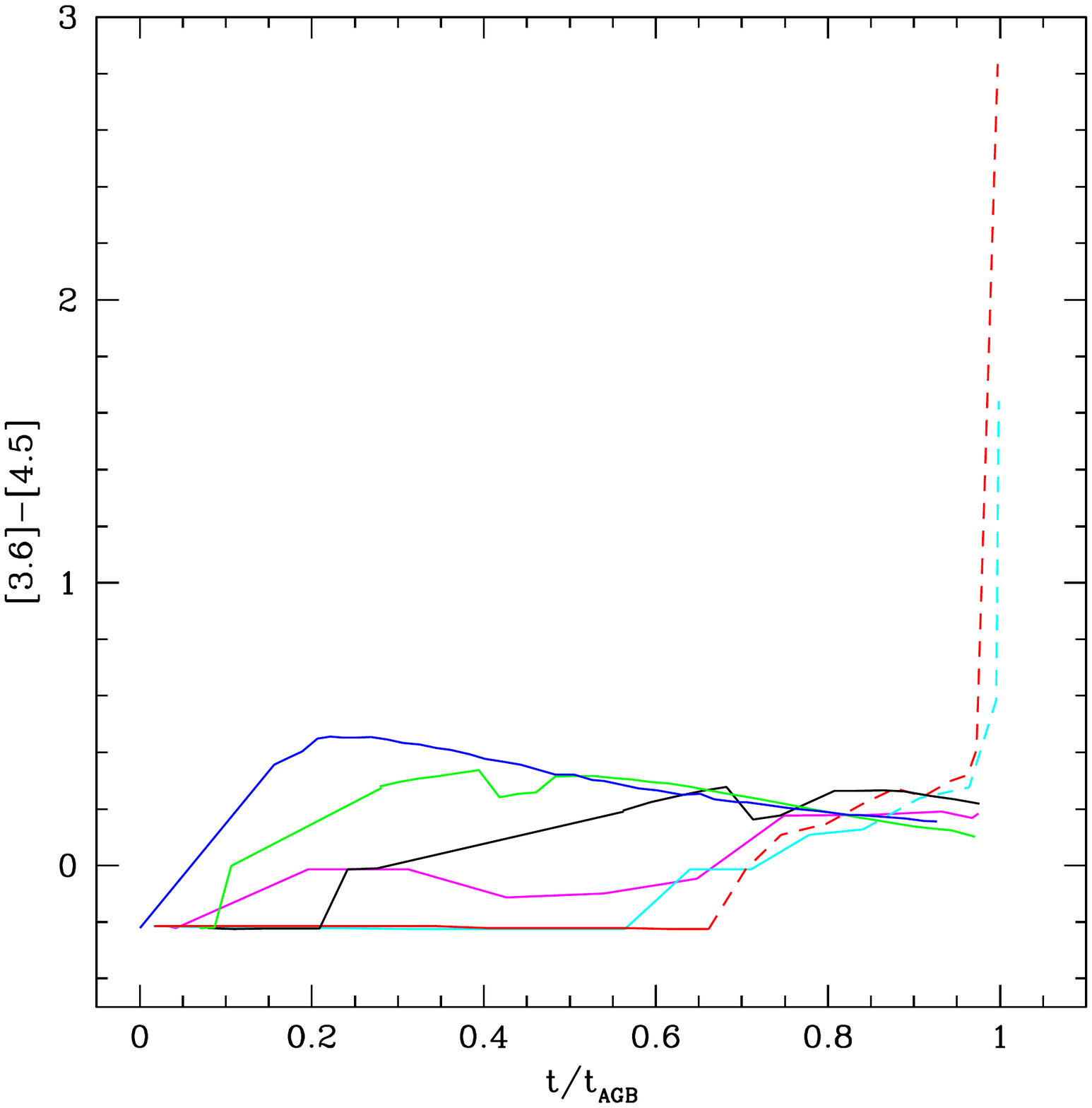}}
\end{minipage}
\begin{minipage}{0.33\textwidth}
\resizebox{1.\hsize}{!}{\includegraphics{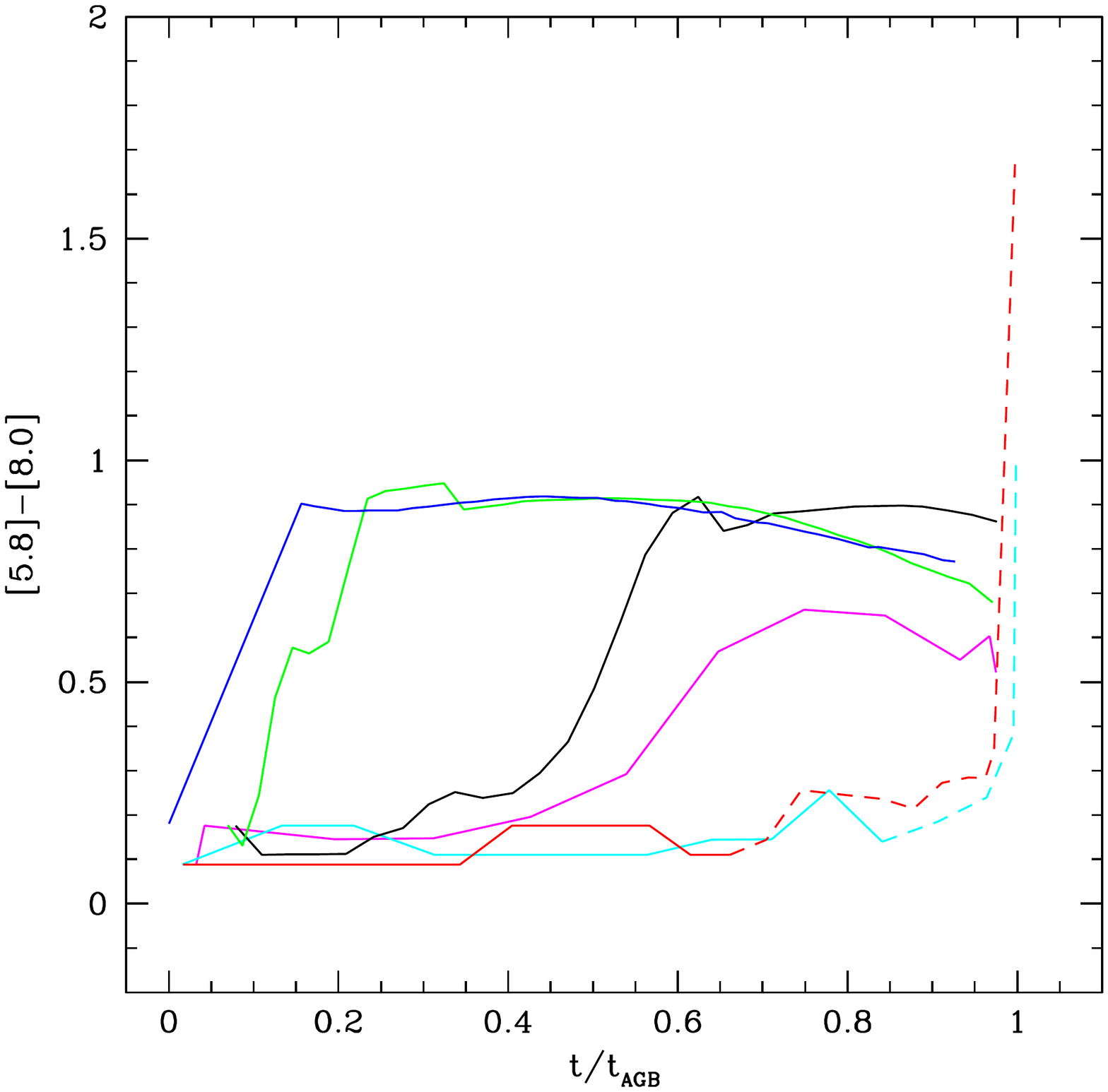}}
\end{minipage}
\begin{minipage}{0.33\textwidth}
\resizebox{1.\hsize}{!}{\includegraphics{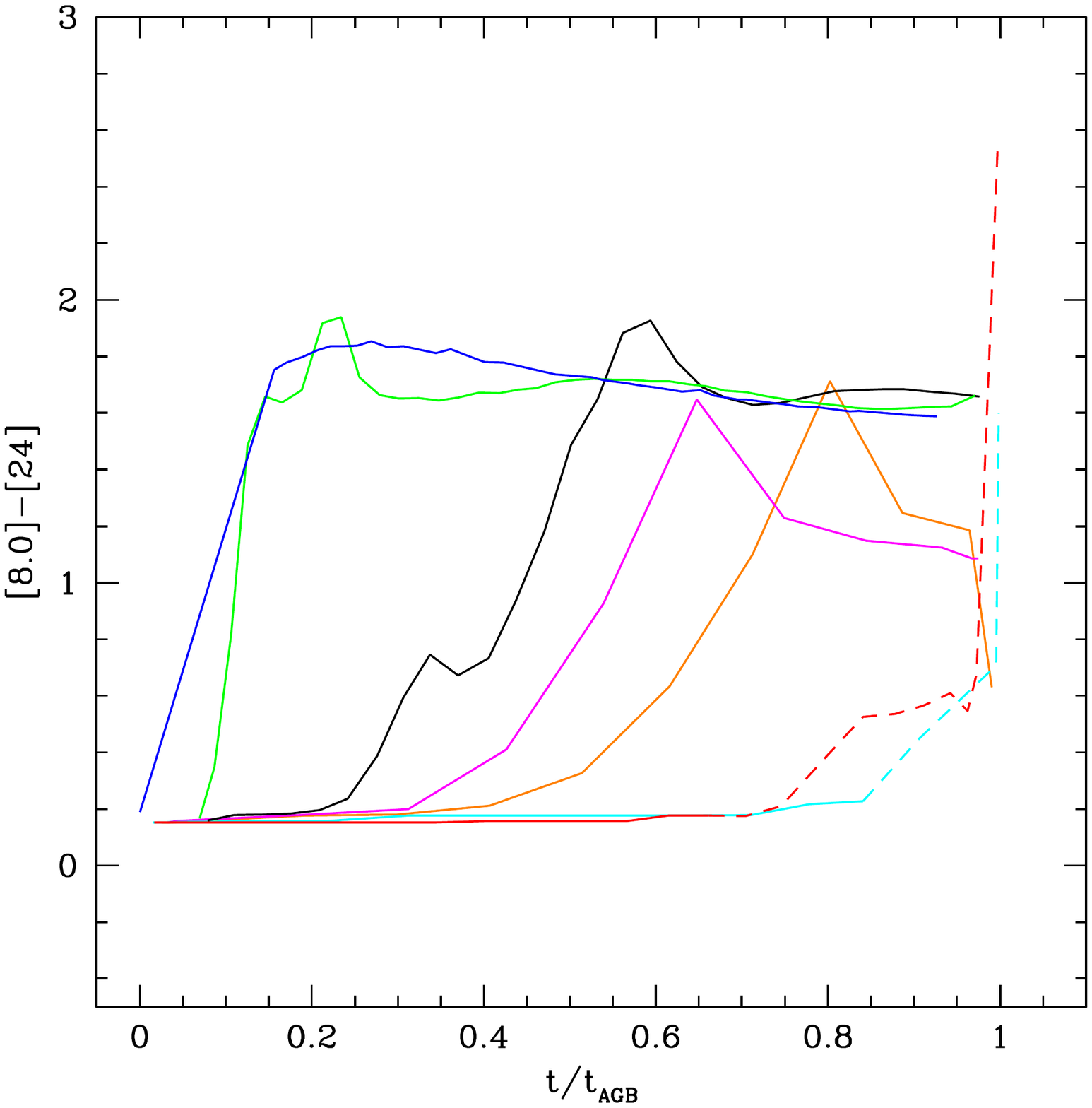}}
\end{minipage}
\vskip-30pt
\caption{The evolution of the colours $[3.6]-[4.5]$ (left panel), $[5.8]-[8.0]$ (middle),
$[8.0]-[24]$ (right) during the AGB phase of models of metallicity $Z=8\times 10^{-3}$
and initial masses $1M_{\odot}$ (magenta), $2M_{\odot}$ (cyan), 
$3M_{\odot}$ (red) $4.5M_{\odot}$ (black), $5.5M_{\odot}$ (green),
$6.5M_{\odot}$ (blue). The orange line in the right panel refers to a
$1.25M_{\odot}$ model. Dashed tracks indicate carbon--rich stars, while solid
lines indicate oxygen--rich stars. Times are normalised to the total duration of the 
AGB phase.}
\label{fcolours}
\end{figure*}

\subsection{The infrared colours of AGBs: predictions from modelling}
\label{ircolours}
Fig. \ref{fcolours} shows the variation of the $[3.6]-[4.5]$, 
$[5.8]-[8.0]$, $[8.0]-[24]$ colors of AGB models with $Z=8\times 10^{-3}$. We consider 
stellar models with initial mass of $1M_{\odot}$ (representing low--mass stars, never reaching 
the C--star stage), $2M_{\odot}$, $3M_{\odot}$ (C--stars), $4.5M_{\odot}$, $5.5M_{\odot}$, 
$6.5M_{\odot}$ (these models experience HBB). 

\begin{figure*}
\begin{minipage}{0.47\textwidth}
\resizebox{1.\hsize}{!}{\includegraphics{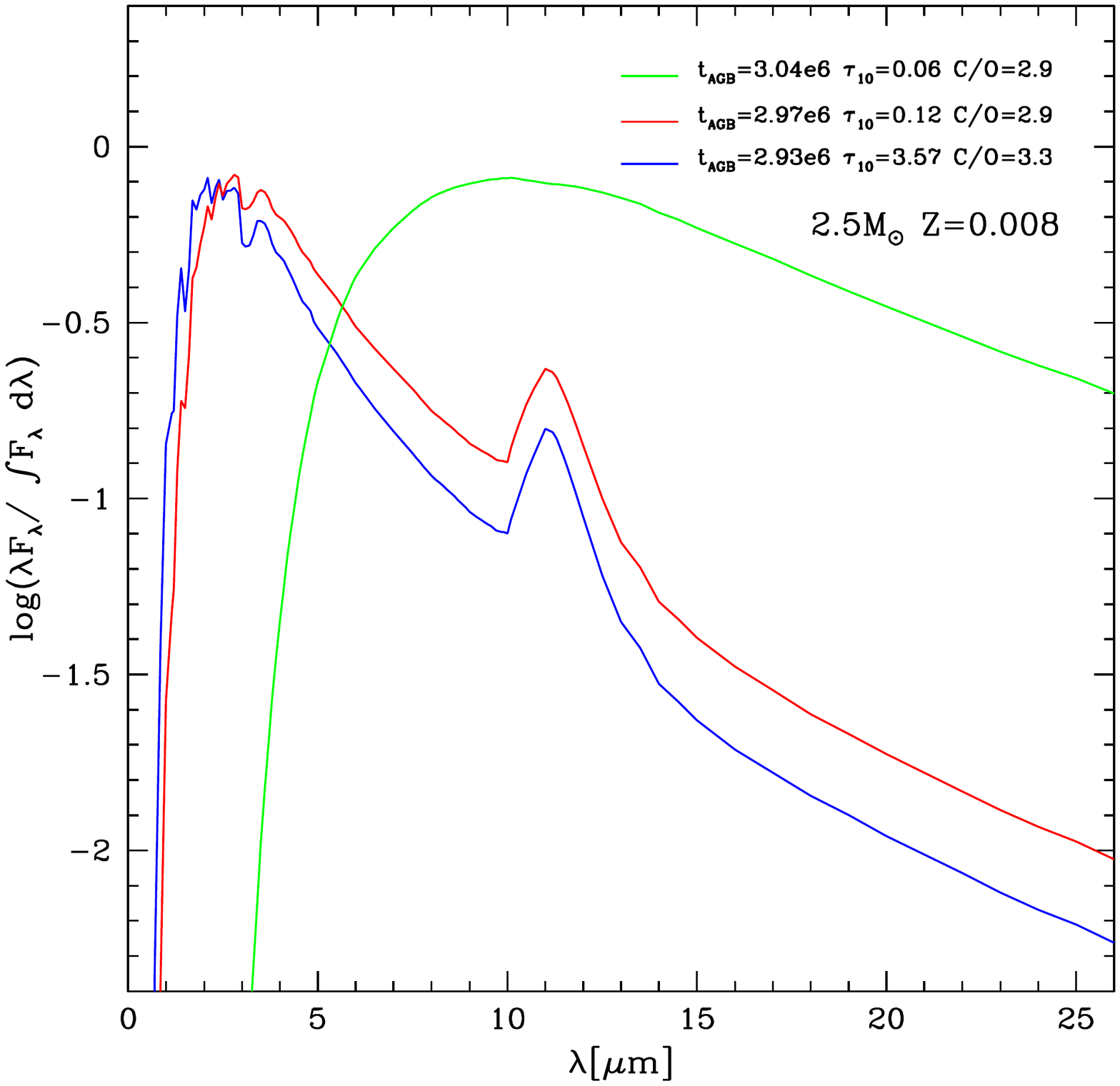}}
\end{minipage}
\begin{minipage}{0.47\textwidth}
\resizebox{1.\hsize}{!}{\includegraphics{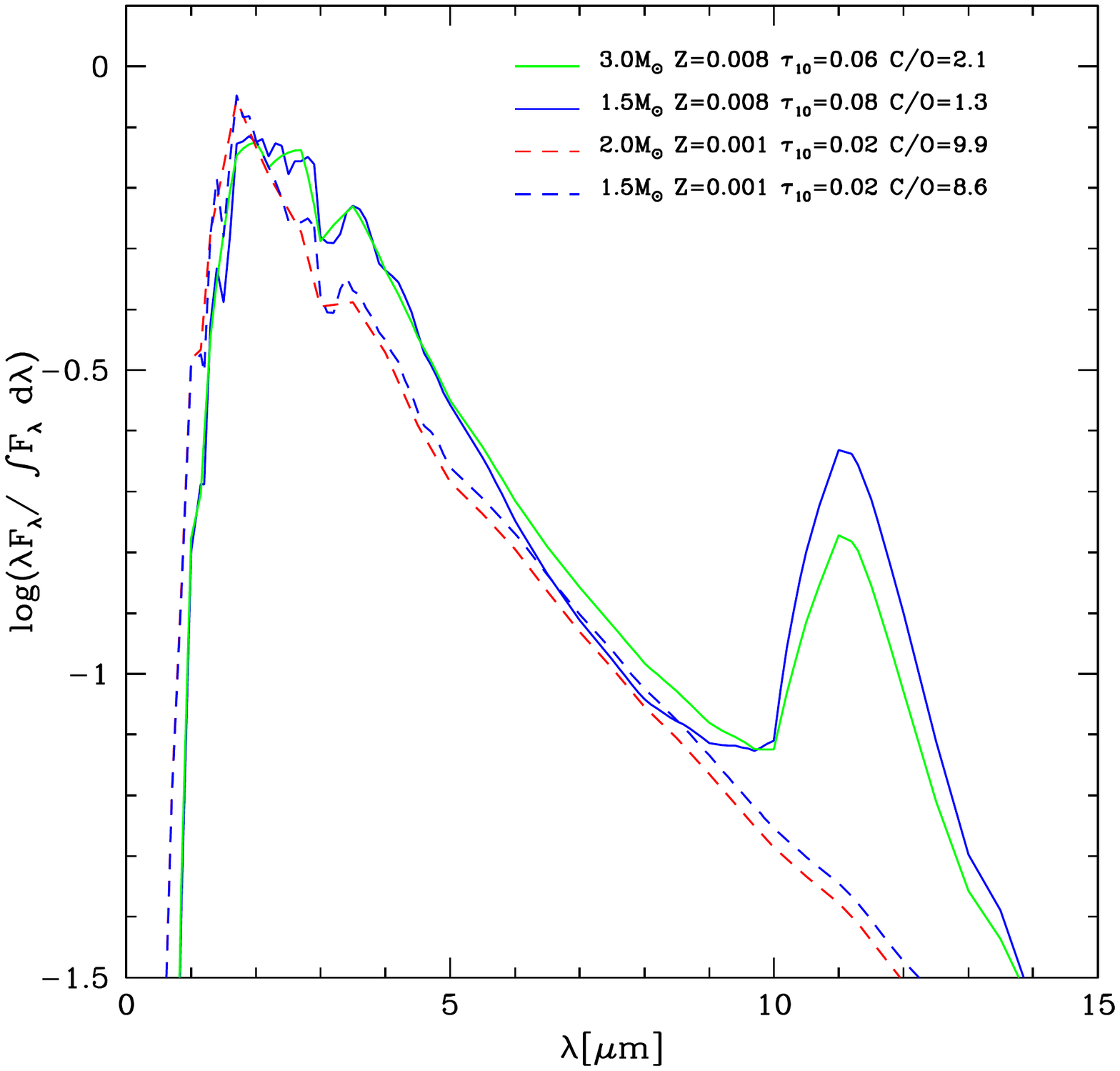}}
\end{minipage}
\vskip-50pt
\caption{Left: The change in the spectral energy distribution of a model of initial
mass $2.5M_{\odot}$ during the AGB phase. Right: The SED of models of various mass
and metallicity, taken at $[3.6]-[4.5]=0.3$.}
\label{fspettri}
\end{figure*}

In the left panel of Fig. \ref{fcolours} we note that for the $2M_{\odot}$ and 
$3M_{\odot}$ models $[3.6]-[4.5]$ starts to increase once the C--star stage is reached, as
a consequence of the progressive enrichment in carbon of the external layers. This phase
encompasses about $\sim 30 \%$ of the total AGB evolution. In the very latest 
evolutionary phases the outermost regions of the star become cooler and cooler, with
effective temperatures of the order of $\sim 2000$ K. During these phases the star 
looses mass with large rates (see Fig. \ref{fmloss8m3}), and great quantities of carbon 
dust are present in the circumstellar envelope; this, in turn, determines a strong 
obscuration of the stellar radiation. 
This is the reason for the fast increase in $[3.6]-[4.5]$, that 
reaches a maximum of $\sim 3$ for the $2.5M_{\odot}$ (not shown) and $3M_{\odot}$ models.
This sequence of events is shown for a $2.5M_{\odot}$ model in the left panel of 
Fig. \ref{fspettri}: we see that the SED is progressively shifted to longer wavelengths as 
the surface carbon increases. Note that in the final stages of the AGB phase (here 
represented by the green line) the emission SiC feature at $11.3\mu$m turns into an 
absorption feature.

\begin{figure*}
\begin{minipage}{0.43\textwidth}
\resizebox{1.\hsize}{!}{\includegraphics{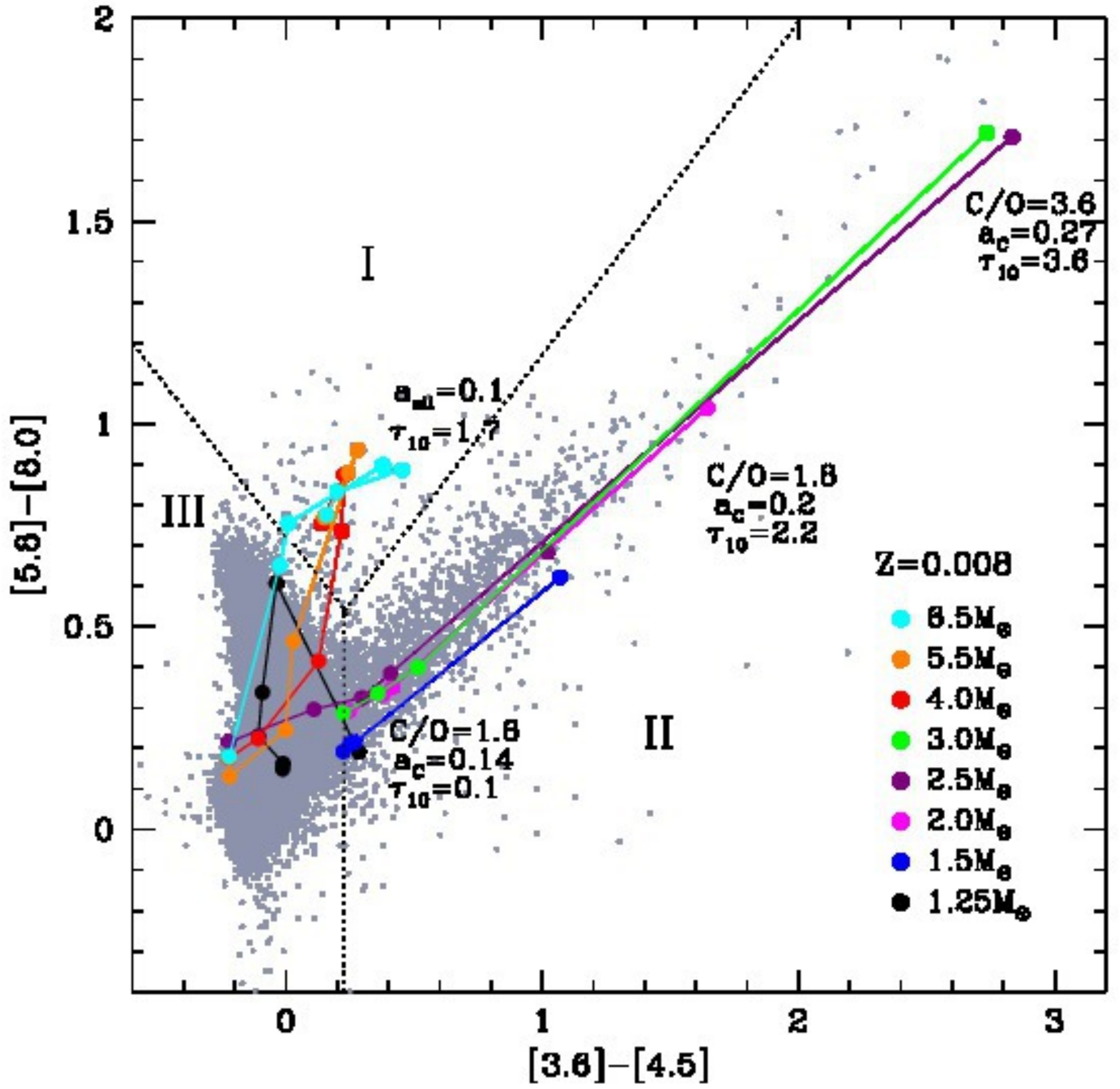}}
\end{minipage}
\begin{minipage}{0.43\textwidth}
\resizebox{1.\hsize}{!}{\includegraphics{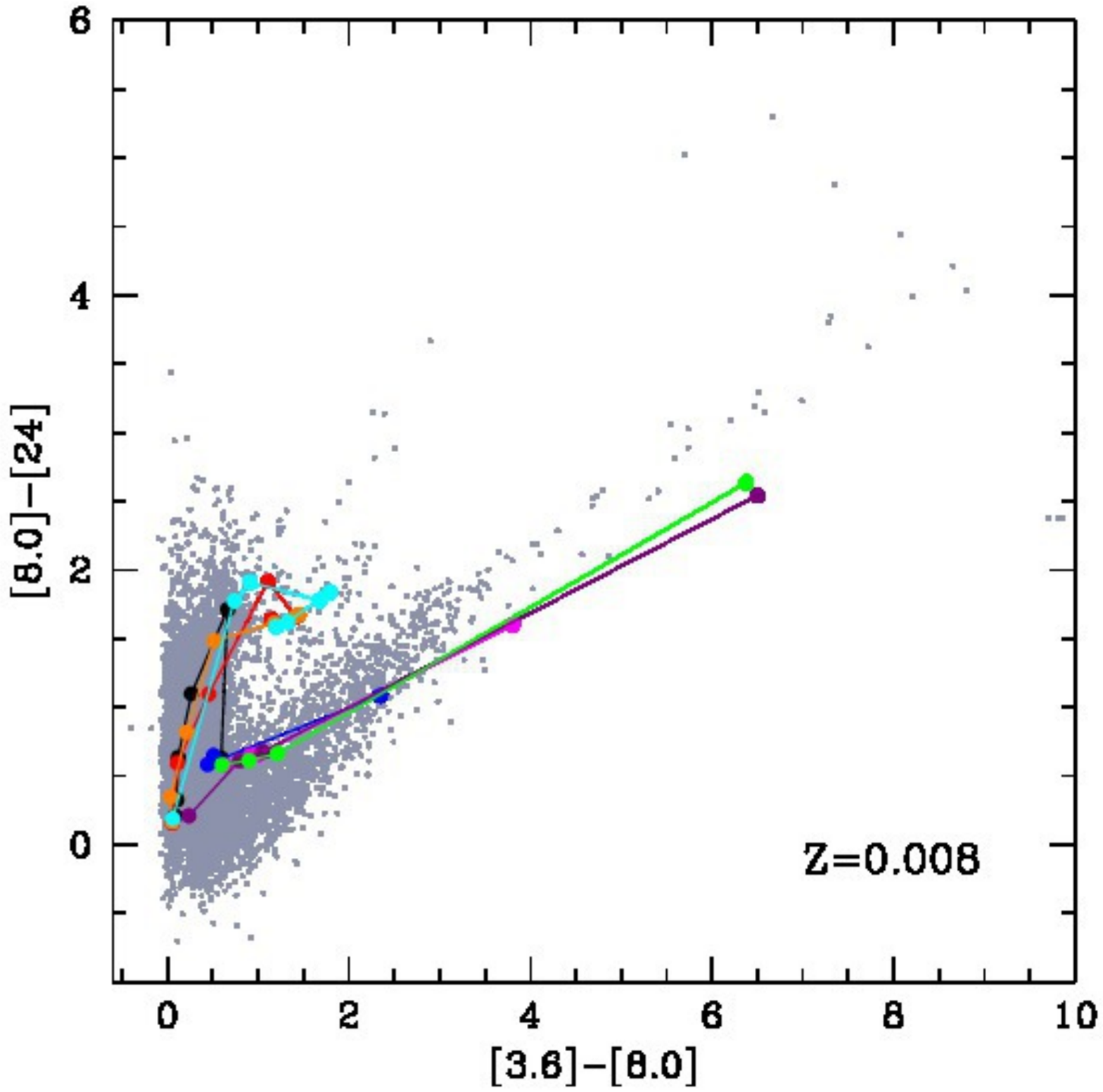}}
\end{minipage}
\vskip-50pt
\begin{minipage}{0.43\textwidth}
\resizebox{1.\hsize}{!}{\includegraphics{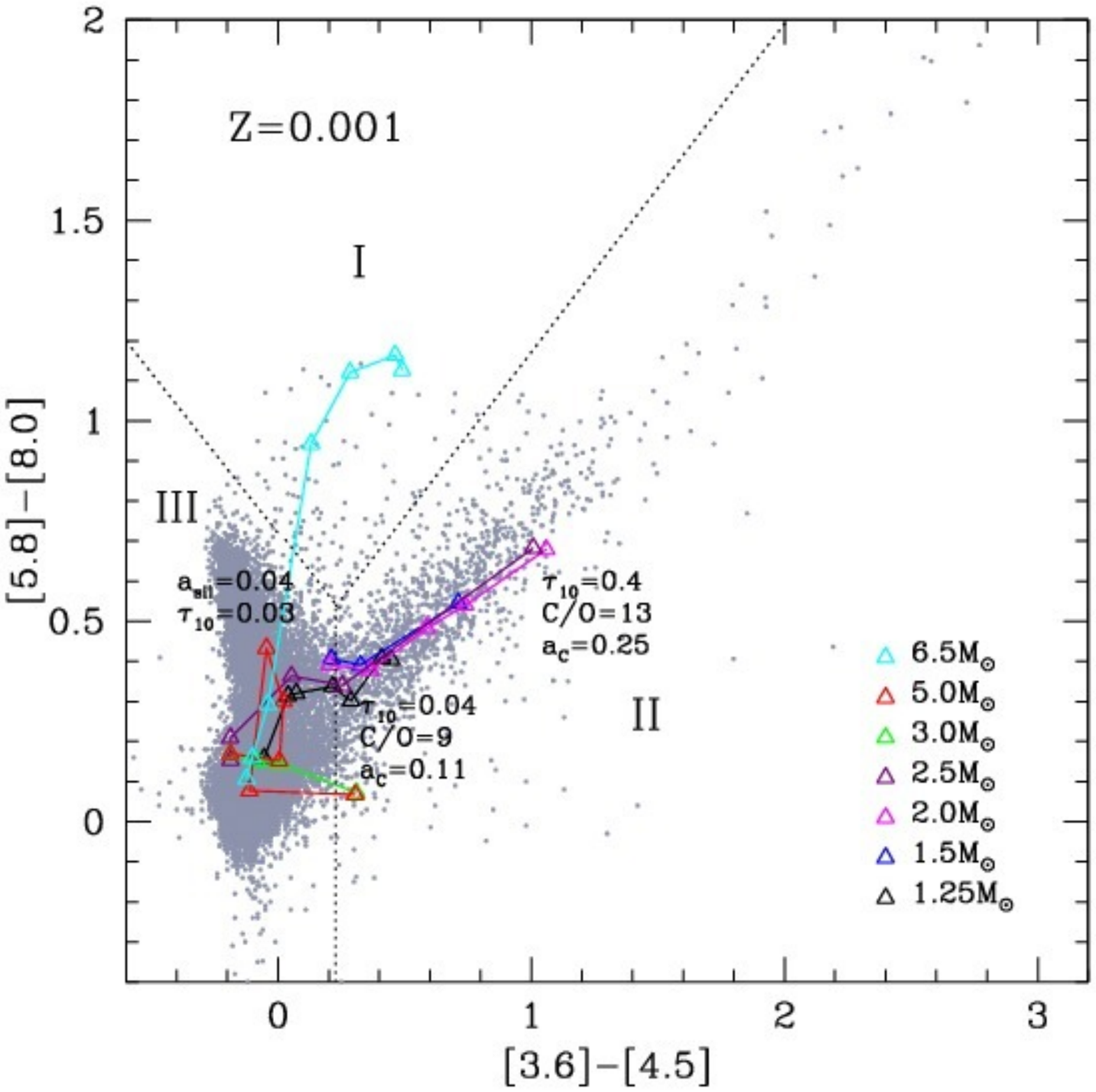}}
\end{minipage}
\begin{minipage}{0.43\textwidth}
\resizebox{1.\hsize}{!}{\includegraphics{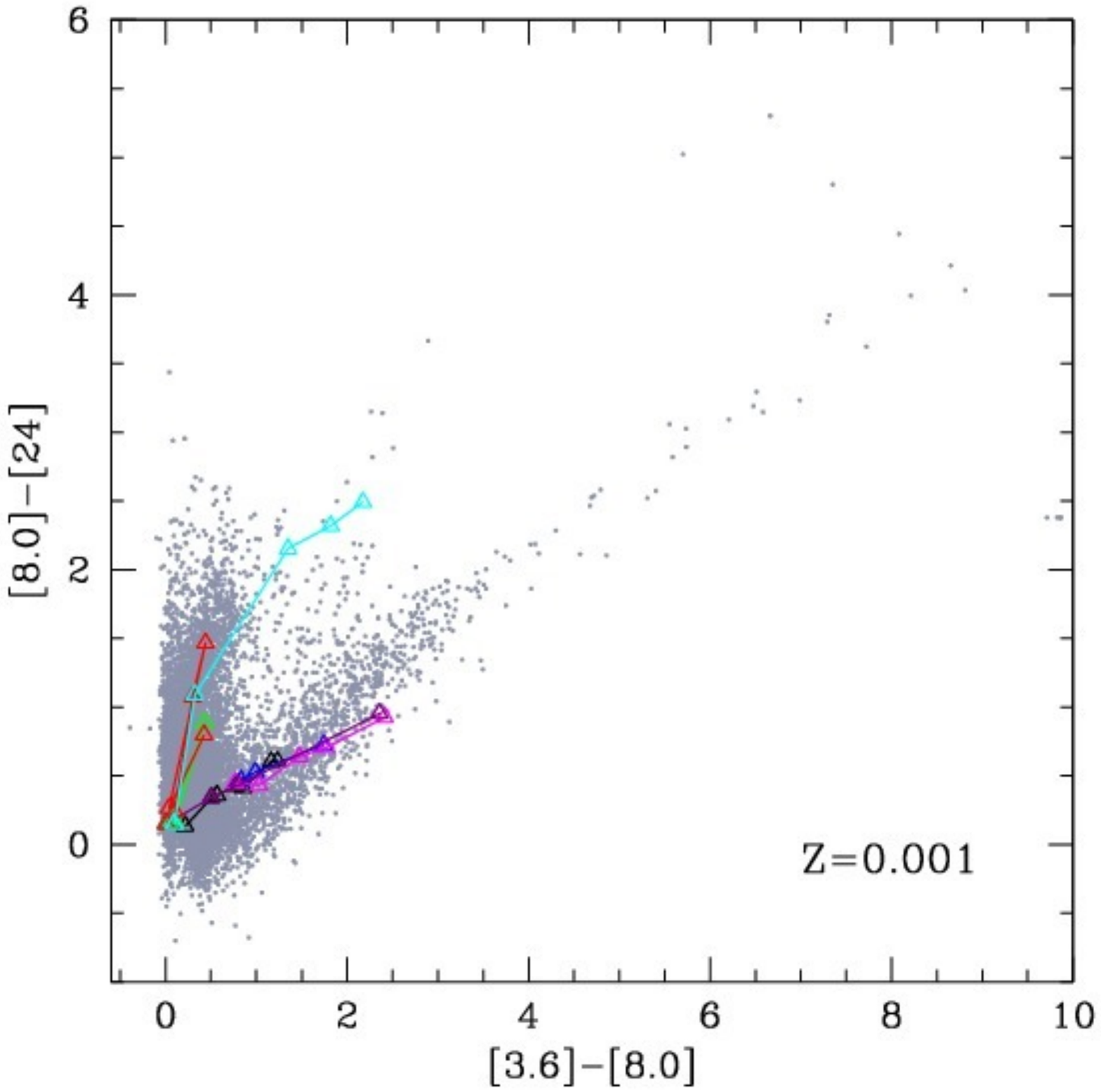}}
\end{minipage}
\vskip-50pt
\caption{Evolutionary tracks of AGB models of metallicity $Z=8\times 10^{-3}$ (top
panels) and $Z=10^{-3}$ (bottom) in the colour--colour ($[3.6]-[4.5], [5.8]-[8.0]$)
(left panels) and ($[3.6]-[8.0], [8.0]-[24]$) planes (right). The values of the
optical depths $\tau_{10}$, of the C/O ratio and of the size (in $\mu$m) of 
carbon grains (indicated with $a_C$) and silicates particles ($a_{sil}$)
along some tracks are indicated. The diagrams in the left panels are divided into 
regions I, II, III, used to classify AGB stars (see text in section \ref{class} for 
details).}
\label{ftracceccd}
\end{figure*}

\begin{figure*}
\begin{minipage}{0.43\textwidth}
\resizebox{1.\hsize}{!}{\includegraphics{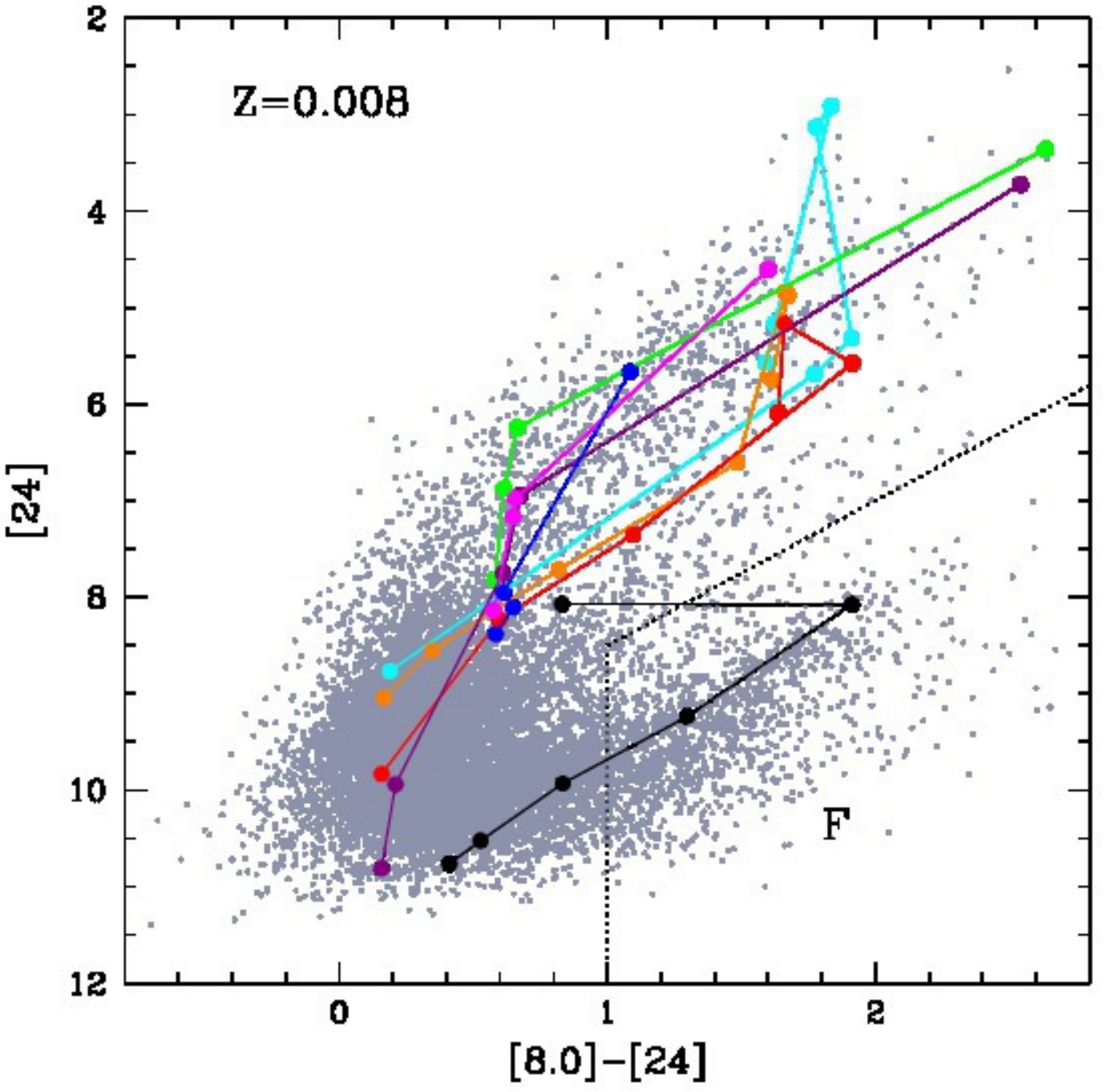}}
\end{minipage}
\begin{minipage}{0.43\textwidth}
\resizebox{1.\hsize}{!}{\includegraphics{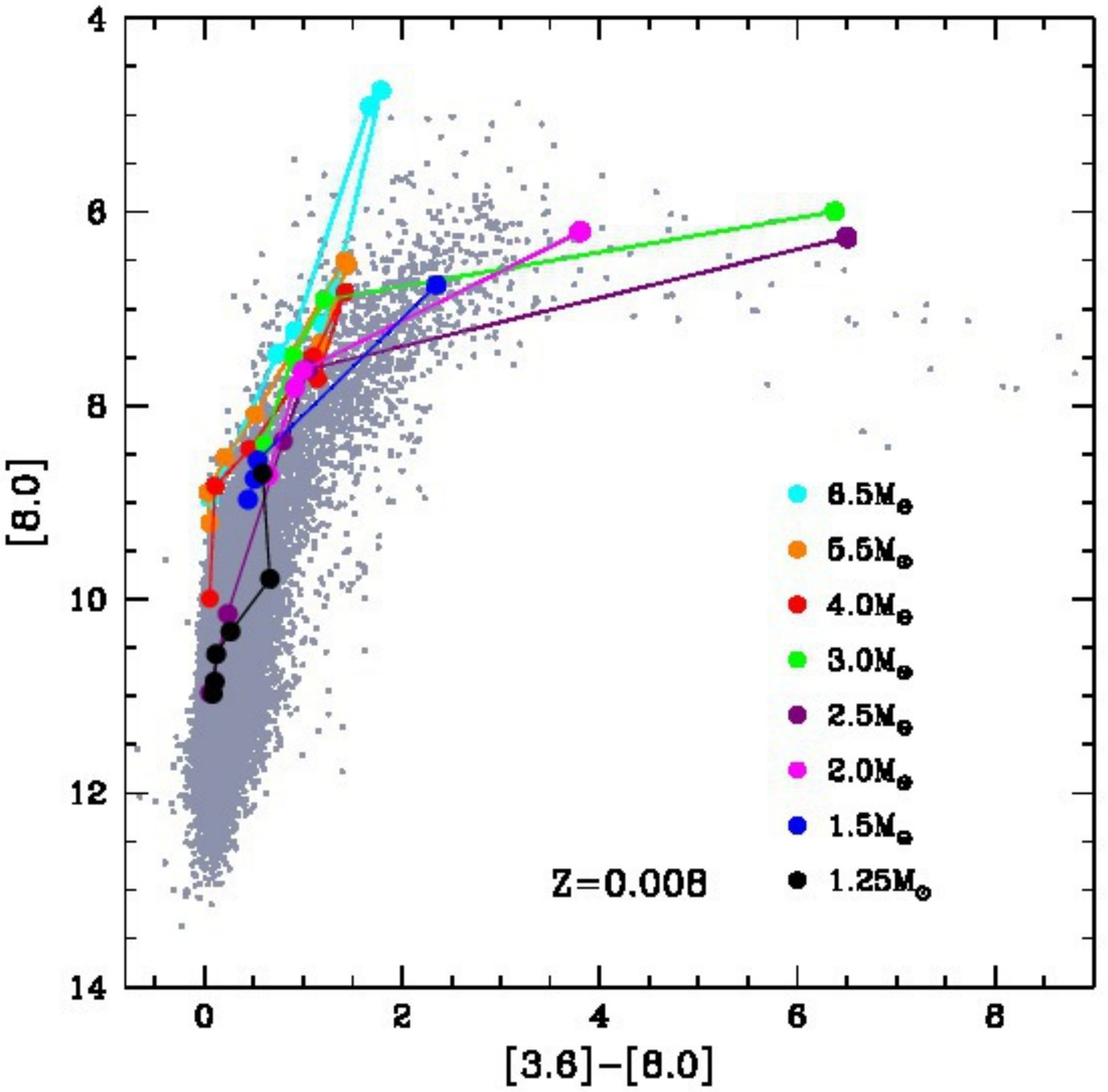}}
\end{minipage}
\vskip-50pt
\begin{minipage}{0.43\textwidth}
\resizebox{1.\hsize}{!}{\includegraphics{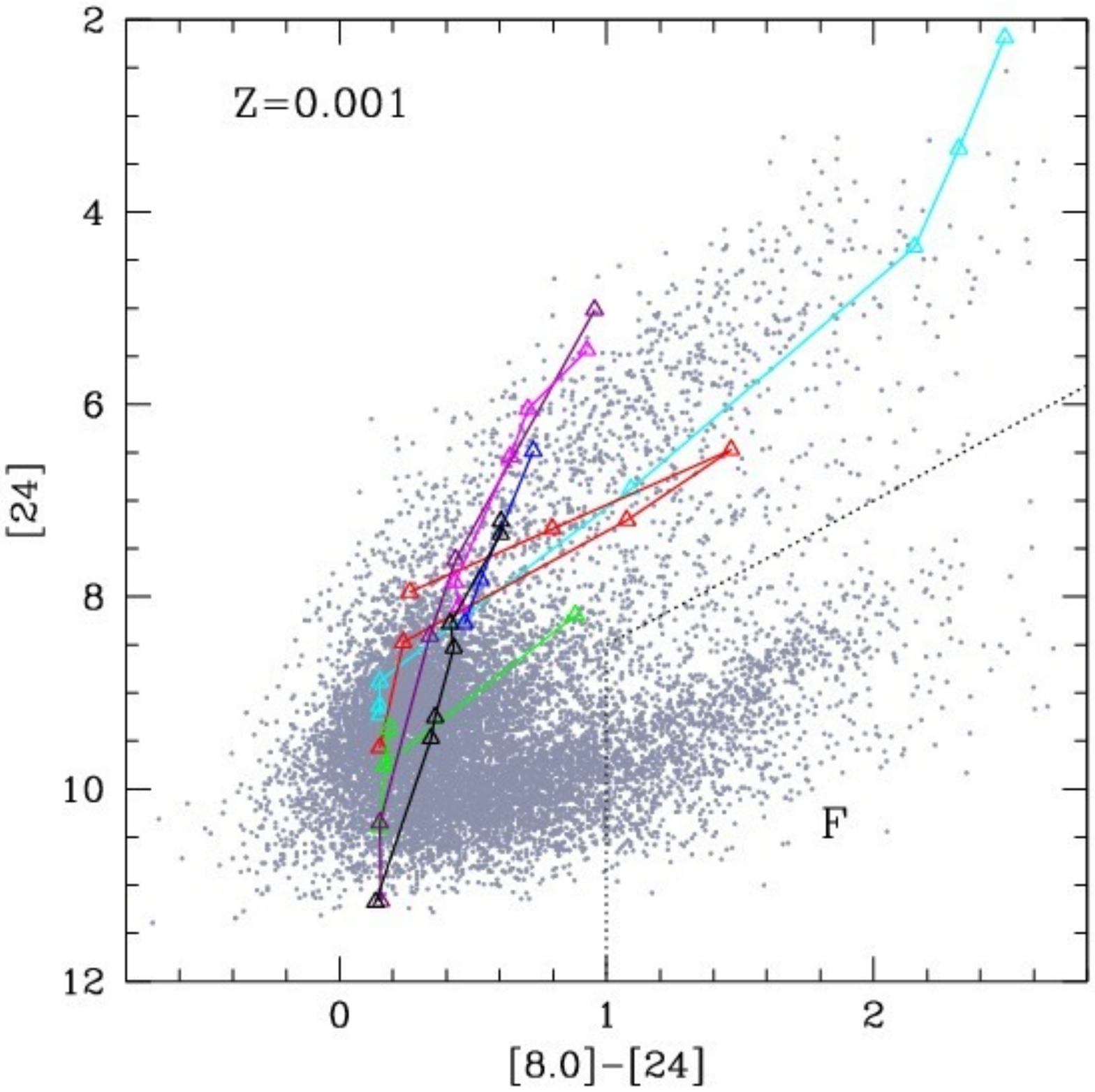}}
\end{minipage}
\begin{minipage}{0.43\textwidth}
\resizebox{1.\hsize}{!}{\includegraphics{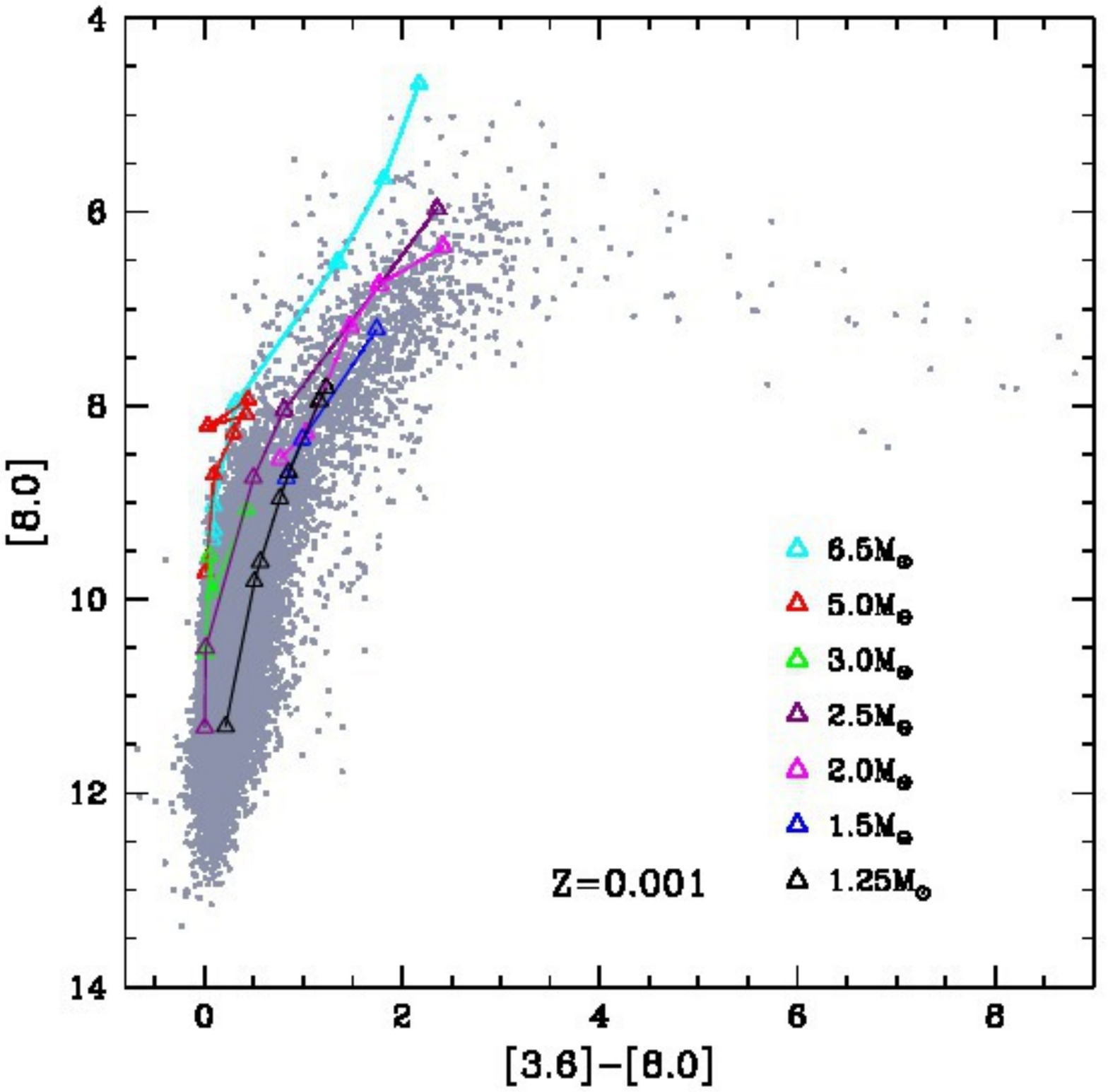}}
\end{minipage}
\vskip-50pt
\caption{Evolutionary tracks of AGB models of metallicity $Z=8\times 10^{-3}$ (top
panels) and $Z=10^{-3}$ (bottom) in the colour--magnitude ($[8.0]-[24], [24]$)
(left panels) and ($[3.6]-[8.0], [8.0]$) diagrams (right). In the left panels we show
the region F, used for the classification of AGBs introduced in section \ref{class}
(see text for details). The colour--coding of the various tracks is the same as 
Fig. \ref{ftracceccd}.}
\label{ftraccecmd}
\end{figure*}

In the $2M_{\odot}$ case the reddest value reached is smaller ($[3.6]-[4.5] \sim 1.6$), 
owing to the lower amount of carbon available in the envelope (see left panel of 
Fig. \ref{fcstar}). The duration of these phases characterised by thermal emission
of dust is within $\sim 5\%$ of the total AGB life.

Models with mass above $3M_{\odot}$ never become carbon stars and follow a different behaviour.
The thermal emission from dust is lower, owing to the smaller extinction coefficients
of silicates in comparison with carbon grains: $[3.6]-[4.5] < 0.5$ in all cases. 
The discussion in Section 3.1 outlined that in these stars the maximum luminosity, when
the star experiences the highest mass loss rates, is reached at an intermediate phase
during the AGB evolution. This is also the phase when the largest quantities of dust is
formed in the circumstellar envelope. Therefore, the trend of $[3.6]-[4.5]$ is not a 
monotonic increase in time: the reddest values are achieved at this phase of maximum dust production, 
when the HBB is experienced. The highest values in the left panel of Fig. \ref{fcolours} correspond to the maximum 
luminosities in Fig. \ref{fhbb}.

In the $1M_{\odot}$ case, the $[3.6]-[4.5]$ colour shows a gradual increase in time, owing to
the progressively higher rate with which silicates form, which, in turn, shifts the SED
to longer wavelength. The thermal emission from dust is small in this 
case, with $[3.6]-[4.5] < 0.2$ for the whole evolution.

\begin{figure*}
\begin{minipage}{0.47\textwidth}
\resizebox{1.\hsize}{!}{\includegraphics{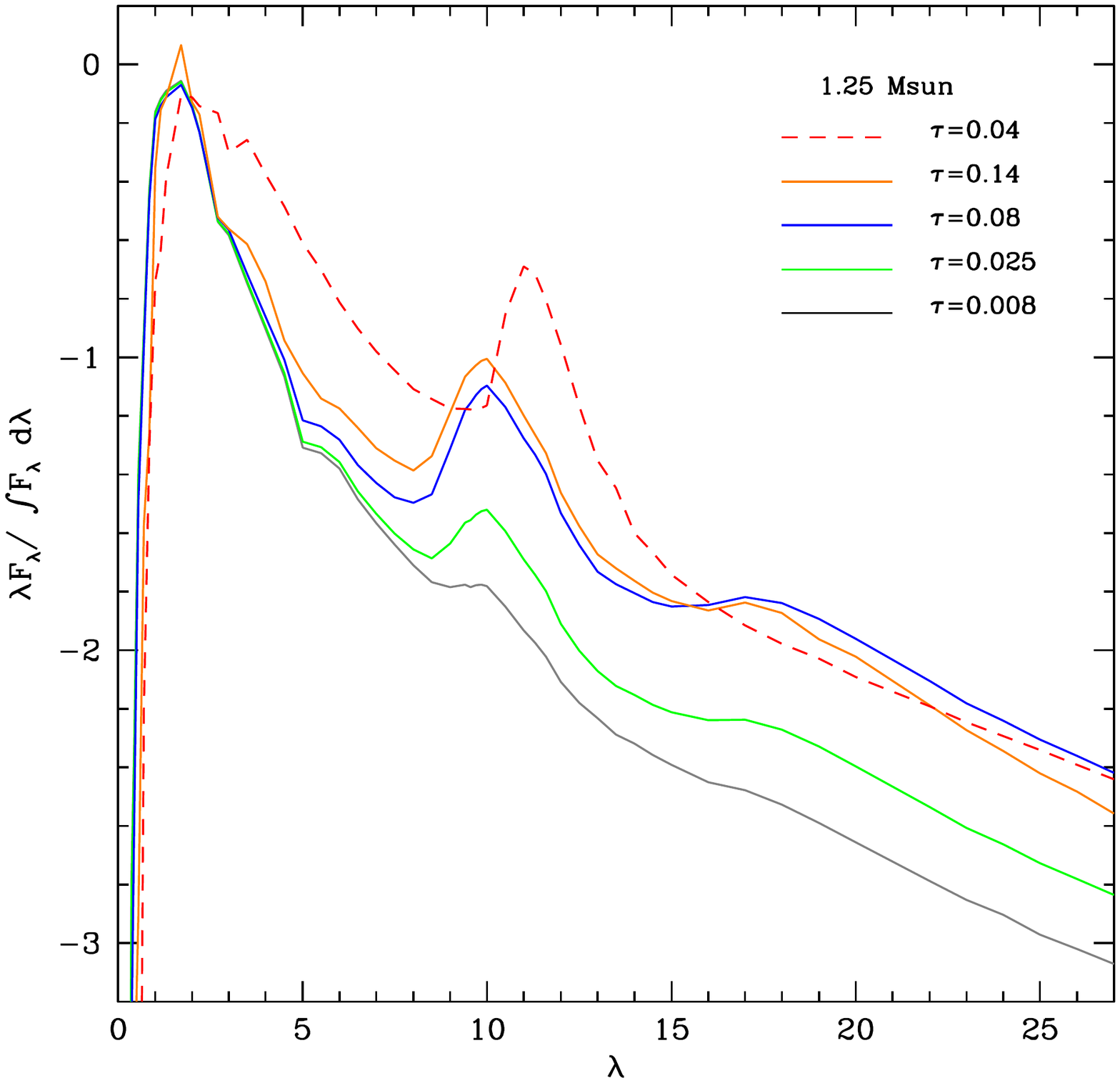}}
\end{minipage}
\begin{minipage}{0.47\textwidth}
\resizebox{1.\hsize}{!}{\includegraphics{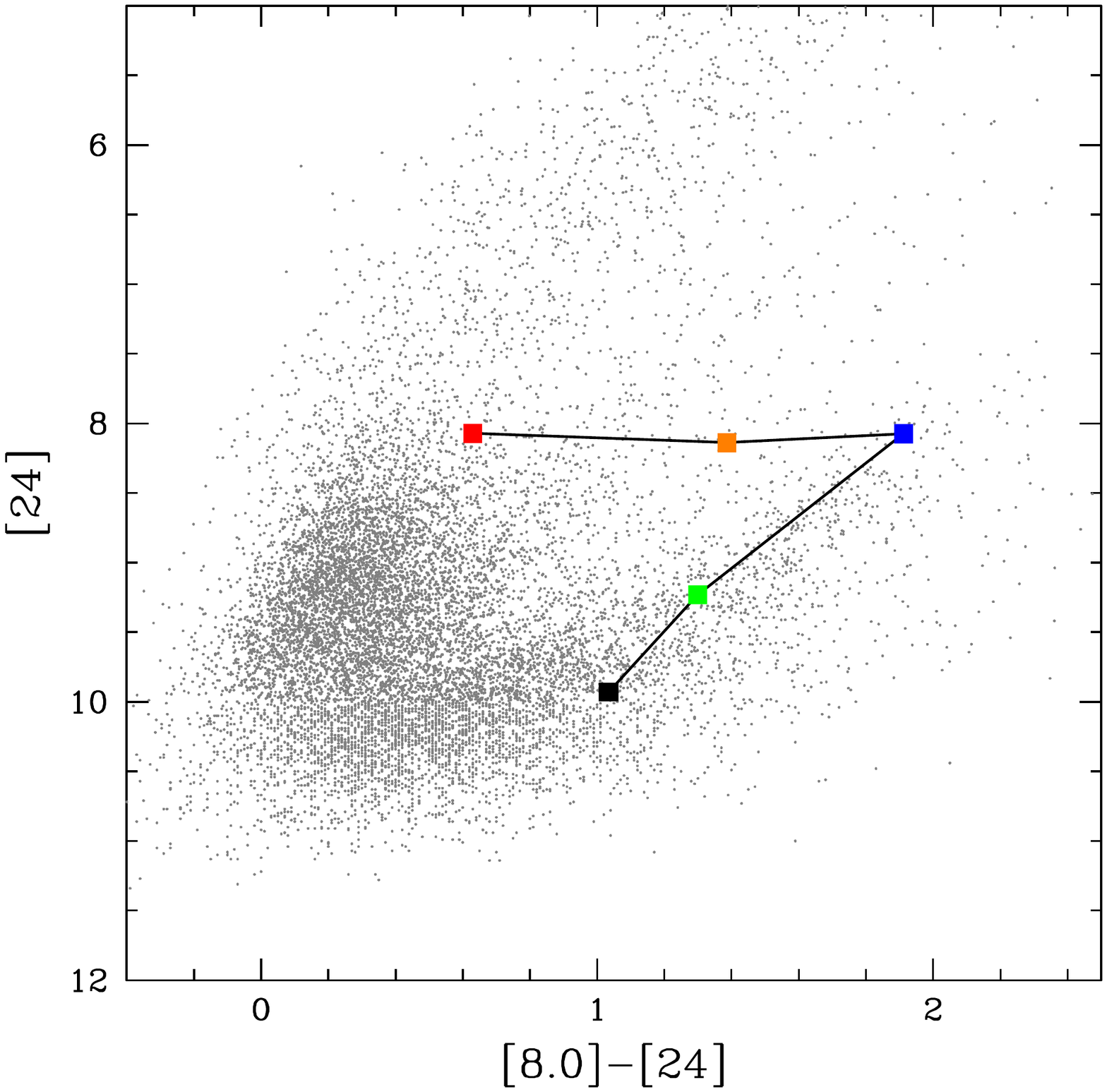}}
\end{minipage}
\vskip-50pt
\caption{Left panel: the synthetic SED of a $1.25M_{\odot}$ model of metallicity 
$Z=8\times 10^{-3}$ in various evolutionary stages. The O--rich phase of the star is 
represented with the solid line, while the dashed line refers to the C--rich stage. 
Right panel: the theoretical track in the [8.0]-[24] vs [24] colour--magnitude 
diagram, where the coloured points refer to the corresponding spectra in the left panel.}
\label{spettri125}
\end{figure*}

The $[5.8]-[8.0]$ colour for the C--star models follows a behaviour similar to $[3.6]-[4.5]$.
In this case the largest values, reached in the latest phases, is $[5.8]-[8.0] \sim 1.8$
for the $2.5M_{\odot}$ and $3M_{\odot}$ models. The difference between the reddest colours 
reached by C-- and oxygen--rich stars is smaller than for $[3.6]-[4.5]$, because the formation of
the silicates feature in O--rich objects at $9.7\mu m$ favours the increase in the $[8.0]$ 
flux, making $[5.8]-[8.0]$ redder. During the phase of strongest HBB of O--rich models,
$[5.8]-[8.0] \sim 1$. The $1M_{\odot}$ model, not experiencing HBB, only reaches 
$[5.8]-[8.0] \sim 0.6$.

The $[8.0]-[24]$ colours also becomes redder and redder as the amount of dust
formed in the envelope gets larger, and the total flux from dust thermal radiation increases. 
C--star models exhibit the same behavior as in the other colours.
The amount of excess carbon with respect to oxygen in the atmosphere 
reaches the highest value at the end of the AGB phase, resulting in a steep rise in the amount of
carbon dust formed. 
The [8.0]-[24] colour becomes red, reaching a final value of $\sim3$. The trend followed by oxygen--rich stars is 
qualitatively different. In models experiencing HBB, like in the other colours, the reddest 
values are reached in conjunction with the phase of maximum efficiency of HBB; 
however, in lower mass models, not experiencing HBB, $[8.0]-[24]$ reaches a maximum value 
slightly below $\sim 2$, and decreases subsequently, when the formation of the silicates 
feature increases the $[8.0] \mu m$ flux.

\subsection{Dusty AGB models: theoretical tracks}
\label{tracce} 
Fig. \ref{ftracceccd} shows the evolutionary tracks of models of different mass in the
colour--colour $([3.6]-[4.5], [5.8]-[8.0])$ (left panels, hereinafter CCD1) and 
$([3.6]-[8.0], [8.0]-[24])$ (right, hereinafter CCD2) planes. Fig. \ref{ftraccecmd} shows
the same tracks in the colour--magnitude $([8.0]-[24], [24])$ (left) and 
$([3.6]-[8.0], [8.0])$ (right) diagrams (hereinafter CMD24 and CMD80). Top panels refer 
to $Z=8\times 10^{-3}$ models, while the tracks of $Z=10^{-3}$ stars are shown in the bottom 
panels. 

The sequences of carbon--stars and oxygen--rich models bifurcate in the CCD1 and CCD2 
planes, the oxygen--rich stars tracing a more 
vertical sequence. This can be seen in the top panels of Fig. \ref{ftracceccd}, 
by comparing the tracks of the $4M_{\odot}$, $5.5M_{\odot}$ and $6.5M_{\odot}$ models, 
with those of their $M \leq 3M_{\odot}$ counterparts. The bifurcation 
between C--rich and oxygen--rich models is also evident in CMD80, shown in the right
panels of Fig. \ref{ftraccecmd}.
This behaviour can be understood on the basis of the discussion in section
\ref{ircolours}: in O--rich stars the formation of the
silicates feature at $9.7\mu m$ leads to a decrease in the $[8.0]$ magnitude:
this provides a straight explanation of the high slope of the corresponding tracks in the 
CMD80 plane, and favours redder $[5.8]-[8.0]$ (see Fig. \ref{fcolours}) and 
$[3.6]-[8.0]$ colours.

Metallicity has important effects on the excursion of the evolutionary tracks in these
planes. 
First, for what concerns oxygen-- rich stars, models of higher metallicity reach redder colours
in the CCD1 and CCD2 planes. The comparison between the tracks of the $Z=10^{-3}$ and 
$Z=8\times 10^{-3}$ models in CCD1 shows that while massive AGBs of the latter population 
evolve up to colours $[5.8]-[8.0] \sim 1$, their lower--Z counterparts, with the exception of 
SAGB models with initial mass above $\sim 6 M_{\odot}$, barely reach 
$[5.8]-[8.0] \sim 0.5$. In the CCD2, higher--Z, massive AGBs evolve to $[8.0]-[24] \sim 2$, 
while their $Z=10^{-3}$ counterparts reach $[8.0]-[24] \sim 1.5$. These
differences originate from the larger quantities of dust formed in the envelope of 
higher--metallicity stars, as a consequence of the larger amount of silicon available.
The analysis of the colour--magnitude diagrams, shown in Fig. \ref{ftraccecmd}, confirms
that oxygen--rich stars with higher--metallicity evolve to redder IR colours. Not only the tracks of the
$Z=8\times 10^{-3}$ models reach higher $[3.6]-[8.0]$ and $[5.8]-[8.0]$ colours, but also the
$8.0\mu$m and $24\mu$m fluxes are larger than their $Z=10^{-3}$ counterparts, as a
consequence of the reprocessing of the stellar radiation by silicates grain in the
circumstellar envelope.

The metallicity of the stars also influences the distribution of carbon stars in the
various planes. As shown in Fig. \ref{ftracceccd}, C--rich objects of different metallicity 
define similar trends, with the difference that the $Z=8\times 10^{-3}$ models 
evolve to redder colours: while for these stars we find that the tracks reach 
$[3.6]-[4.5] \sim 3$, $[3.6]-[8.0] \sim 7$, $[8.0]-[24] \sim 2.5$, $[5.8]-[8.0] \sim 7$, 
in the $Z=10^{-3}$ case, despite the carbon excess reached is larger (see Fig. \ref{fcstar}),
we have $[3.6]-[4.5] < 1$, $[3.6]-[8.0] < 2.5$, $[8.0]-[24] < 1$, $[5.8]-[8.0] < 2$. 
This is because lower--Z stars evolve at higher effective temperatures, pushing 
the dust forming layer far away from the stellar surface, in a region of smaller density, where 
dust formation occurs with a lower efficiency. These arguments 
outline the delicate interplay between the surface carbon abundance and the temperature 
of the external regions in determining the amount of carbon dust formed.

The evolution of the $Z=8\times 10^{-3}$ models of initial mass $M\sim 1.25, 1.5M_{\odot}$
deserves particular attention, as it will be also important in the interpretation of the
observations. As shown in the left panel of Fig. \ref{fcstar}, 
these stars evolve as oxygen--rich objects for most ($\sim 70-80 \%$) of their AGB life,
and eventually become carbon--stars. They do not reach extremely red colours, because 
their envelope is lost before great amounts of carbon are accumulated at the surface.
Their tracks in the various planes present turning points, associated to the transition
from M-- to C--stars. This behaviour is particularly evident in the CMD24 plane (see 
top--left panel of Fig. \ref{ftraccecmd}), where the track corresponding to the 
$M\sim 1.25M_{\odot}$ model (indicated with a black line) first moves to the red, then turns 
into the blue. The right panel of Fig. \ref{spettri125} shows in more details the
excursion of the track, whereas in the left panel we show the SED of the same model, in different
evolutionary phases. In the first part of the AGB evolution the optical depth increases, owing
to the larger and larger quantities of silicates formed in the circumstellar envelope.
Consequently, the silicates feature at $9.7\mu$m becomes more prominent during the
evolution (see the various SEDs shown in the left panel of Fig. \ref{spettri125}).
After becoming C--star, the optical depth decreases, and the star evolves to the blue.
This peculiar behaviour of low--mass AGBs is restricted to $Z=8\times 10^{-3}$ models,
because lower--Z stars produce smaller quantities of silicates, thus their tracks are
bluer (see bottom--left panel of Fig. \ref{ftraccecmd}). 

The track of the $1.5M_{\odot}$ $Z=8\times 10^{-3}$ model in the CCD1 moves to the red as the surface carbon 
increases (see blue line in Fig. \ref{ftracceccd}). However, compared to the lower--Z model with the same mass, or to
the models with the same metallicity but higher masses, the track occupy the lower side of
the CCD1: for a given $[3.6]-[4.5]$, the $[5.8]-[8.0]$ is bluer. To understand this
trend, we show in the right panel of Fig. \ref{fspettri} the SED of models of various
mass and metallicity, in the evolutionary stage when $[3.6]-[4.5] \sim 0.3$. 
A clear difference among models of different metallicity is the SiC emission feature, 
practically absent in the $Z=10^{-3}$ models, owing to scarcity of silicon in the envelope.
We also note that the optical depth of the 
$1.5M_{\odot}$ model ($\tau_{10}=0.08$) is slightly larger than the $3M_{\odot}$ star
($\tau_{10}=0.06$), despite in the latter case the surface carbon is larger. 
The higher carbon content in the $3M_{\odot}$ case favours larger quantities of dust 
particles in the regions of the envelope at temperatures $\sim 1100$K, 
the threshold value to allow condensation of gas molecules into solid carbon grains. However, 
this effect is more than counterbalanced by the higher acceleration experienced by the wind:
compared to the $1.5M_{\odot}$ model, the profile of density is steeper, thus a smaller
contribution to the overall value of $\tau_{10}$ is given by the outermost layers of
the $3M_{\odot}$ model, in comparison with its smaller mass counterpart.

In the comparison between the dust composition of the circumstellar envelopes of the two 
models, we have a higher quantity of SiC in the $1.5M_{\odot}$ case; conversely,
owing to the larger carbon available at the surface, the $3M_{\odot}$ envelope is dominated 
by carbon--dust.
The SED of the $1.5M_{\odot}$ model (blue line in the right panel of Fig. \ref{fspettri}),
compared to the corresponding SED of the $3M_{\odot}$ star (green line) reflects more the 
shape of the SiC feature: the relative flux is higher at $\sim 11\mu$m, and is declining 
more steeply at shorter wavelengths, in the $5-8 \mu$m region of the spectrum.
While the $[3.6]-[4.5]$ colour is not affected by these differences, the $[5.8]-[8.0]$
results bluer in these models, which motivates their different position in the CCD1.

\begin{figure*}
\begin{minipage}{0.47\textwidth}
\resizebox{1.\hsize}{!}{\includegraphics{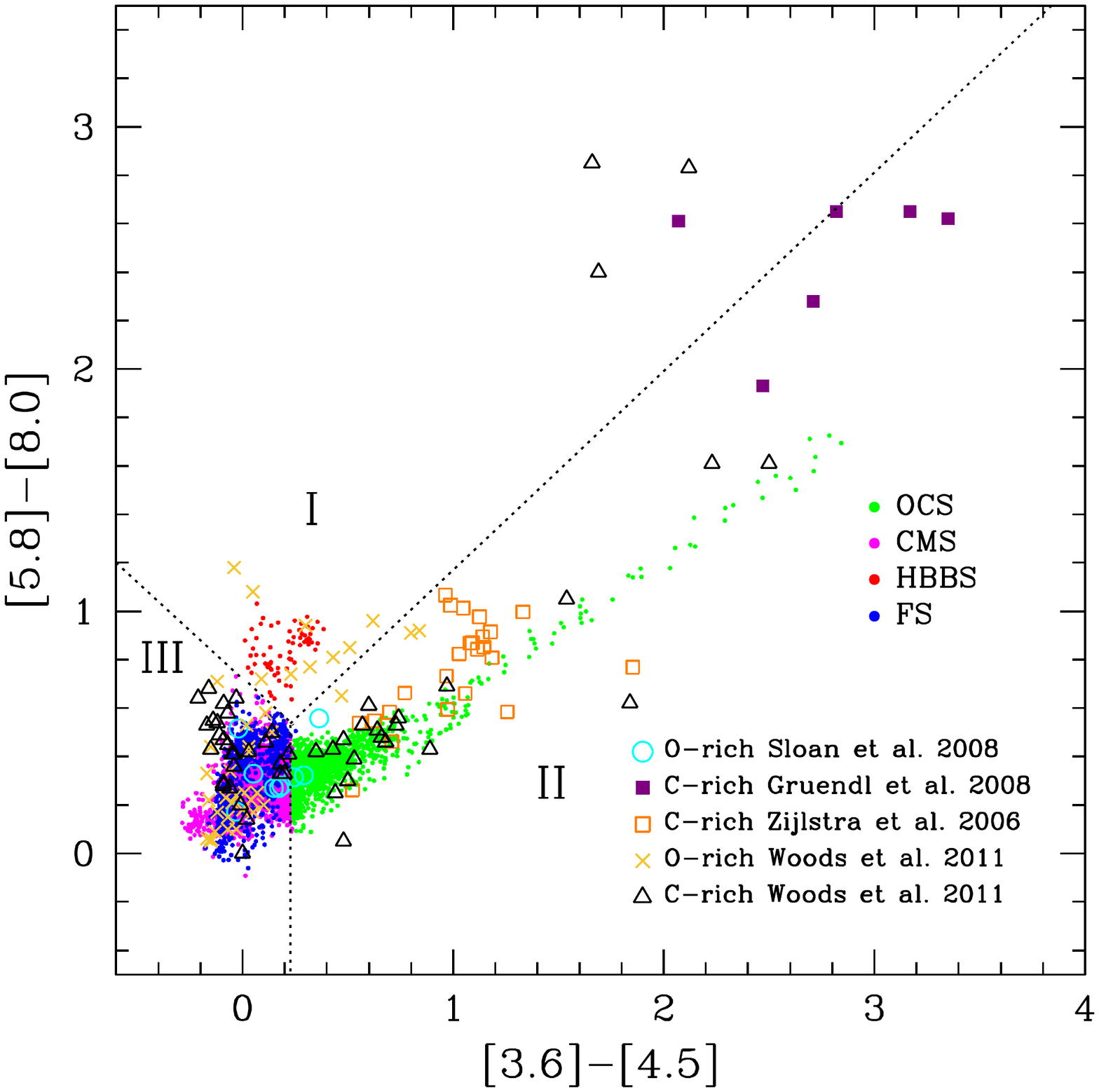}}
\end{minipage}
\begin{minipage}{0.47\textwidth}
\resizebox{1.\hsize}{!}{\includegraphics{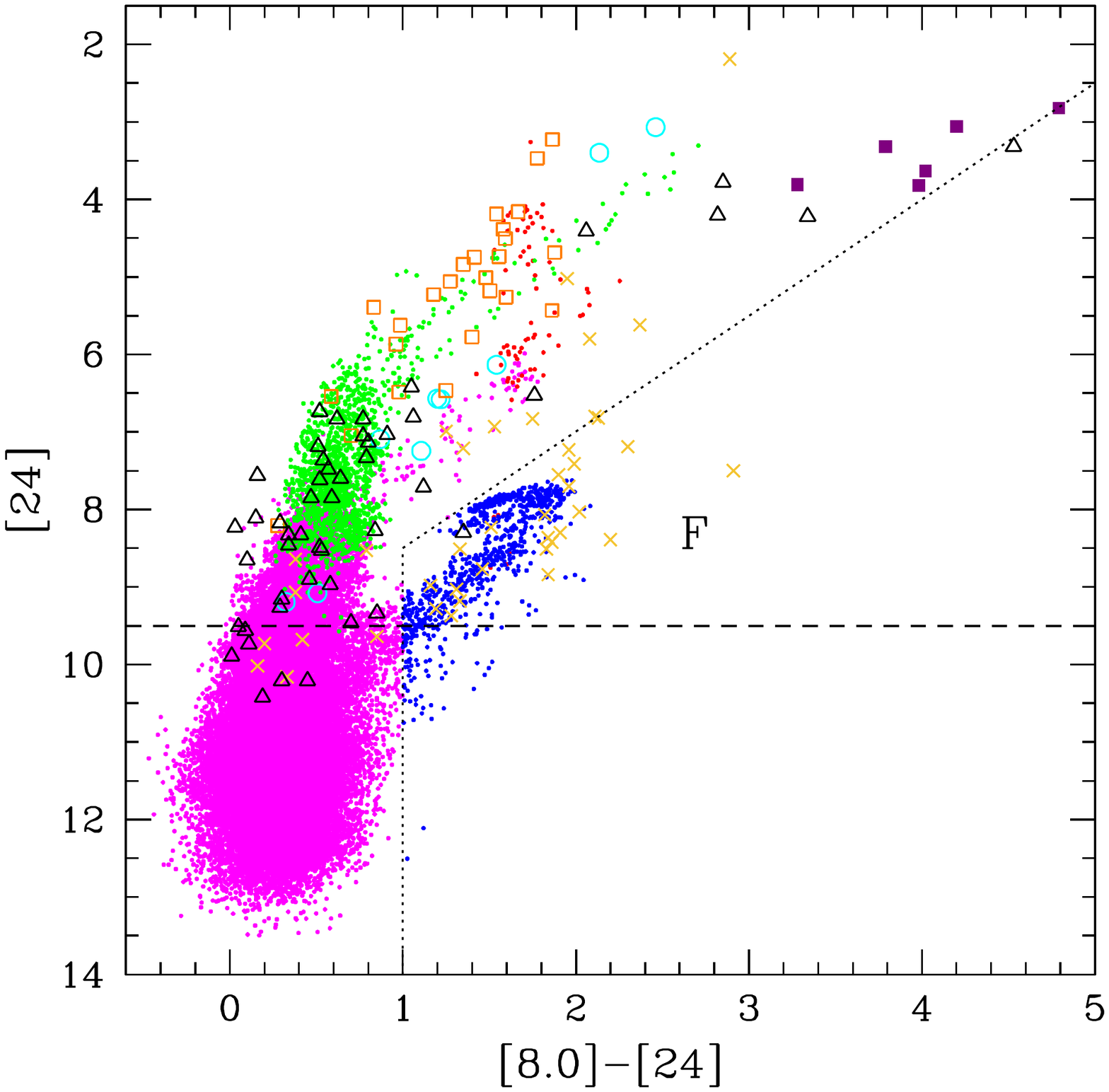}}
\end{minipage}
\vskip-50pt
\caption{The distribution of the synthetic diagrams in the $[3.6]-[4.5]$ vs $[5.8]-[8.0]$ 
(left) and $[8.0]-[24]$ vs $[8.0]$ (right) planes. Different colours refer to the classification 
presented in section \ref{class}: OCS (green), HBBS (orange), FS (blue) and CMS (magenta). 
Spectroscopically confirmed C--stars from \citet{gruendl08}, \citet{zijlstra06},
\citet{woods11} and O--rich stars from \citet{sloan08} and \citet{woods11} are also
shown. The dotted line in the right panel represents the cut applied to the samples at 
$[24] < 9.5$, to obtain a completeness $\sim 100\%$ of the data.}
\label{synpop}
\end{figure*}

\section{Which classification for AGB stars?}
\label{class}
The results discussed in the previous section showed that AGB stars 
populate different regions in the colour--colour and colour--magnitude planes, 
depending on their mass, metallicity, and optical depth.
The evolutionary tracks are the outcome of complex computations of the AGB evolution and
of the dust formation process, and are extremely sensitive to a number of physical
inputs, such as convection, mass loss, treatment of convective borders, the entire
description of the dust formation process.
Assessing the reliability of these models demands comparison with the observations,
now possible thanks to several surveys of the LMC population of AGB stars, 
described in next section. This approach will hopefully help reducing the uncertainties
affecting the afore mentioned physical mechanisms.
To undertake this analysis, we need to identify groups of stars with specific properties,
that populate selected regions in the colour--colour and colour--magnitude planes
obtained with the Spitzer bands. 

We use the tracks presented and discussed in the previous section to propose a 
classification of AGB stars in the LMC into four groups, each characterized by specific
evolutionary properties, and occupying well defined regions in the CCD1, CCD2, CMD80, 
CMD24 planes. 

This classification is based on the following results, shown in Fig. \ref{ftracceccd}
and \ref{ftraccecmd}: 
\begin{itemize}

\item{The region in the upper side of the CCD1, zone I in the left panels of 
Fig. \ref{ftracceccd}, is populated exclusively by massive AGBs experiencing HBB,
surrounded by silicates, with $\tau_{10}>0.1$; we will refer to these models 
as Hot Bottom Burning Stars (HBBS).}

\item{The zone II in the CCD1 is populated by carbon stars with SiC particles and
carbon dust in their envelopes, and optical depth $\tau_{10} > 0.02$. We will refer
to these models as Obscured Carbon Stars (OCS).}

\item{The only tracks evolving in zone F in the CMD24 plane are those corresponding to
low--mass stars, $M \sim 1.25M_{\odot}$, in the phases preceding the
C--star phase. We will call these models F stars (FS) in the following sections.
}

\item{Region III in the CCD1 is crossed by tracks of various mass and metallicity, both
oxygen-- and carbon--rich. These models are not found to be significantly obscured.
We will refer to them as C-- and oxygen--rich stars (CMS), and they encompass all the models
not belonging to any of the three previous groups.}
\end{itemize}

Both the observed sources and the models produced by the synthetic
modelling will be classified according to the criteria given above, following their
position in the CCD1 and CMD24 planes.

In the following, we will describe the observational sample that will be used for
our analysis.

\section{AGB stars in the LMC: observations and theoretical predictions}

\subsection{Historical identification and classification of AGBs in the LMC}
\label{histclass}
The first works aimed at the identification of the AGB population in the LMC
were presented by \citet{blanco78}, \citet{richer83}, \citet{frogel90}.
More recently, \citet{cioni00a} attempted a classification of the AGB sample of
the LMC based on IJKs data of over one million point sources in the direction of the LMC, 
included in the DENIS catalogue. This classification is based 
on the fact that all stars brighter than the tip of the red-giant branch must be AGB stars, and in the 
colour--magnitude (I-J, I) diagram they are separated by a diagonal line from younger 
and foreground objects \citep{cioni00b}.

A similar criterion was followed to identify AGBs in the colour--magnitude diagram 
(J-Ks, Ks), obtained with 2MASS \citep{skrutskie98} data. \citet{cioni06} identified
a region in this diagram, enclosed by two lines, that should include the AGB stars in
the sample \citep[see eq. 1 and 2 and Fig. 1 in][]{cioni06}. The total sample was further
split into O--rich and C--rich candidates; the two spectral classes were discriminated by means 
of a straight line, whose expression is given in Eq. 4 in \citet{cioni06}. 

A considerable step forward in the study of dust obscured evolved stars in the LMC came
with the data from IRAC and MIPS, mounted onboard of Spitzer. In particular, the SAGE 
Survey \citep{meixner06} produced photometric data taken with IRAC of over six 
million stars. This allowed a considerable progress in the studies of dust--enshrouded
stars, because the IRAC and MIPS filters are centered in the spectral region
where most of the emission from optically thick circumstellar envelopes occurs.

\citet{blum06}, studying the obscured objects, identified a 
sequence of stars in the $J-[3.6] > 3$ portion of the colour--magnitude 
(J-[3.6], [3.6]) diagram, that were classified as "extreme" AGB stars \citep[see Fig. 3 
in][]{blum06}. 

The classification of LMC AGBs into C--rich, O--rich candidates, introduced by
\citet{cioni06}, completed by the extreme stars by \citet{blum06}, was subsequently used 
in all the more recent investigations \citep{srinivasan09, srinivasan11,riebel10,riebel12}.

\begin{figure*}
\begin{minipage}{0.43\textwidth}
\resizebox{1.\hsize}{!}{\includegraphics{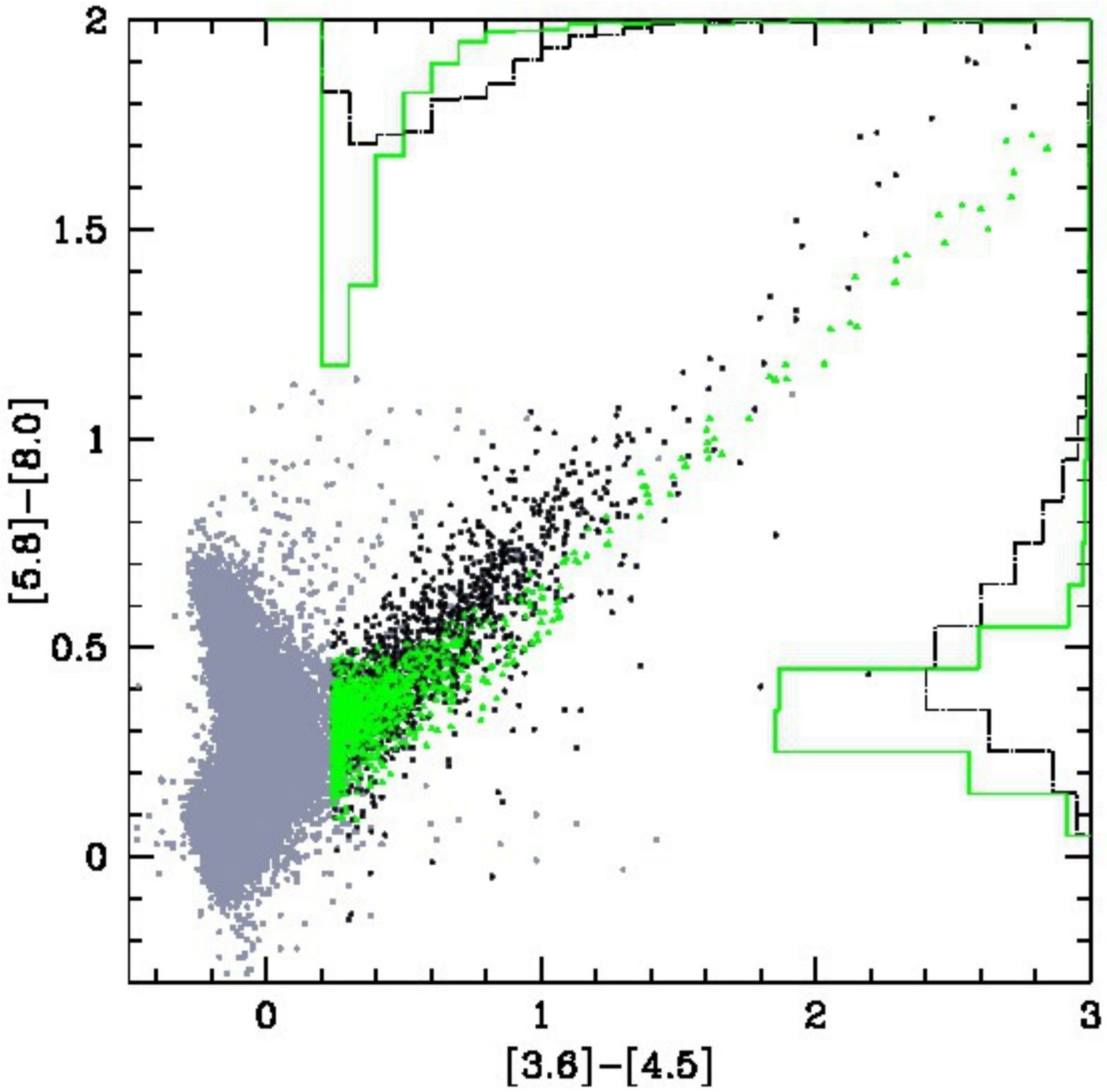}}
\end{minipage}
\begin{minipage}{0.43\textwidth}
\resizebox{1.\hsize}{!}{\includegraphics{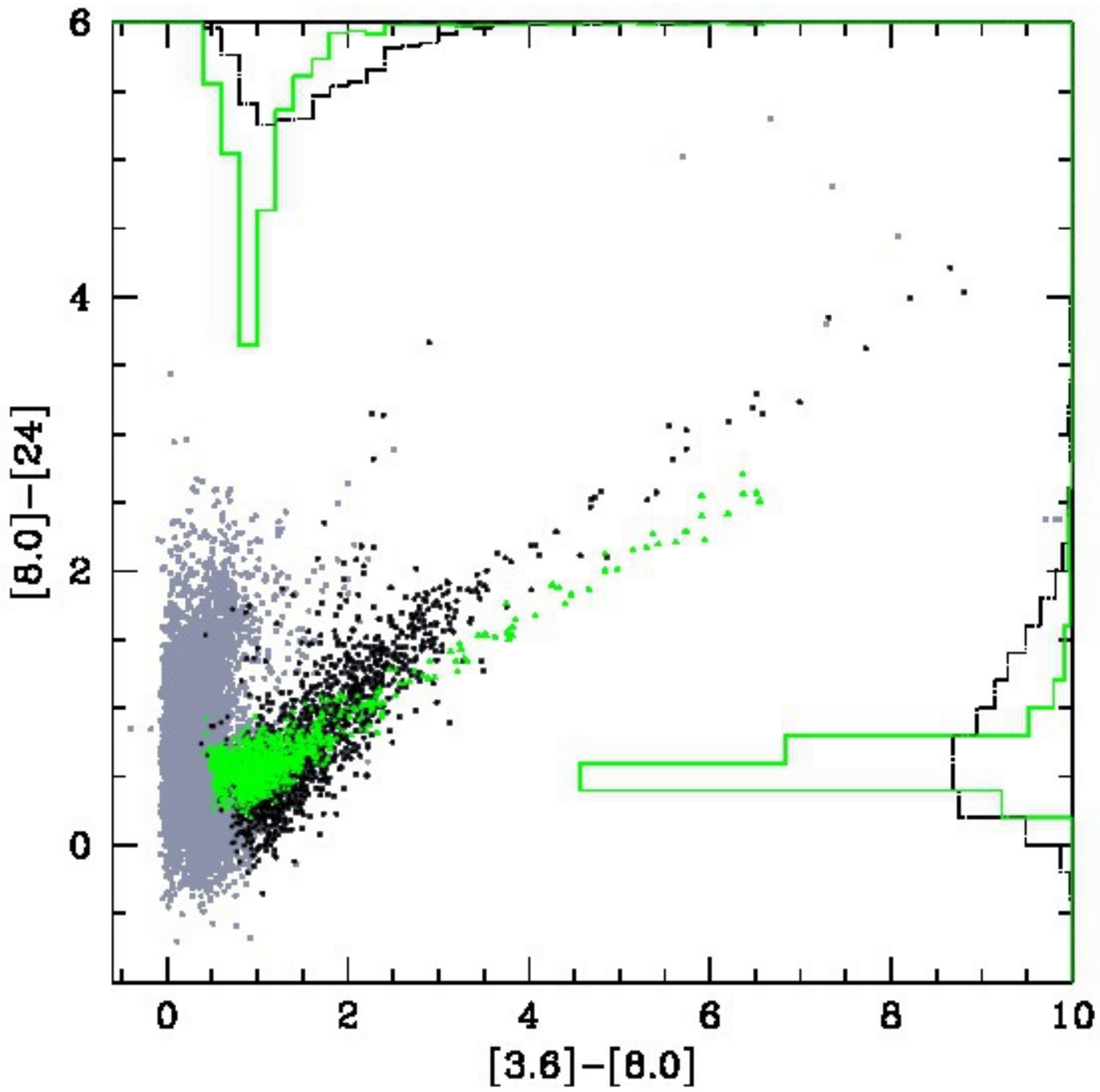}}
\end{minipage}
\vskip-50pt
\begin{minipage}{0.43\textwidth}
\resizebox{1.\hsize}{!}{\includegraphics{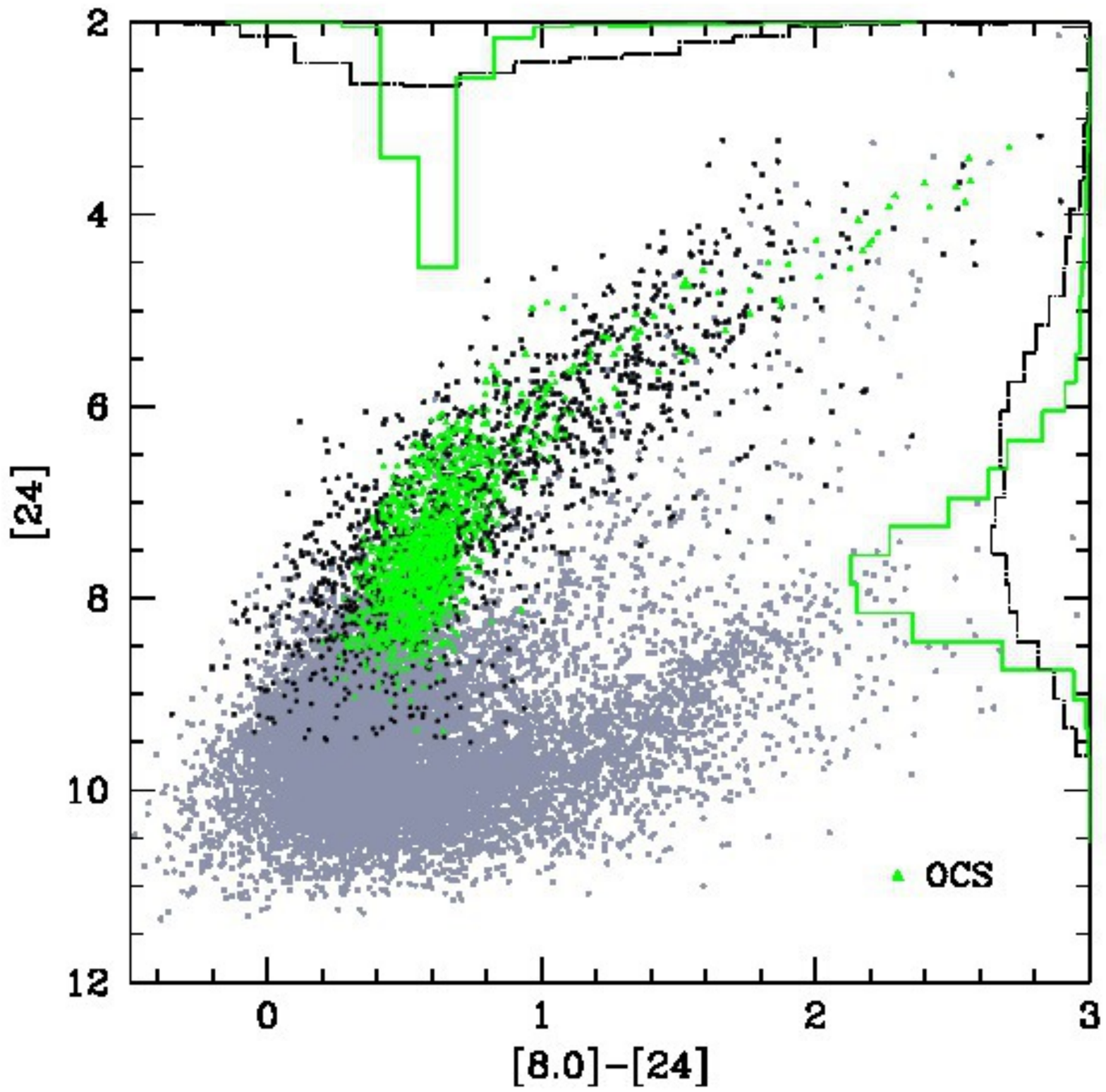}}
\end{minipage}
\begin{minipage}{0.43\textwidth}
\resizebox{1.\hsize}{!}{\includegraphics{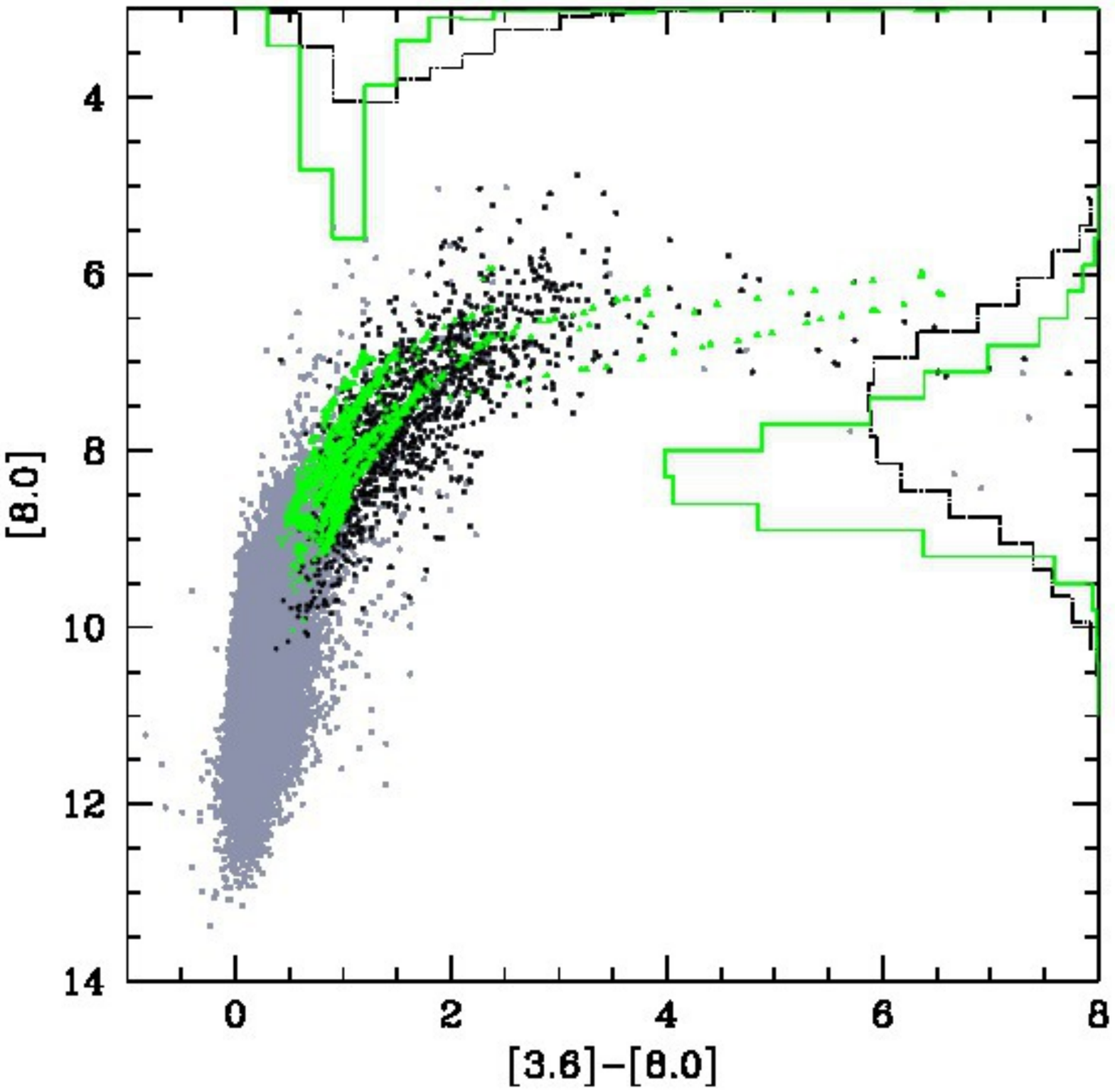}}
\end{minipage}
\vskip-30pt
\caption{The distribution in the colour--colour ($[3.6]-[4.5], [5.8]-[8.0]$) 
(top--left panel) and  ($[3.6]-[8.0], [8.0]-[24]$) (top--right) diagrams, and in the
colour--magnitude ($[8.0]-[24], [24]$) (bottom--left), ($[5.8]-[8.0], [8.0]$)
(bottom--right) planes of the sample of OCS discussed in section \ref{secocs}. 
The whole sample of stars by \citet{riebel12} is shown with grey points; the black
dots indicate the stars in the \citet{riebel12} sample that fall in region II of the
CCD1, and are thus classified as OCS according to the criteria given in Section \ref{class}.
The green dots indicate results from synthetic modelling.  
Note that the statistical analysis is limited to stars with $[24]<9.5$.}
\label{focs}
\end{figure*}

\begin{figure*}
\begin{minipage}{0.43\textwidth}
\resizebox{1.\hsize}{!}{\includegraphics{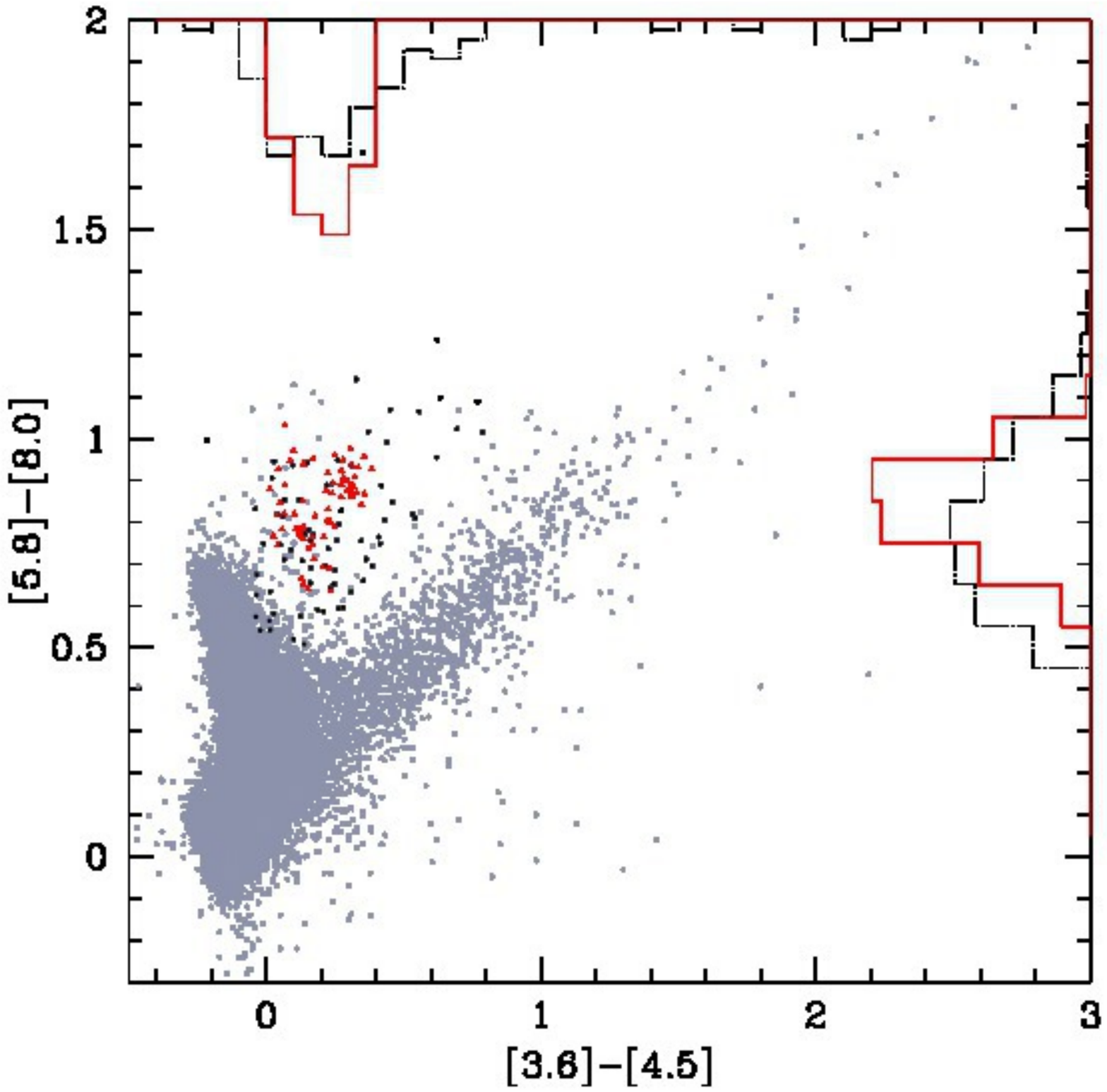}}
\end{minipage}
\begin{minipage}{0.43\textwidth}
\resizebox{1.\hsize}{!}{\includegraphics{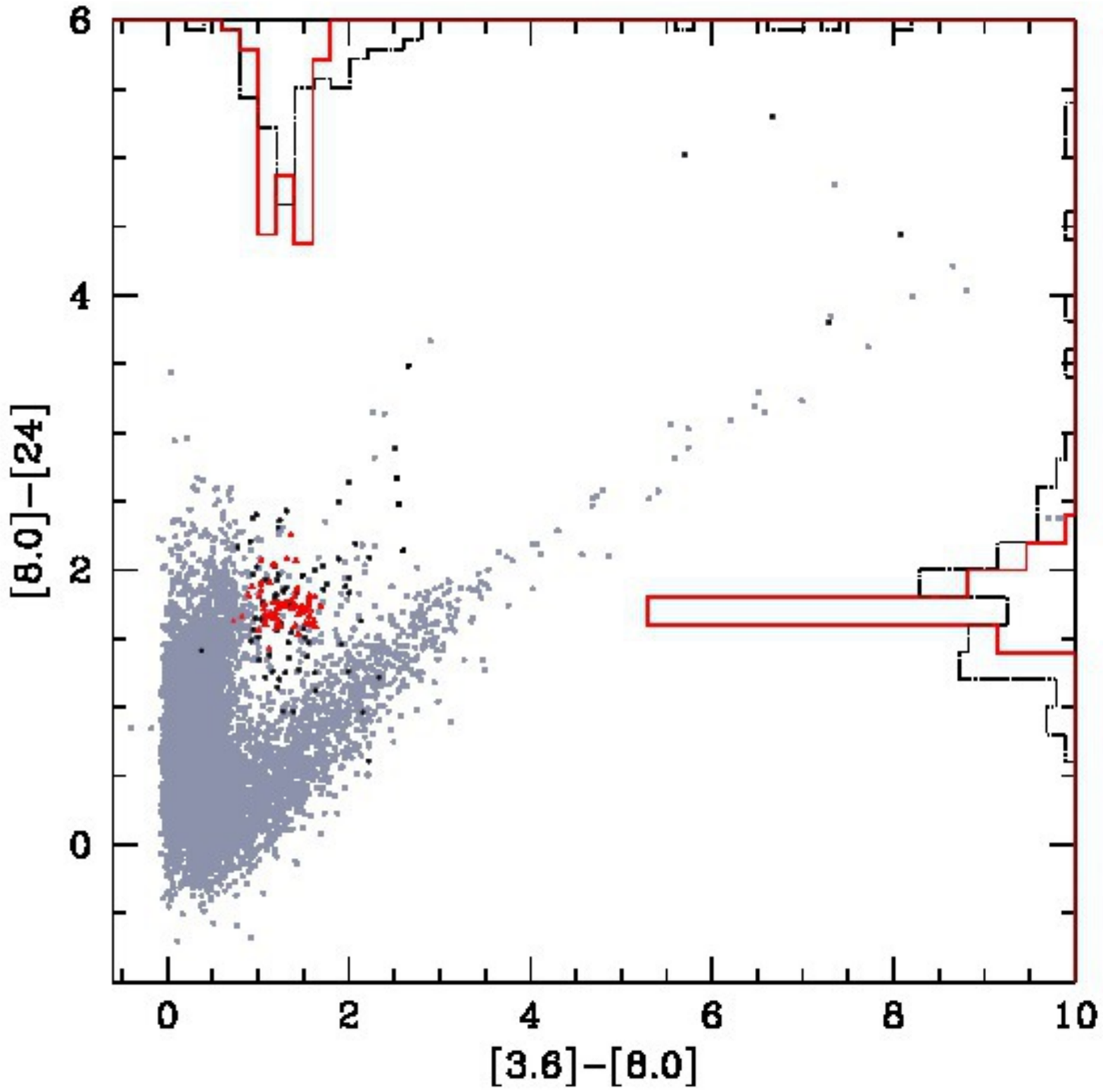}}
\end{minipage}
\vskip-50pt
\begin{minipage}{0.43\textwidth}
\resizebox{1.\hsize}{!}{\includegraphics{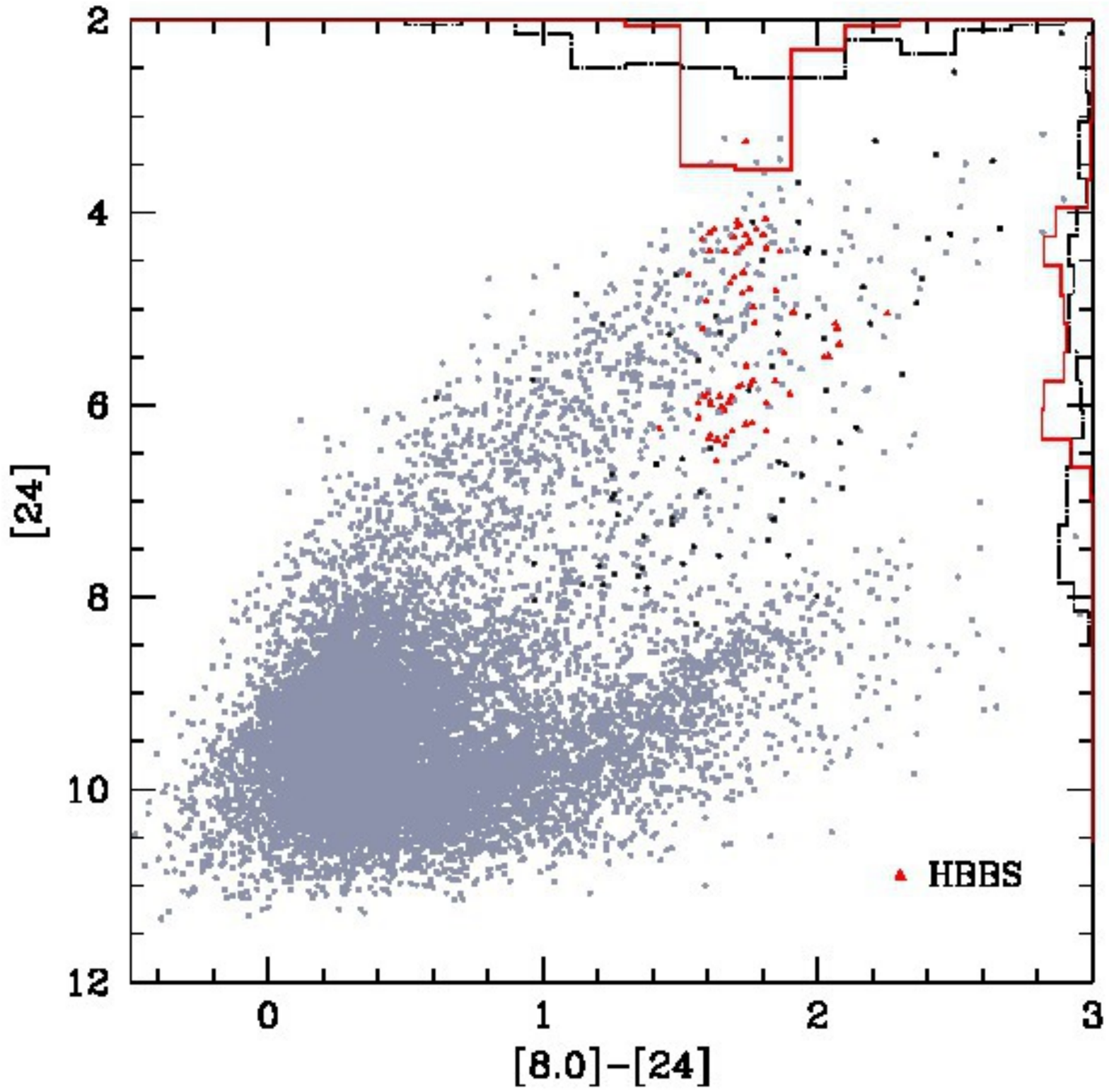}}
\end{minipage}
\begin{minipage}{0.43\textwidth}
\resizebox{1.\hsize}{!}{\includegraphics{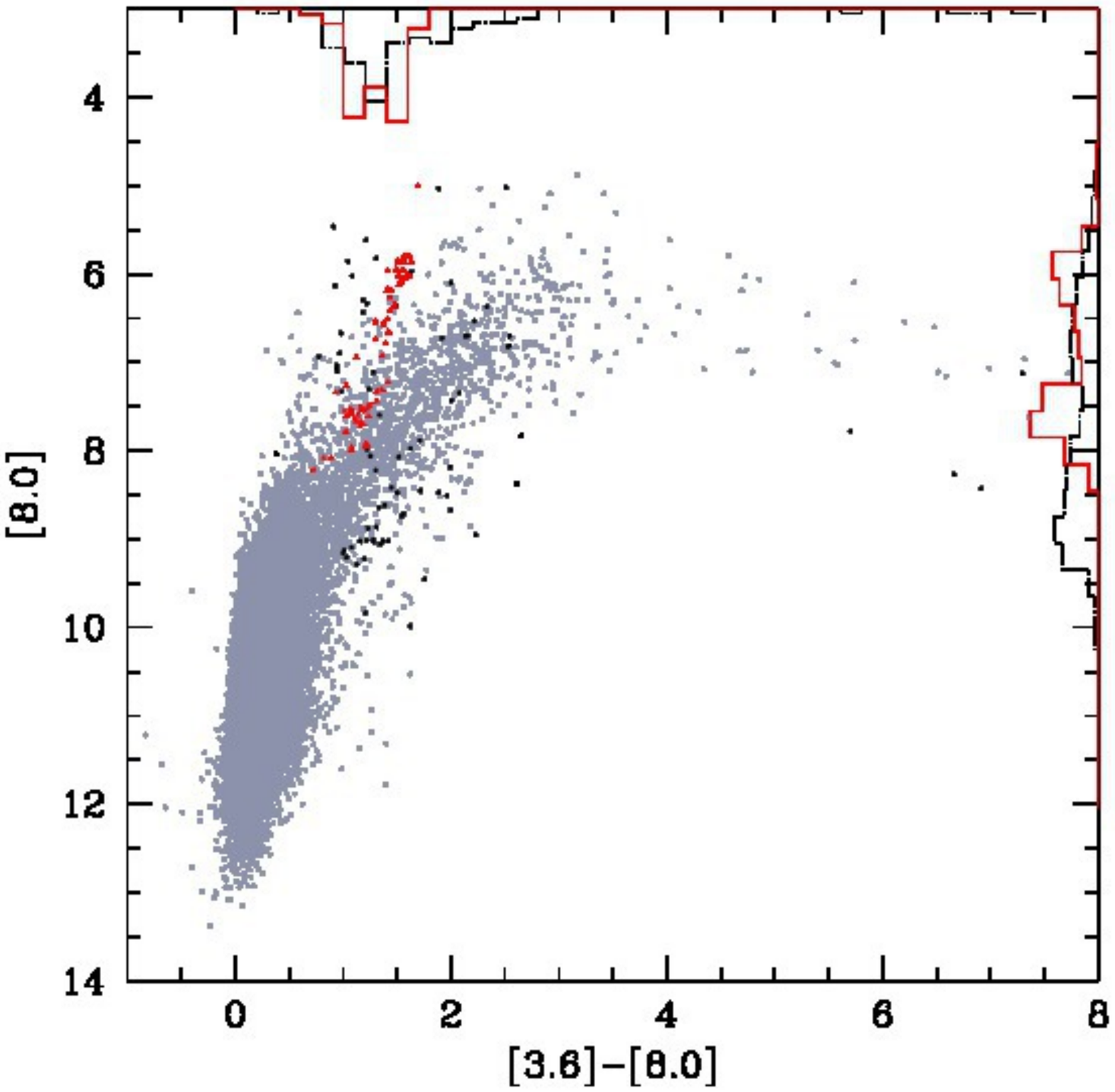}}
\end{minipage}
\vskip-30pt
\caption{The same as Fig. \ref{focs}, but referring to the stars classified as HBBS
in section \ref{class}. The results from synthetic modelling are shown in red.}
\label{fhbbs}
\end{figure*}

\begin{figure*}
\begin{minipage}{0.43\textwidth}
\resizebox{1.\hsize}{!}{\includegraphics{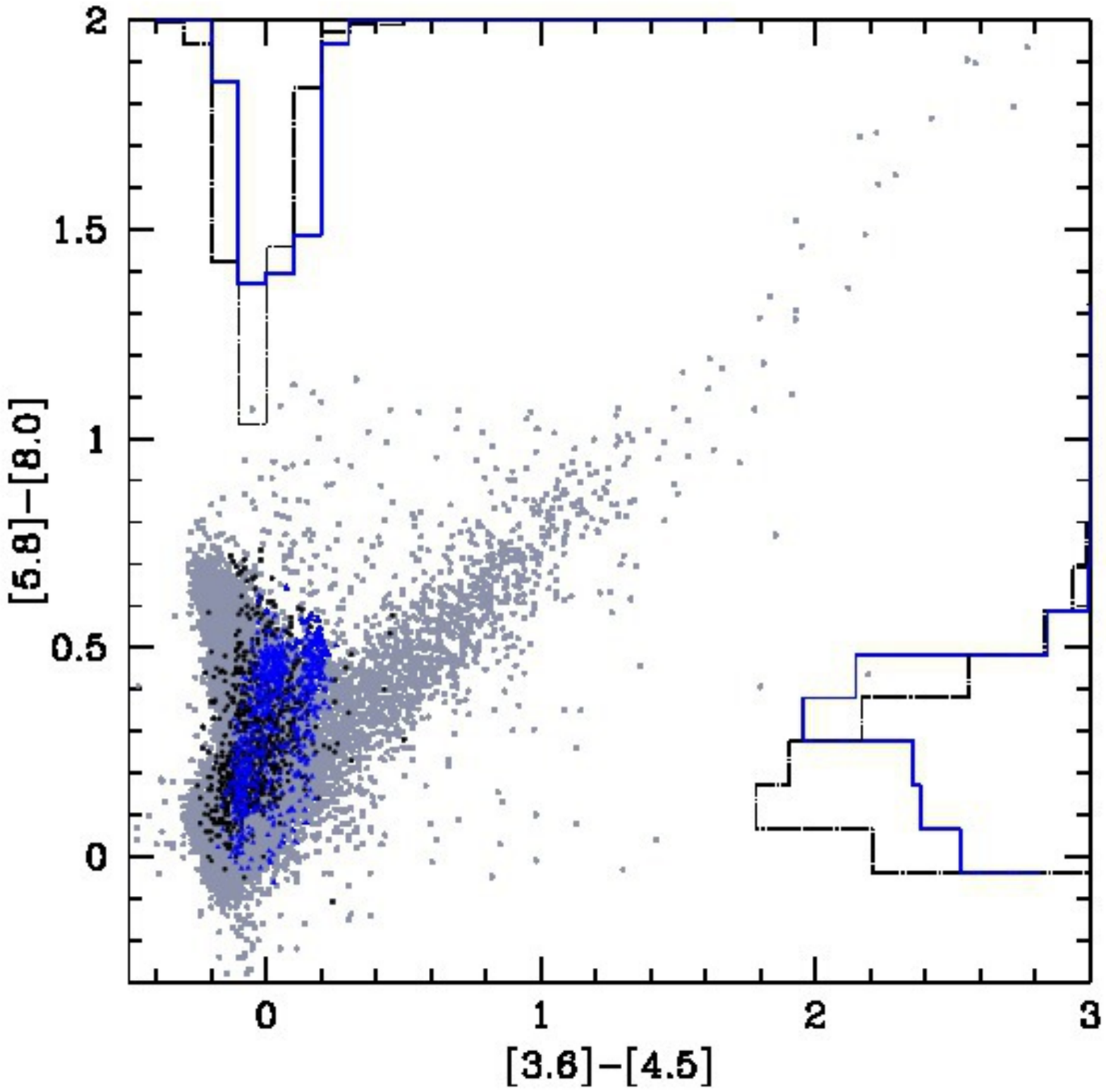}}
\end{minipage}
\begin{minipage}{0.43\textwidth}
\resizebox{1.\hsize}{!}{\includegraphics{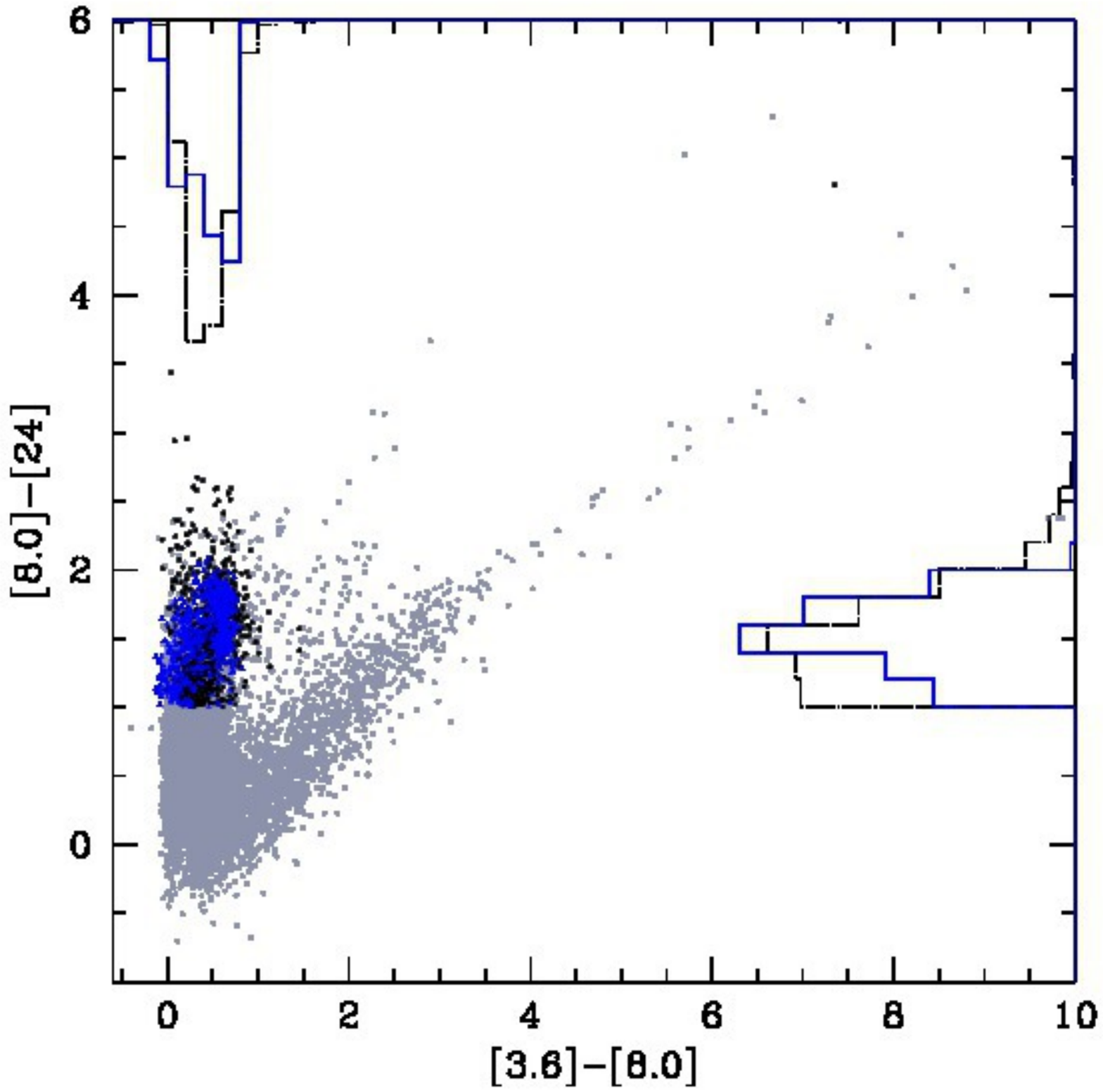}}
\end{minipage}
\vskip-50pt
\begin{minipage}{0.43\textwidth}
\resizebox{1.\hsize}{!}{\includegraphics{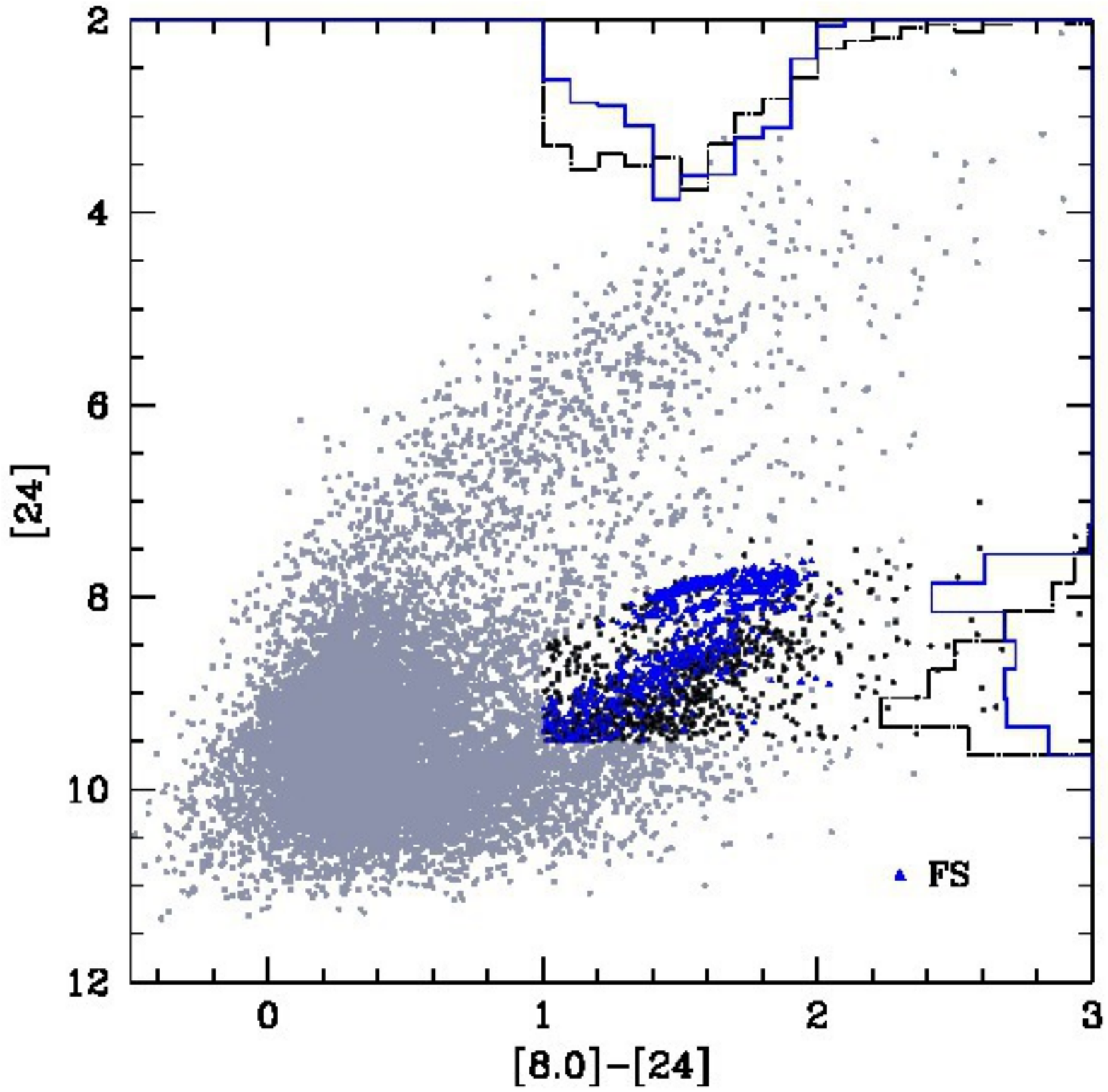}}
\end{minipage}
\begin{minipage}{0.43\textwidth}
\resizebox{1.\hsize}{!}{\includegraphics{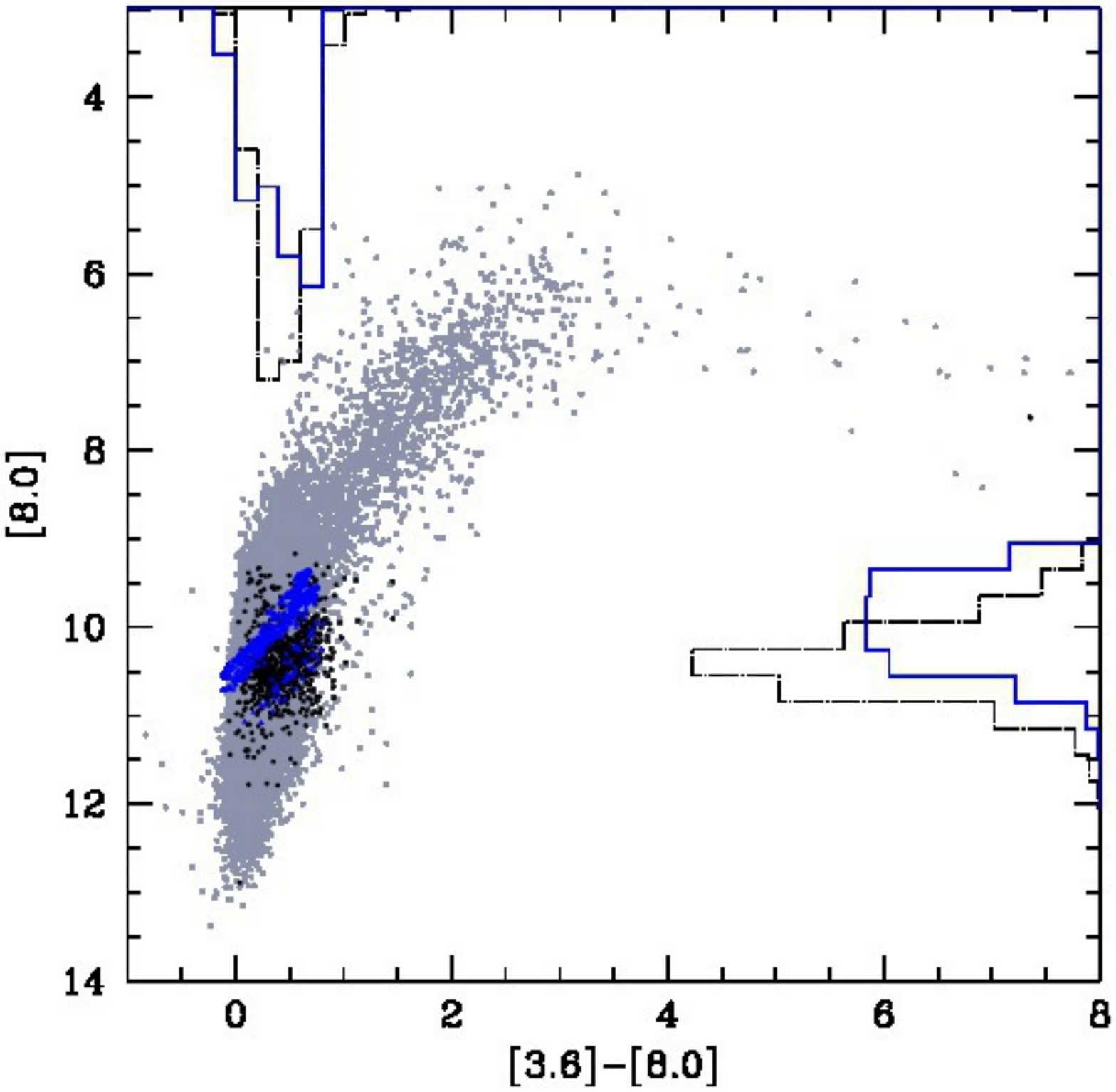}}
\end{minipage}
\vskip-30pt
\caption{The same as Fig. \ref{focs}, for the FS stars populating the finger
identified by \citet{blum06}. The models are plotted in blue.}
\label{ffs}
\end{figure*}

\begin{figure*}
\begin{minipage}{0.43\textwidth}
\resizebox{1.\hsize}{!}{\includegraphics{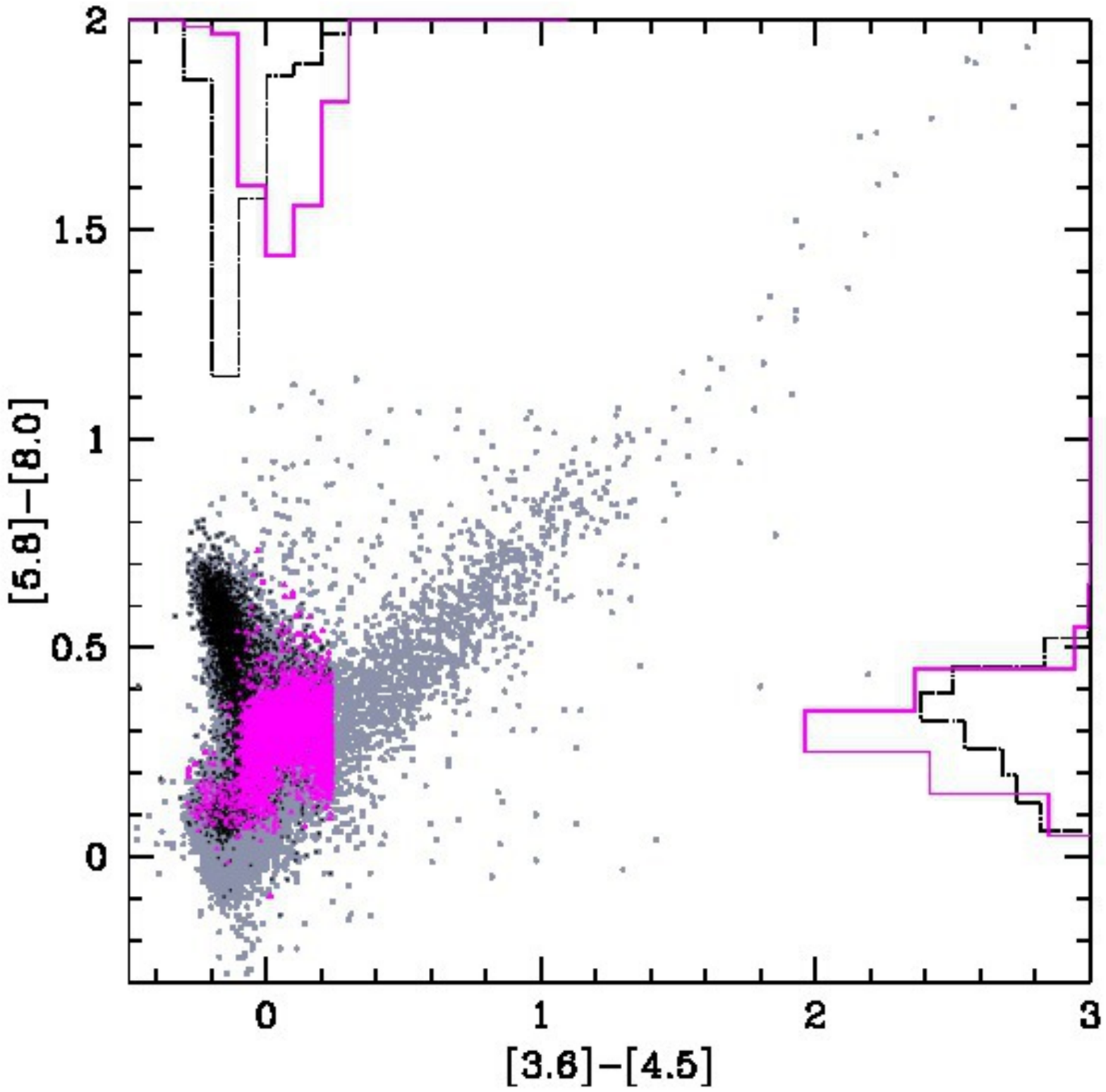}}
\end{minipage}
\begin{minipage}{0.43\textwidth}
\resizebox{1.\hsize}{!}{\includegraphics{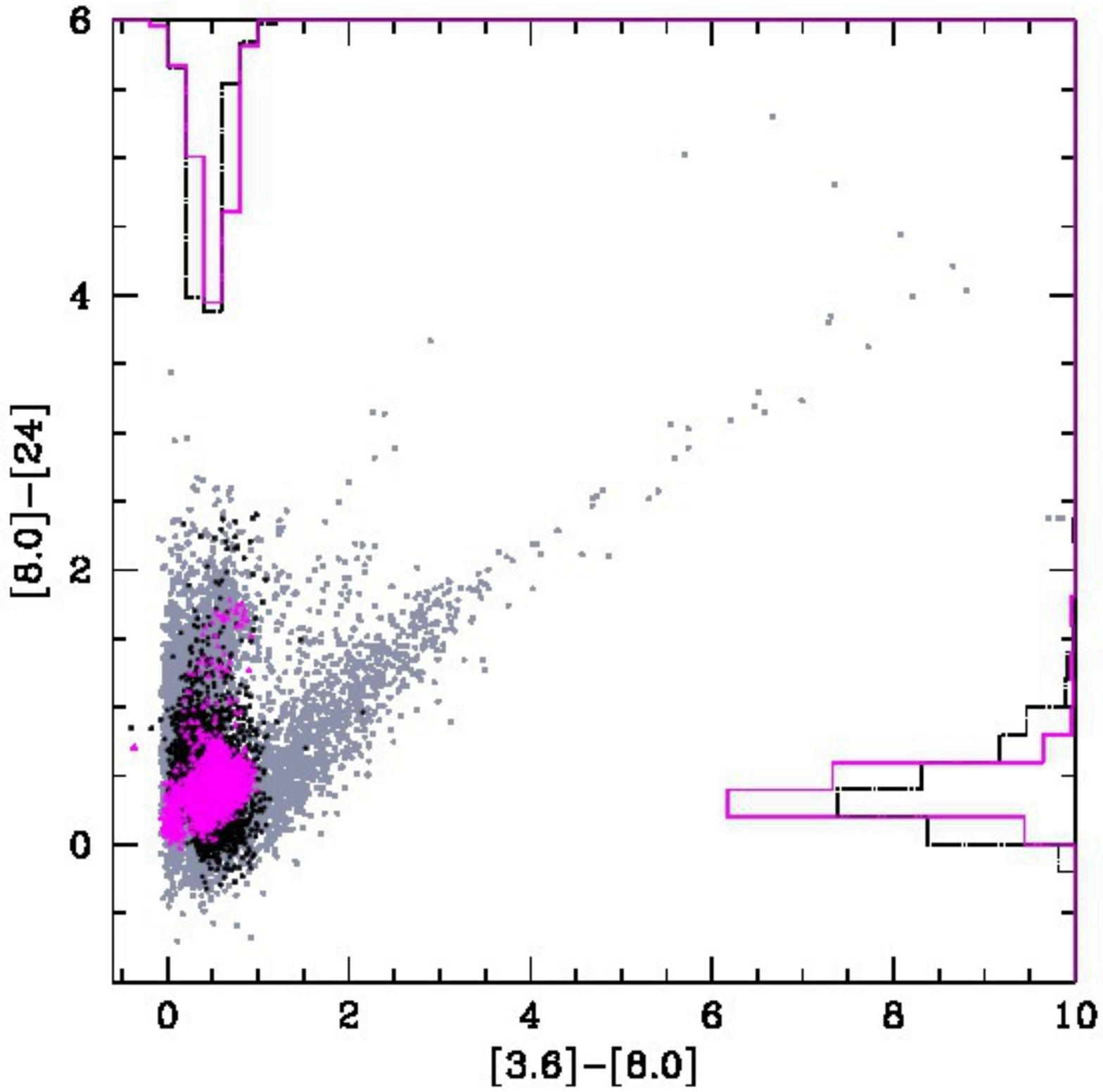}}
\end{minipage}
\vskip-50pt
\begin{minipage}{0.43\textwidth}
\resizebox{1.\hsize}{!}{\includegraphics{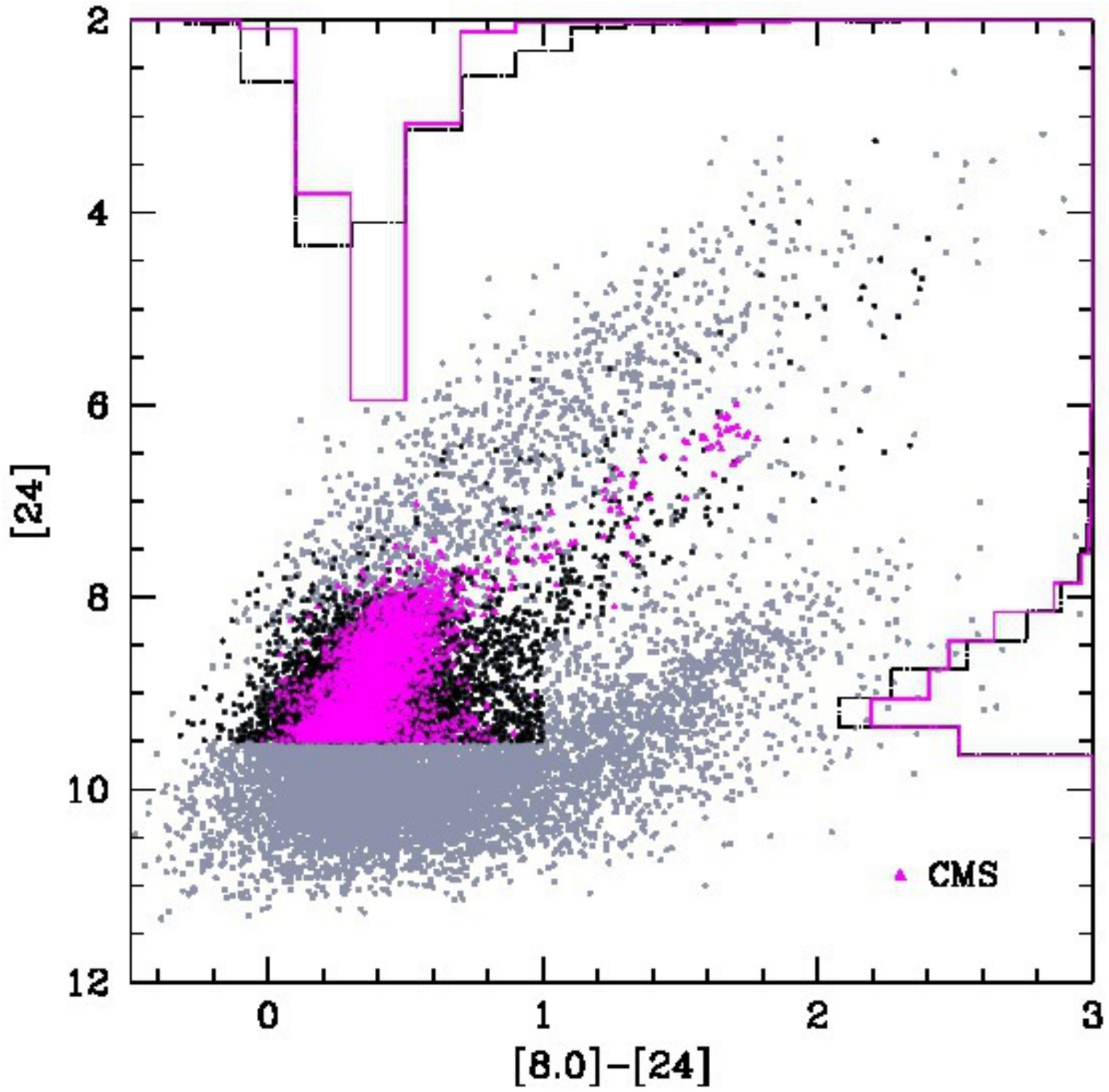}}
\end{minipage}
\begin{minipage}{0.43\textwidth}
\resizebox{1.\hsize}{!}{\includegraphics{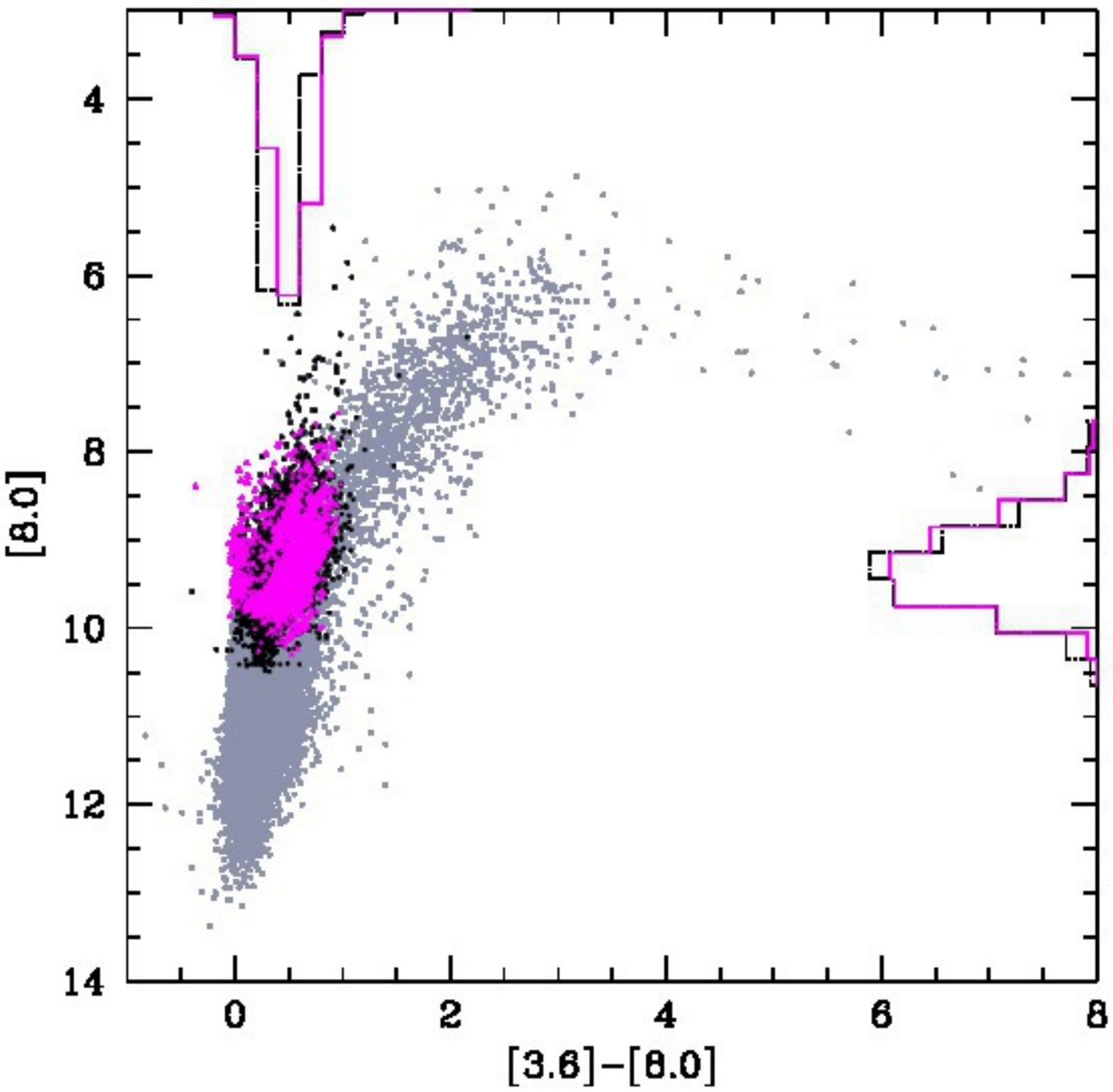}}
\end{minipage}
\vskip-30pt
\caption{The distribution in the colour--colour and colour--magnitude diagrams of the
stars classified as CMS in section \ref{class}. Results from synthetic modelling are
shown in magenta.}
\label{fcms}
\end{figure*}

\subsection{Our selection of the sample}
\label{newsample}
We base our analysis on data available from the SAGE 
survey \citep{meixner06}, particularly the magnitudes in the 3.6, 4.5, 5.8 and 8.0$\mu$m 
IRAC bands, and the 24$\mu$m MIPS band. 

\citet{riebel10} extracted from the SAGE catalogue, containing $\sim 6.5$ million sources,
a list of $\sim 30,000$ evolved stars with high
quality infrared photometry, $\sim 17,000$ of which were classified as AGB stars. From this
sample we selected the objects whose 24 $\mu$m flux is available, ruling out the sources
for which $[24] > 9.5$. This choice allows a full statistical analysis, because
the completeness of the data approaches $100 \%$ at $[24] = 9.5$; also, the $3\sigma$ error
on [8.0]-[24] is above 0.5 mag above this limit \citep{sargent11}, rendering unreliable any
comparison with the observations. After the afore mentioned cut at $[24] < 9.5$, we
are left with a final sample consisting of $\sim 6,500$ stars. With 
this choice we exclude from our analysis the vast majority of low--luminosity, oxygen--rich stars 
present in the original sample by \citet{riebel10}. However, accounting for these objects 
would add only a little contribution to the present investigation, and, more 
importantly, their contribution to the overall dust production is expected to be 
small (below $\sim 5\%$).

What makes LMC an ideal target for studies of stellar populations is the high-galactic 
latitude, that minimizes the foreground contamination. The analysis by \citet{cioni06} 
shows that the (J-Ks, Ks) criterion, adopted to select the AGB sample, is affected by a very 
modest contamination by Galaxy foreground. 2MASS selected sources probably include genuine 
RGB stars at the faintest magnitudes of M-type candidates (Fig. 1 in \citet{cioni06});
however, those sources are excluded from our selected sample, owing to the cut at $[24]<9.5$. 

Concerning distant objects, in the (J-[3.6],[3.6]) diagram (see Fig. 3 
in \citet{blum06}), the locus defined by the external galaxies does not overlap with the
region occupied by AGBs. While distant galaxies are found at $[3.6] > 12.5$, AGBs
populate the brighter part of the diagram, in the regions at $[3.6] < 12$, thus preventing 
a relevant contamination by these objects.  
The detailed analysis by \citet{boyer11} shows that little contamination by foreground
and background sources is expected for the AGB sample in the MCs, with a contamination
of the O--rich AGB sample estimated to be $2.5\%$. 

The reddest sources in our list overlap with the region of the CMD24 also populated by Young
Stellar Objects (YSOs). \citet{whitney08} isolated regions in the colour--magnitude 
($[8.0]-[24]$, $[8.0]$) diagram expected to be populated by YSOs (see Figure 3 in their 
paper). This separation includes a stringent cut at $[8.0]-[24]=2.2$ and 
$[8.0] > 11 + 1.33 \times ([8.0]-[24])$, to exclude AGB stars. We find that $\sim 10$ 
sources in our catalogue occupy this region of the diagram, that represent less then 0.3\% 
of our entire sample.

 \citet{riebel10} classified the AGB stars on the basis of the IR colours, dividing
the total sample among "oxygen--rich", "carbon--rich" and "extreme" objects. This classification
was based on the prescriptions given by \citet{cioni06}, discussed in the previous section. 
Among the stars from \citet{riebel10} included in our selected sample of 
$\sim 6500$ objects, we find that $23\%$ are O--rich, $55\%$ are C--rich and $22\%$ are 
extreme stars. The relative 
fraction of O--rich stars is much smaller than in the original sample analysed by 
\citet{riebel10}, because of our choice of focusing our attention on the sources with 
$[24] < 9.5$, thus ruling out many low--luminosity, oxygen--rich stars.

\subsection{Synthetic diagrams in the Spitzer bands}
\label{diagrams}
The comparison among the models and the observations is based on the analysis of the
CCD1, CCD2, CMD24 and CMD80 diagrams.
This is the best choice to test our theoretical framework, because most of the emission
from dust--enshrouded stars occurs in the infrared bands. The observed distribution of
stars in the various diagrams is compared with the synthetic diagrams, obtained on the
basis of our tracks.

To construct the synthetic diagrams we used the SFH of the LMC given by 
\citet{harris09}, that also provide the relative distribution of stars among the
different metallicities, as shown in Fig. \ref{fsfr}. In the present analysis we assume 
that the $Z=2.5\times 10^{-3}$ stars share the same properties of their $Z=10^{-3}$ 
counterparts: we thus consider three metallicities, namely $Z=10^{-3}$, 
$Z=4\times 10^{-3}$, $Z=8\times 10^{-3}$. 

We trace the history of star--formation rate and the stellar metallicities
 since the formation of the LMC, 15 Gyr ago. The time steps used are:
10Myr for the epoch ranging from 100Myr ago to now; 100Myr for the epoch
going from 1Gyr to 100Myr ago; 1Gyr for the epochs previous to 1Gyr ago.

At each time step we extract randomly a number of stars, distributed among the three
metallicities considered, determined by the following factors: a) the value of the star formation
rate; b) the relative percentages of stars of different metallicities; c) the 
duration of the entire AGB phase of the star that has just completed the core helium 
burning phase in the epoch considered. We used a Salpeter's IMF, with index $x=2.3$.
The outcome of this work consists in a series of points extracted along the tracks of 
the various masses considered; for each point the infrared magnitudes are obtained by 
calculating a synthetic spectrum, as described in section \ref{spectramodel}.

Fig. \ref{synpop} shows the results of our simulations, with the expected distribution 
of stars in the CCD1 (left panel) and CMD24 (right). Following the classification 
introduced in section \ref{class}, 
the stars have been coloured according to the group they belong to: OCS, defined as
the stars falling in region II in the CCD1 (left panel of Fig. \ref{synpop}), are shown 
in green; HBBS, populating region I in the CCD1 (left panel of Fig. \ref{synpop}), are 
indicated in red; FS stars, defined as the stars in the F zone of the CMD24
(right panel of Fig. \ref{synpop}), are shown in blue; finally, magenta points indicated
CMS stars, defined as the objects in zone III in the CCD1, not belonging to the FS group. 
In the same figure we also show
the spectroscopically confirmed C--stars by \citet{gruendl08}, \citet{zijlstra06},
\citet{woods11}, and O--rich sources by \citet{sloan08} and \citet{woods11}.

The outcome of this synthetic approach is the simulation of the whole AGB sample
in the LMC. However, coherently with the criterion for selecting the sample given in
section \ref{newsample}, we will use in the statistical analysis described in the
following sections only the stars extracted with $[24]<9.5$.

\begin{table}
\begin{center}
\caption{Percentage of the stars belonging to the four groups in which we divided the 
observed sample and synthetic population of AGBs used in the present analysis (see text in 
section \ref{class} for details).}
\begin{tabular}{l|c|c|c|c}
\hline
 & OCS & HBBS & FS & CMS  \\ 
\hline
observed &  19  &  1  & 12 &  68  \\
expected &  22  &  1  & 11 &  66  \\
\hline
\end{tabular}
\end{center}
\label{tabpop}
\end{table}

\section{Understanding the observations of AGB stars in the LMC}
The analysis presented in this section is based on the comparison between observations of AGB
stars in the LMC and the theoretical predictions, obtained by the synthetic modelling
described in the previous section.

Following the classification introduced in section \ref{class}, we first compare the star counts
in the zones I, II, III, F shown in Fig. \ref{ftracceccd}, \ref{ftraccecmd} and \ref{synpop} with 
those observed. We check consistency among the number of objects in each group, by
comparing the observed and expected distributions of colours and magnitudes. 

The goal of the present analysis is twofold. On one hand we test the reliability 
of our theoretical understanding of the physics of AGBs, in terms of their 
evolutionary properties and the dust composition of their envelopes. At the same time,
this approach allows a characterization of the stars observed, to determine their 
age, metallicity, surface chemical composition, dust present in their circumstellar
envelope.

The results are shown in Table 2, where we report the observed and predicted fractions
of stars in each group. The overall agreement is very good. For each of the four groups introduced in 
section \ref{class},
Fig. \ref{focs}--\ref{fcms} show the comparison between the observed and expected 
distribution of stars in the CCD1 (top--left panels), CCD2 (top--right), 
CMD24 (bottom--left), CMD80 (bottom--right) diagrams. In each panel we show the 
observed points, present in the sample used here, extracted from \citet{riebel10},
as black points. The stars from our simulation falling in each group are indicated
with coloured points, using the same coding as in Fig. \ref{synpop}. For completeness,
we also show as grey points in Fig. \ref{focs}--\ref{fcms} all the stars in the original 
sample by \citet{riebel10}.

We now discuss separately the stars in each group.

\subsection{Obscured carbon stars}
\label{secocs}
OCS are defined as the stars populating region II in the CCD1, shown in 
Fig. \ref{ftracceccd} and \ref{synpop}. In Fig. \ref{focs} we show the observed points 
belonging to this group in the CCD1, CCD2, CMD24 and CMD80 as black dots; the stars from 
our simulations and classified as OCS are indicated in green. 
According to our modelling, no oxygen--rich star is expected to evolve into this region of the CCD1 
(see left panels of Fig. \ref{ftracceccd}), thus the OCS group is entirely composed of 
C--stars. This is in agreement with the interpretation of the colour--colour 
$([3.6]-[4.5], [5.8]-[8.0])$ diagram of the LMC given by \citet{srinivasan11} (see their
Figure 7). The authors identified the stars in the diagonal band traced by OCS as
carbon stars; the sources with the redder colours correspond to the objects with the
larger optical depth.

In the CCD1, OCS trace a diagonal band, with $0.2 < [3.6]-[4.5] < 3$, 
$0.2 < [5.6]-[8.0] < 1.6$. The optical depth increases along this sequence, ranging 
from $\tau_{10}=0.01$ for the least obscured dust objects, at $[3.6]-[4.5] \sim 0.2$, to 
$\tau_{10}=3.5$, for the stars surrounded by optically thick envelopes, with 
$[3.6]-[4.5] \sim 3$. A similar behaviour is followed in the CCD2, with 
$[3.6]-[8.0]$ ranging from $\sim 0.5$ to $[3.6]-[8.0] \sim 7$.

Based on the arguments discussed in section \ref{cstarmod}, we interpret OCS as an
evolutionary sequence. Once they achieve the carbon--star stage, the stars become 
progressively more obscured and reddened (see the dashed lines in Fig. \ref{fcolours})
because: a) as a consequence of repeated TDU episodes, the surface 
carbon abundance increases (see Fig. \ref{fcstar}), thus increasing the density of carbon 
molecules available for condensation; b) the increase in carbon leads to cooler temperatures 
\citep{marigo02}, which makes the dust formation region closer to the surface of
the star, in a higher density region.

The evolution of the amount of dust formed around C--rich AGBs and the sequence of
the different dust layers present are discussed in details in previous papers by our group 
\citep[see, e.g.,][]{paperIV}, where an interested reader can also find a detailed 
discussion on the size of the particles of the individual species present in the 
circumstellar envelope (see in particular Figure 5 in \citet{paperIV} and Figure 1 
in \citet{flavia14}).
These stars are surrounded by two dusty layers: a) a more internal region, $\sim 2$
stellar radius away from the surface of the star, with SiC grains of $\sim 0.05-0.08 \mu$m
size\footnote{This results depends on the metallicity, because the abundance of the 
key--element to form SiC, i.e. silicon, scales with Z \citep{nanni13b, paperIV}.};
b) a more external zone, $\sim 5R_*$ from the surface, with SiC and 
solid carbon particles. The latter grains determine most of the obscuration of the
radiation coming from the star; their dimension ranges from $\sim 0.05 \mu$m
in the less obscured OCS ($0.01 < \tau_{10} < 0.1$), up to $\sim 0.2 \mu$m in the most heavily 
dust obscured objects  ($\tau_{10} > 3$) \citep{nanni13b, paperIV}.

\begin{figure*}
\begin{minipage}{0.33\textwidth}
\resizebox{1.\hsize}{!}{\includegraphics{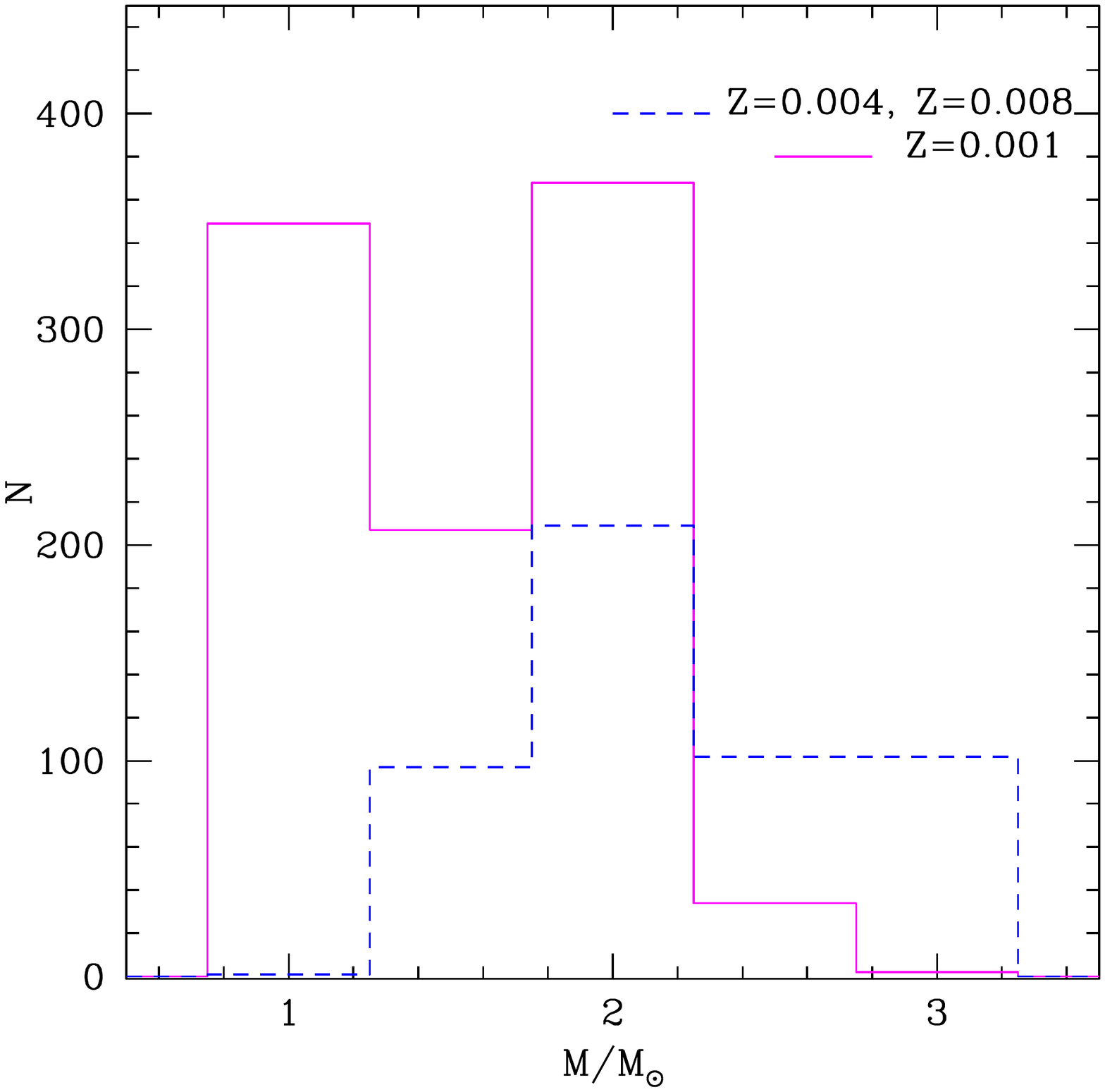}}
\end{minipage}
\begin{minipage}{0.33\textwidth}
\resizebox{1.\hsize}{!}{\includegraphics{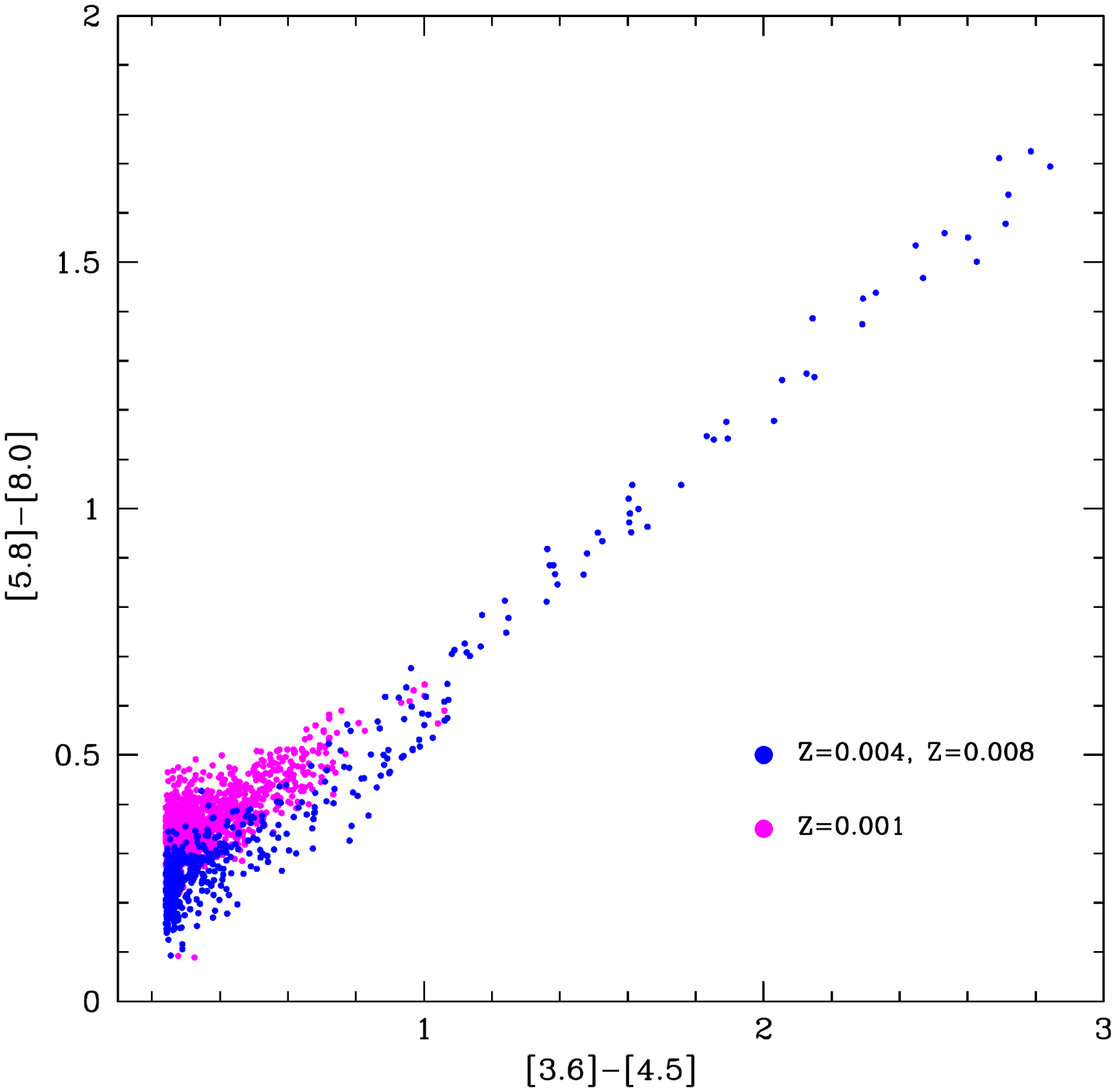}}
\end{minipage}
\begin{minipage}{0.33\textwidth}
\resizebox{1.\hsize}{!}{\includegraphics{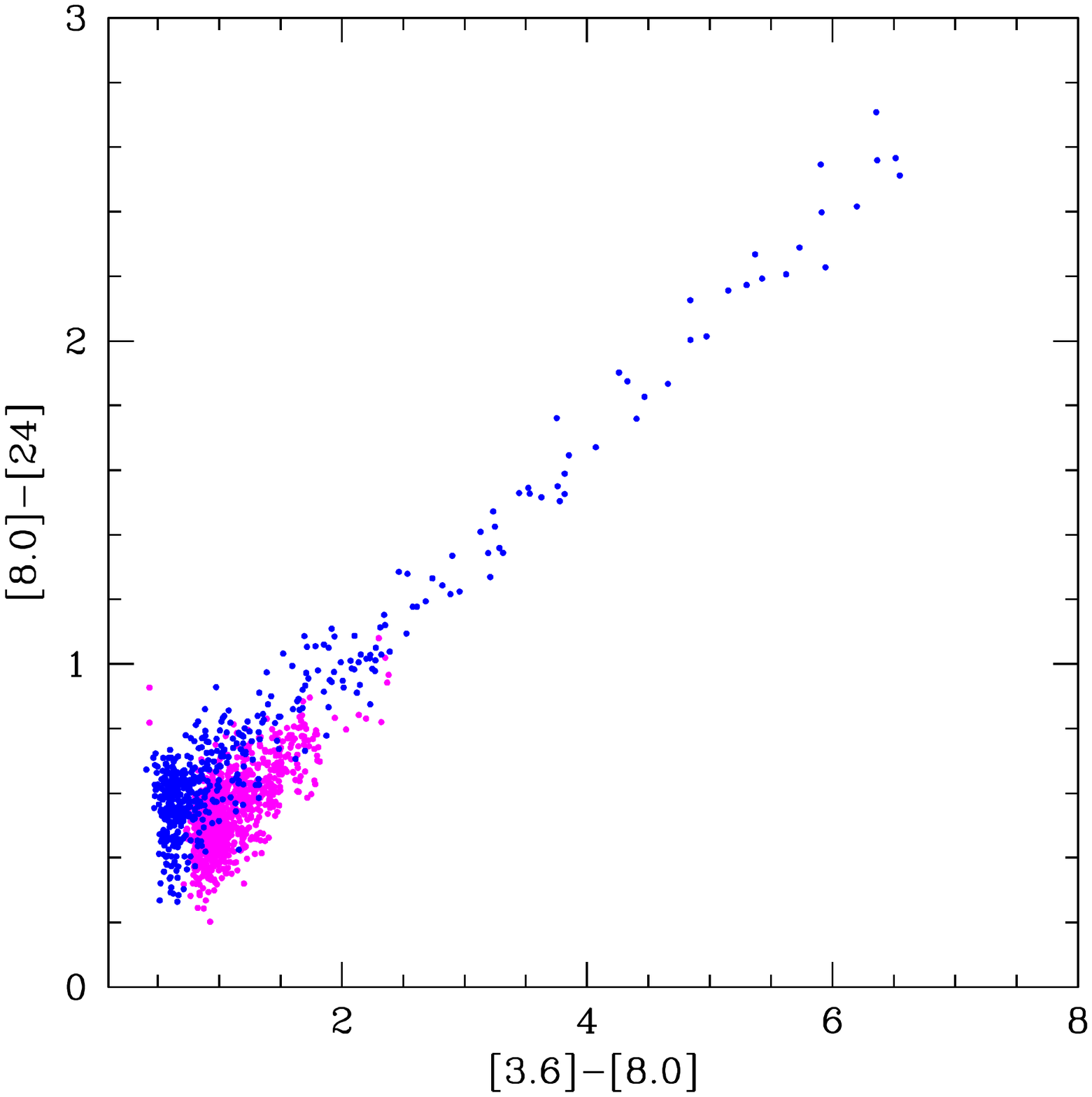}}
\end{minipage}
\vskip-30pt
\caption{Left panel: the distribution of OCS in term of initial mass and metallicity. 
Magenta, solid line refers to $Z=10^{-3}$, while the blue, dashed line represents 
$Z \geq 4 \times 10^{-3}$. Middle and right panels: the expected population of dust obscured, 
C--rich stars in the colour--colour $([3.6]-[4.5], [5.8]-[8.0])$ (middle) and
$([3.6]-[8.0], [8.0]-[24])$ (right) planes. Magenta points refer to the $Z=10^{-3}$ 
population, whereas $Z=4-8\times 10^{-3}$ stars are indicated with blue dots.}
\label{focsz}
\end{figure*}

The progenitors of OCS are stars of initial mass in the range $1M_{\odot} < M \leq 3M_{\odot}$, 
formed $3 \times 10^8 - 3\times 10^9$ years ago (see Table 1). Younger
(and more massive) objects experience HBB, thus not reaching the C--star stage.
The left panel of Fig. \ref{focsz} shows the predicted distribution
of OCS in terms of initial mass and metallicity of the progenitors. The majority of OCS are
the descendants of low--mass stars with masses $M \sim 1-2M_{\odot}$, formed during the
burst of SFH in the LMC that occurred $\sim 2$ Gyr ago \citep{harris09}. These objects are 
mainly low--metallicity (Z below $4\times 10^{-3}$) stars, that spend $\sim 70\%$ 
of their AGB lifetime as carbon-stars (see right panel of Fig. \ref{fcstar}). A not 
negligible tail ($\sim 10\%$ of the total number of stars extracted, classified as OCS) 
of higher Z ($Z>4 \times 10^{-3}$) stars of mass 
$2M_{\odot} \leq M \leq 3M_{\odot}$ is evident in Fig.\ref{focsz}. The latter group of stars
are the descendants of objects formed during the peak in the SFH occurring $\sim 5\times 10^8$ 
year ago (see Table 1 for the evolutionary timescales of the individual 
masses), when the majority of the stars have a metallicity $Z>4 \times 10^{-3}$, as
shown in Fig. \ref{fsfr} \citep{harris09}.

Though smaller in number, this group of more massive OCS, of metallicity 
$Z \geq 4\times 10^{-3}$, play a relevant role in the 
interpretation of the observed CCD1 and CCD2, as they are the only stars expected to evolve
redder than $[3.6]-[4.5] > 1$ and $[3.6]-[8.0] > 2.5$ (see left and middle panels 
of Fig. \ref{fcolours})
We will refer to these models as Extremely Obscured Carbon Stars (EOCS). 
The reason for this is twofold: a) stars more massive than $\sim 2M_{\odot}$ 
experience a high number of TDU episodes, thus they accumulate great quantities
of carbon in the surface regions, which reflects on a high--efficiency formation of
solid carbon particles; b) as shown in Fig. \ref{fcstar}, the degree of obscuration 
reached by lower--Z models is smaller, because they evolve at larger surface temperatures, 
which causes the dusty layer to form at larger distances and smaller densities. 
The evident decrease in the number of stars on the reddest portion of the CCD1 and
CCD2 (populated by the EOCS) is partly due to the
fact that only a limited range of masses of the more metal--rich component is expected to 
evolve in those zones of the planes; a further reason is the short duration of the most
obscured phase (Fig. \ref{fcolours}), a consequence of the strong increase in the 
rate of mass loss during these evolutionary stages.

An obvious difference in the dust composition among models of different metallicity is
the quantity and size of SiC grains formed. In the $Z=10^{-3}$ case, owing to the
scarcity of silicon, the formation of SiC (if any) is extremely modest, whereas in 
stars with $Z=4-8\times 10^{-3}$ the contribution of
SiC to the thermal emission of dust (which can be expressed via the optical
depth $\tau_{10}$) is far from being negligible. The silicon
available in the surface of the star is smaller than carbon, thus solid carbon is produced
in larger quantities than SiC; however, SiC particles form closer to the surface of the star, in a relatively higher temperature region, thus providing an important contribution 
to $\tau_{10}$.
The variation in the SiC/C ratio is the reason for the spread in the observed 
$[5.8]-[8.0]$ colours of OCS in the region $[3.6]-[4.5] < 1$ of the CCD1. This observational
evidence (see, e.g., the left--top panel of Fig. \ref{focs}) is confirmed by our 
simulations. The middle panel of Fig. \ref{focsz} shows the 
metal distribution of stars in this zone of the CCD1, determined by our synthetic modelling: 
models of low-- and high--metallicity are indicated, respectively, with magenta and blue
points. This plot shows that, for a given $[3.6]-[4.5]$, low--metallicity stars assume 
redder $[5.8]-[8.0]$ colours, while more metal--rich objects populate the lower portion of the CCD1.
Looking at the theoretical tracks in this plane, shown in the left panels of 
Fig. \ref{ftracceccd}, we see that the latter region of the CCD1 is reproduced by the track 
of low--mass stars belonging to the more metal--rich population, in the phases following 
the beginning of the carbon--star phase. Spectral observations of carbon stars located in
magenta regions will find a weaker SiC feature, compared with blue region with
$[3.6]-[4.5] < 1$.
The spread in the observed sequence of OCS vanishes for $[3.6]-[4.5] > 1$:
as shown in Fig. \ref{ftracceccd}, this stems from the lack of metal poor
stars in this zone of the CCD1, that is populated only by stars of
metallicity $Z\geq 4\times 10^{-3}$ (compare the tracks of $Z=10^{-3}$ and
$Z=8\times 10^{-3}$ models in the left panels of Fig. \ref{ftracceccd}).

The situation in the CCD2 is similar: the diagonal band traced by OCS in this plane 
exhibit an intrinsic width, that becomes progressively smaller for $[3.6]-[8.0] > 3.5$.
The right panel of Fig. \ref{focsz} shows that even in this case the spread is associated 
to the metal content of the star, metal--rich objects populating the upper side of the 
CCD2.

Among the stars in the sample defined in section \ref{newsample}, $22\%$ are found
to be in region II in the CCD1 shown in the left panels of Fig. \ref{ftracceccd}, that 
according to our interpretation is populated by OCS. This is in nice agreement with our 
prediction ($20\%$, see Table 2), confirming that the overall duration of the C--rich 
phase for the masses involved in this process is well predicted by our models. In terms of 
the colour--distribution of the OCS, we see in the top--left and top--right panels of 
Fig. \ref{focs} that our distribution of OCS is excessively peaked towards the less dust
obscured objects, indicating that the transition to the highly--obscured phase is too slow. 
This is presumably due to the large sensitivity on the effective temperature of the mass 
loss rate adopted \citep{wachter02, wachter08}, that makes the whole residual envelope to 
be lost rapidly as the external regions of the star become carbon rich, possibly indicating 
the need of a softer dependance of $\dot M$ on $T_{eff}$.

In the CMDs shown in Fig. \ref{focs} (bottom panels) the OCS trace a diagonal band, 
which, consistent with their distribution in the CCDs, can be interpreted 
as an evolutionary sequence, towards higher degrees of obscuration. During the carbon--rich 
phase, the total luminosity changes little at a given initial 
mass \citep[see middle panel of Fig. 1 in][]{flavia14}.
Hence, the increase of the $8.0\mu$m and $24\mu$m flux are not due to the 
increase of the luminosity, rather to the increase in the dust optical depth, 
that makes more efficient the absorption of optical and near infrared photons emitted by 
stars and the re--emission at mid--infrared wavelengths. 

Unlike the CCD1 and CCD2, in the CMD24 plane the OCS population is partially 
overlapped to the bright oxygen--rich stars, thus inhibiting the possibility of discrimination among 
the two samples. This can be seen by the comparison of the position of the OCS and HBBS
in the CMD24 plane, shown, respectively, in the bottom--left panels of Fig. \ref{focs} and 
Fig. \ref{fhbbs}. The tracks of OCS and HBBS shown in the left panels of 
Fig. \ref{ftraccecmd} further support this conclusion.

As shown in the bottom--right panels of Fig. \ref{focs} and Fig. \ref{fhbbs}, and confirmed 
by the tracks shown in the right panels of Fig. \ref{ftraccecmd}, the situation is more clear 
in the CMD80, where the two sequences are separated. The observations show that the
majority of stars with $[3.6]-[8.0]>1.5$, $[8.0]<8$ are carbon stars 
\citep{matsuura09, woods11}, and we agree that this region is mainly occupied with high 
mass-loss rate C--stars. Indeed we see in Fig. \ref{synpop} that our OCS models nicely fit 
the position of spectroscopically identified carbon--rich stars sampled by 
\citet{zijlstra06} in the CCD1 and CMD24. The 
same models also reproduce the IR colours of the C--rich sample by \citet{woods11}. 
However, a few stars belonging to the sample by \citet{gruendl08} and \citet{woods11}
are barely reproduced by our models: this suggests the need for an improvement in the description 
of the star+dust systems of C--rich stars in the very latest evolutionary phases. 
OCS constitute the vast majority of dust obscured stars in the CMD24 and CMD80 planes, 
in the regions $[8.0] < 8$ and $[24] < 6$; this is evident in the distribution of stars in
Fig. \ref{focs}.

Concerning the distribution among the various metallicities, we find that, similarly
to the CCD1 and CCD2, only metal--rich stars evolve to the redder regions of the 
CMD24 and CMD80, as can be seen in Fig. \ref{ftraccecmd}, showing the evolutionary
tracks in these planes. We identify the two regions with $[3.6]-[8.0] > 2$ and 
$[8.0]-[24] > 1$ as populated by the most obscured OCS belonging to the more metal--rich 
populations.

\subsection{Stars experiencing Hot Bottom Burning}
In section \ref{class} we defined HBBS as the stars in region I in the CCD1, shown in
Fig. \ref{ftracceccd}. This zone is populated by a group of stars clustering
around $[3.6]-[4.5] \sim 0.2$, $[5.8]-[8.0] \sim 0.8$, detached from the rest of the LMC
population of AGBs. The observed stars falling in the HBBS region are shown with black 
dots in Fig. \ref{fhbbs}; the red points indicate results from our simulation. 
We interpret these sources (see the tracks shown in Fig. \ref{ftracceccd}) as 
the descendants of massive AGBs, with mass initially above $3M_{\odot}$, experiencing strong 
HBB at the base of the convective envelope. 
According to the arguments presented in section \ref{dustmodel}, the circumstellar
envelope of HBBS hosts a more internal ($\sim 2R_*$ away from the surface of the star),
largely transparent zone, populated by alumina grains, and a more external region,
$\sim 10R_*$ far from the surface, where silicates grains form and grow. The latter dust 
species is the most relevant in determining the degree of obscuration of the star.

The occurrence and the strength of HBB is the key--quantity to determine the amount of
silicates formed. The degree of obscuration is considerably smaller than
their C--rich counterparts ($\tau_{10}$ is below 2 in all cases): this difference
stems from the much higher availability of carbon molecules in the envelope of carbon
stars, compared to the abundance of silicon in the outer regions of oxygen--rich stars.

We see in the left panels of Fig. \ref{ftracceccd} that the 
theoretical tracks of HBBS in the CCD1 and CCD2 are more vertical than those of OCS.
In the CCD1, the reason is the prominent silicate feature at $\sim 9.7 \mu$m, which 
determines an increase in the $\sim 8.0 \mu$m flux, thus rendering the  $[5.8]-[8.0]$ 
colour extremely red. This effect is clearly evident in the middle panel of Fig. \ref{fcolours}, 
showing that, in comparison with C--stars, oxygen--rich stars experiencing HBB evolve at redder 
$[5.8]-[8.0]$, which reach the highest values once strong HBB conditions are experienced. 
For what concerns CCD2, the higher slope of the tracks of HBBS compared to OCS
stems from the optical properties of silicates, that reprocess the radiation emitted from
the central star, with a substantial emission at mid--infrared wavelengths.
 
HBBS formed during the burst in the SFH in the LMC occurring $\sim 10^8$ years ago, as
shown in Fig. \ref{fsfr}. These sources descend from stars with initial mass in the range 
$3.5M_{\odot} \leq M \leq 7.5M_{\odot}$\footnote{The strength of HBB
experienced by intermediate mass AGBs is extremely sensitive to the convection model
used \citet{vd05}. The models presented in this work are based on the FST treatment, 
that favours strong HBB in all stars more massive than $\sim 3M_{\odot}$. In AGB models
based on the traditional Mixing Length Theory, HBB is found in a narrower range of masses
(see the detailed discussion in \citet{ventura13} on this argument)}. They belong to the more
metal--rich population, because of the small percentage of low--Z stars 
formed in these epochs \citep{harris09}. Also, stars of $Z<4\times 10^{-3}$, with the exception of massive SAGBs,
produce only a modest quantity of dust \citep{paperII}, thus they are not expected to
evolve into the region in the CCD1 plane (region I in the left panels of Fig. \ref{ftracceccd}) 
populated by HBBS (see bottom--left panel of Fig. 
\ref{ftracceccd}).

The paucity of objects in the HBBS region (they account for $\sim 1\%$ of the total sample,
see Table 2) stems not only from the intrinsically small number of stars formed in 
the relevant range of mass, but also as a consequence of HBB: as shown in 
Fig. \ref{fhbb}, HBB produces a fast increase in the luminosity of the star, that, in 
turn, favours a rapid loss of the residual external mantle. The limitation of the HBBS 
population to the metal--rich component is a further reason for the small number of HBBS 
observed. 

In the colour--magnitude ($[3.6]-[8.0]$, $[8.0]$) diagram (see bottom--right panels 
of Fig. \ref{focs} and \ref{fhbbs}) the HBBS define an almost vertical sequence, separated 
from OCS. The reason is once more the silicate feature, that renderes the $[8.0]\mu$m 
flux of HBBS brighter than OCS at a given $[3.6]-[8.0]$. The observations 
have shown that high mass--loss rate oxygen--rich AGB stars, though minority in number, 
contaminate the region $[3.6]-[8.0]>1.5$, $[8.0]<8$ \citep{matsuura09, woods11}. Our 
models do not predict such a population, indicating that the excursion of the
theoretical tracks in this plane is too vertical, with no bending towards redder
$[3.6]-[8.0]$ colours.
Our massive AGB models experience large mass loss rates, strongly favouring the formation 
of silicates. We therefore rule out that this effect originates from 
the description of the AGB evolution. The discrepancy among the observations and
the theoretical predictions suggests a problem in the shape of the synthetic
spectra in the region of the silicates feature, that would affect the theoretical $[8.0]\mu$m 
flux. More detailed explorations, using different set of the optical constants of silicates, 
are needed to confirm this hypothesis.

The same separation among OCS and HBBS is not clear in the ($[8.0]-[24]$, $[24]$) 
plane, as can be seen in the bottom--left panels of Fig. \ref{focs} and \ref{fhbbs}.

As shown in Fig. \ref{synpop}, by looking at the theoretical distribution of stars 
in the CCD1 and CMD24 planes, a fraction of the AGB stars spectroscopically classified as 
O--rich in the sample by \citet{woods11} have similar colours of the HBBS population.
 \citet{sargent11} suggested that obscured oxygen--rich stars populate the region 
in the CCD1 at $[3.6]-[4.5] \sim 0.2-0.3$, $[5.8]-[8.0] \sim 0.8-1$, where, according to our 
interpretation, should evolve stars experiencing HBB (see their Fig. 5).
In their analysis, the authors presented a 
wide exploration of the various parameters relevant for the determination of the 
infrared colours (effective temperature, optical depth, inner border of the dusty region,
etc.): the grid of models for oxygen--rich stars was shown first to extend in the direction
of redder $[3.6]-[4.5]$ and $[5.8]-[8.0]$, then, after reaching the position occupied by
HBBS, to turn to much redder $[3.6]-[4.5]$, with $[5.8]-[8.0]$ becoming bluer
\citep[see Fig. 5 in][]{sargent11}.

Our theoretical sequences of oxygen--rich stars experiencing HBB follow a similar path (see the
orange track in the left panel of Fig. \ref{synpop}); however,
our models do not extend beyond the HBBS zone, because we find $\tau_{10} < 2$ in all 
cases, at odds with \citet{sargent11}, that explored the range $10^{-4} < \tau_{10} < 26$.

Concerning the interpretation of the CMDs, we stress that for HBBS, unlike OCS, the spread
in the $8.0\mu$m and $24\mu$m fluxes partly reflects a difference in the overall  luminosity
of the stars observed (see Fig. \ref{fhbb}). The HBBS on the upper side of the two CMDs
($[8.0]$ and $[24] \sim 6$) correspond to the more massive stars ($M\sim 5-6M_{\odot}$), experiencing 
the strongest HBB, with scarce contamination from TDU. This offers the opportunity of 
testing observationally this interpretation, because these sources should show--up the 
typical signatures of proton capture nucleosynthesis, with $^{13}C/^{12}C \sim 0.3$, 
and $C/N \sim 0.02-0.03$. This test could also be used do distinguish massive AGBs from 
Red Super Giant stars, that occupy the same regione in CMD24 and CMD80. The
use of lithium is not straightforward in this context, because 
the lithium--rich phase is rapidly terminated once $^3He$ is consumed in the envelope 
\citep{sackmann92, mazzitelli99}. 

\subsection{Stars in the "finger" identified by \citet{blum06}}
In a paper focused on the infrared color--magnitude diagrams of evolved stars in the LMC,
\citet{blum06} noticed in the colour--magnitude ($[8.0]-[24]$, $[24]$) diagram a sequence of
O--rich candidates defining a prominent finger, spanning a range of $24\mu$m excess  
of $\sim 2$ mag. These stars, shown with black dots in Fig. \ref{ffs}, were identified by 
the authors as a faint population of dusty sources with significant mass loss. 

We refer to this group of stars, populating the region F \citep[here we use the original 
definition by][]{blum06} in the left panels of Fig. \ref{ftraccecmd}, as FS. 
According to our interpretation, this region is populated by low--mass M stars, with
metallicity in the range $4\times 10^{-3} < Z < 8\times 10^{-3}$, of 
initial mass slightly above $1M_{\odot}$, in the AGB phases immediately before becoming 
C--stars. The track of a $M=1.25M_{\odot}$ model of metallicity $Z = 8\times 10^{-3}$, 
evolving into the F region, is shown as a black line in the top--left panel of Fig. \ref{ftraccecmd}. 
The evolution of the same $1.25M_{\odot}$ model, in terms of the excursion of the evolutionary
track in the CMD24 and the SED at some selected evolutionary phases, is shown in 
Fig. \ref{spettri125}. The initial excursion to the red, shown in the right panel of
Fig. \ref{spettri125}, is due to the larger and larger quantities of dust produced in
the circumstellar envelope, while the following turn to the blue, at $[8.0]-[24] \sim 1.8$, 
coincides with the beginning of the C--star phase.
The CMD24 is by definition the best plane where the FS population can be distinguished
from the other AGBs in the sample used here. However, inspection of Fig. \ref{focs}--\ref{fcms}
suggests that also in the CCD2 plane FS populate a well identified, almost vertical region, 
with no OCS and HBBS, and with a limited number of CMS (see next section).
In the CCD1 and CMD80, FS are overlapped to the CMS stars, that will be discussed
in the following section.

FS are the descendants of low--mass ($M \leq 1.5M_{\odot}$) stars 
with metallicity $Z \geq 4\times 10^{-3}$, formed a few Gyr ago; in the bottom--left panel 
of Fig. \ref{ftraccecmd} we see that metal--poor objects do not evolve into the F region, 
owing to the small amount of silicate dust formed in their surroundings.
The nice agreement between the predicted and observed number of FS stars (see Table 2)
confirms the relative duration of the O--rich phase in these low--mass models, as also the 
total duration of the AGB phase, once they reach the C--star stage.
The comparison with the observations in Fig. \ref{synpop} shows that our models of
FS stars nicely fit in the CMD24 plane the position of a fraction of AGB stars in the 
sample by \citet{woods11}, classified as O--rich. 
A word of caution is needed here. While these results, particularly the relative number
of FS stars and the reddest points reached in the ($[8.0]-[24]$, $[24]$) plane 
during their evolution, can be used to further confirm the reliability of the AGB models,
the same does not hold for the description of the dust formation process.
Unlike the cases so far examined, here the wind is not expected to suffer a great
acceleration under the effects of radiation pressure acting on dust particles; this
renders the results partly dependent on the assumptions concerning the initial velocity
with which the gas particles enter the condensation zone. 

On the side of the surface elemental abundance, we expect that these stars reflect 
 substantially the initial composition of the gas from which they formed, considering 
that neither TDU nor HBB modified their composition. Only a slight increase in the helium
and nitrogen content and a decrease in the carbon mass fraction are expected, as a
consequence of the first dredge--up.

\begin{figure*}
\begin{minipage}{0.33\textwidth}
\resizebox{1.\hsize}{!}{\includegraphics{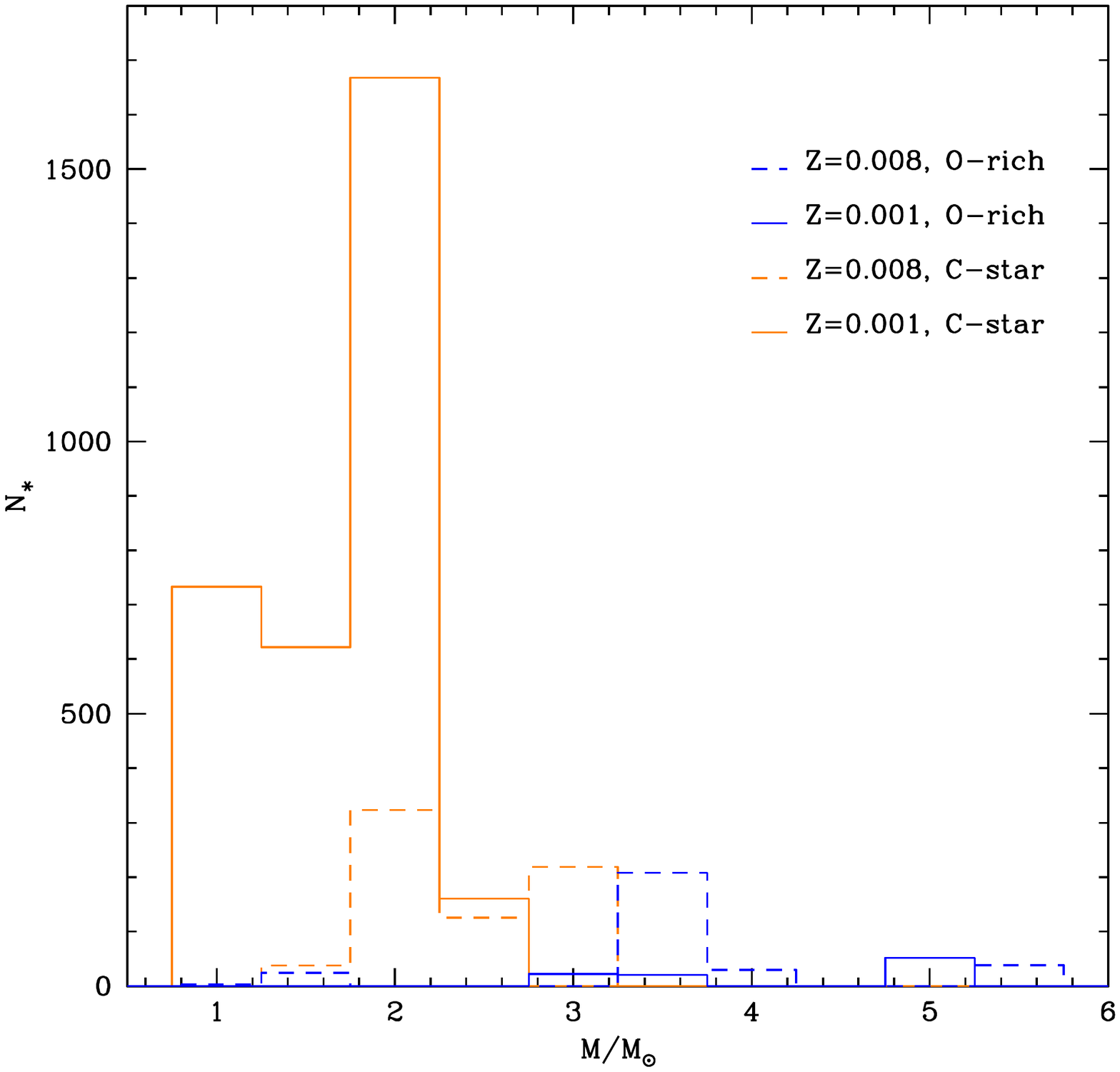}}
\end{minipage}
\begin{minipage}{0.33\textwidth}
\resizebox{1.\hsize}{!}{\includegraphics{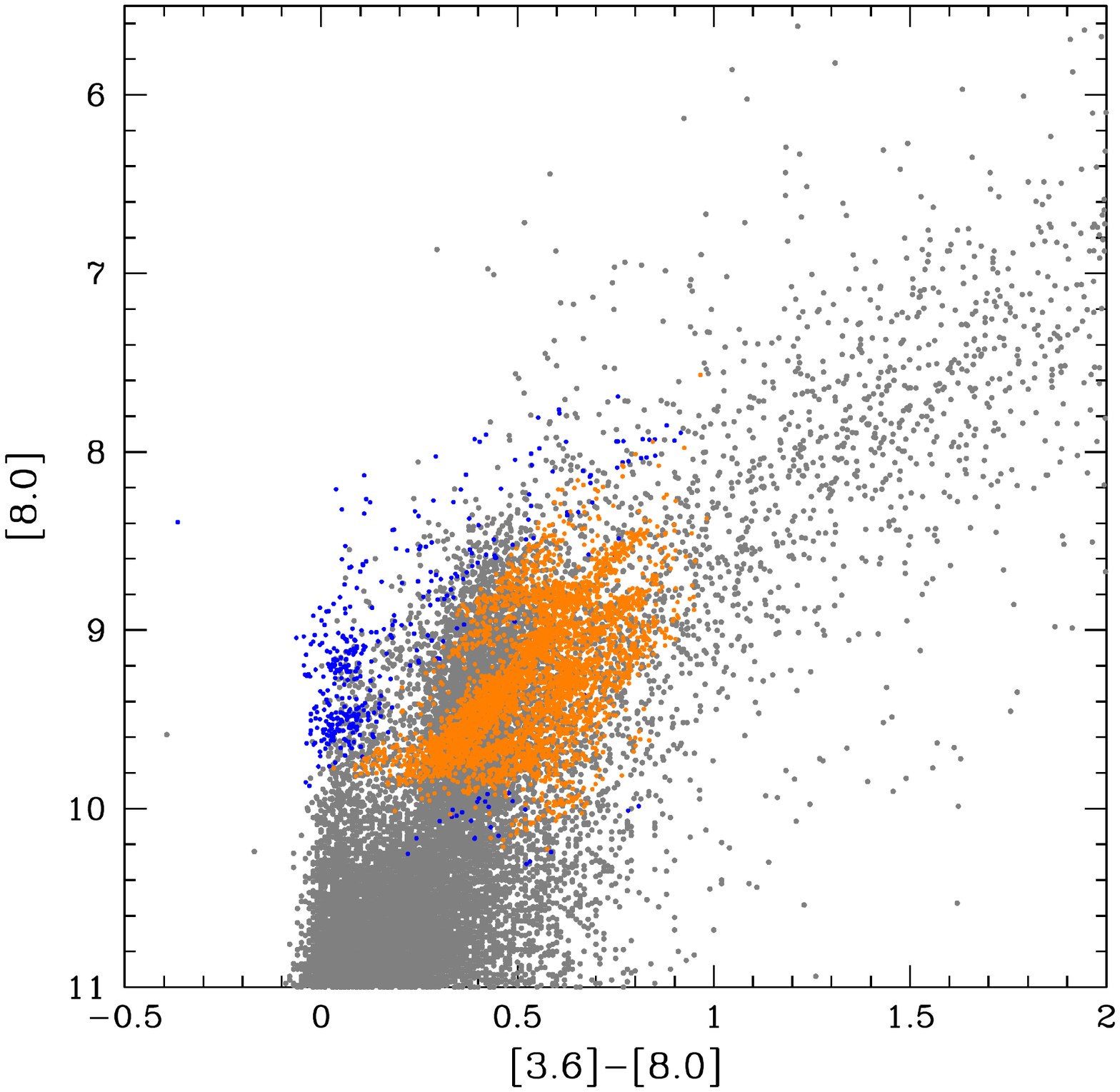}}
\end{minipage}
\begin{minipage}{0.33\textwidth}
\resizebox{1.\hsize}{!}{\includegraphics{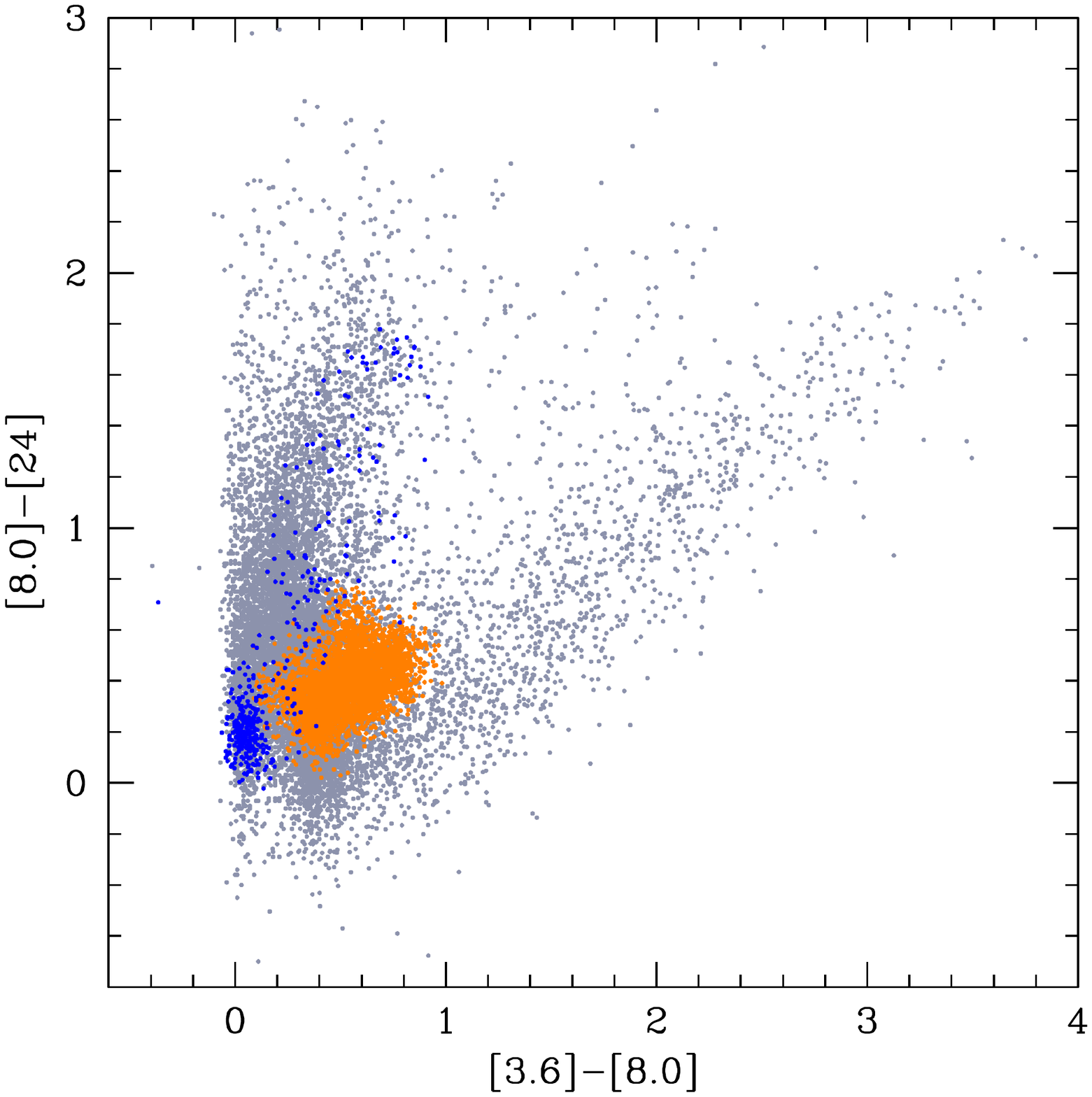}}
\end{minipage}
\vskip-30pt
\caption{Left: the distribution of CMS in term of initial mass and metallicity 
(dashed-line for $Z=8\times 10^{-3}$ and solid line for $Z=10^{-3}$), divided in O--rich 
(blue) and C-stars (orange). Middle: the expected population of CMS in [3.6]-[8.0] 
vs [8.0] plane. Blue points refer to the O--rich population, whereas C--stars
are indicated with orange dots. Right: the CMS population in the colour--colour 
($[3.6]-[8.0],[8.0]-[24]$) diagram, with the division among carbon-- and oxygen--stars.}
\label{mucchio}
\end{figure*}

\begin{table*}
\begin{center}
\caption{Main properties of the dust obscured AGBs in the LMC} 
\begin{tabular}{c|l|c|c|c|c|c|c}
\hline
  & Class & Z & Age (Gyr) & $M/M_{\odot}$ & $\tau_{10}$ & dust (size)& surface elemental abundance  \\ 
\hline
OCS & Extreme ($95\%$), & $10^{-3} (70\%)$  & 0.4-3 & 1-3 & 0.02-1 & C (0.07-0.15$\mu$m) & $10^{-3}<X(C)<0.03$ \\ 
    & C($5\%$) & $4-8\times 10^{-3} (30\%)$ & & & & SiC ($ < 0.08 \mu$m) & $3\times 10^{-5}<X(N)<10^{-3}$ \\
    & & & & & & & $5\times 10^{-4}<X(O)<5\times 10^{-3}$ \\
\hline    
EOCS & Extreme & $4-8\times 10^{-3}$ & 0.5 & 2.5-3 & 1-3 & C (0.15-0.25$\mu$m) & $5\times 10^{-3}<X(C)<0.015$  \\
    & & & & & & SiC (0.08$\mu$m) & $4\times 10^{-4}<X(N)<8\times 10^{-4}$ \\
    & & & & & & & $4\times 10^{-3}<X(O)<5\times 10^{-3}$ \\
\hline    
HBBS & Extreme ($20\%$),  & $8\times 10^{-3}$ & 0.1-0.3 & 3.5-6 & 0.1-1 & Sil. (0.1$\mu$m) & CNO processed  \\
& C($20\%$), O($60\%$) & & & & & Al$_2$O$_3$ (0.05-0.07$\mu$m) & \\
\hline
FS & O & $4-8\times 10^{-3}$ & 2-10 & 1-2 & 0.01-0.1  & Sil. (0.07-0.09 $\mu$m) & oxygen--rich star  \\
& & & & & & Al$_2$O$_3$ (0.003$\mu$m) & \\
\hline
CMS & O($10\%$), C($90\%$) & $10^{-3} (80\%)$ & Any & 1-6 & $< 0.1$ & various & various  \\
& & $4-8\times 10^{-3} (30\%)$ & & & & &  \\
\hline
\end{tabular}
\end{center}
\label{tabelLMC}
\end{table*}

\begin{figure*}
\begin{minipage}{0.33\textwidth}
\resizebox{1.\hsize}{!}{\includegraphics{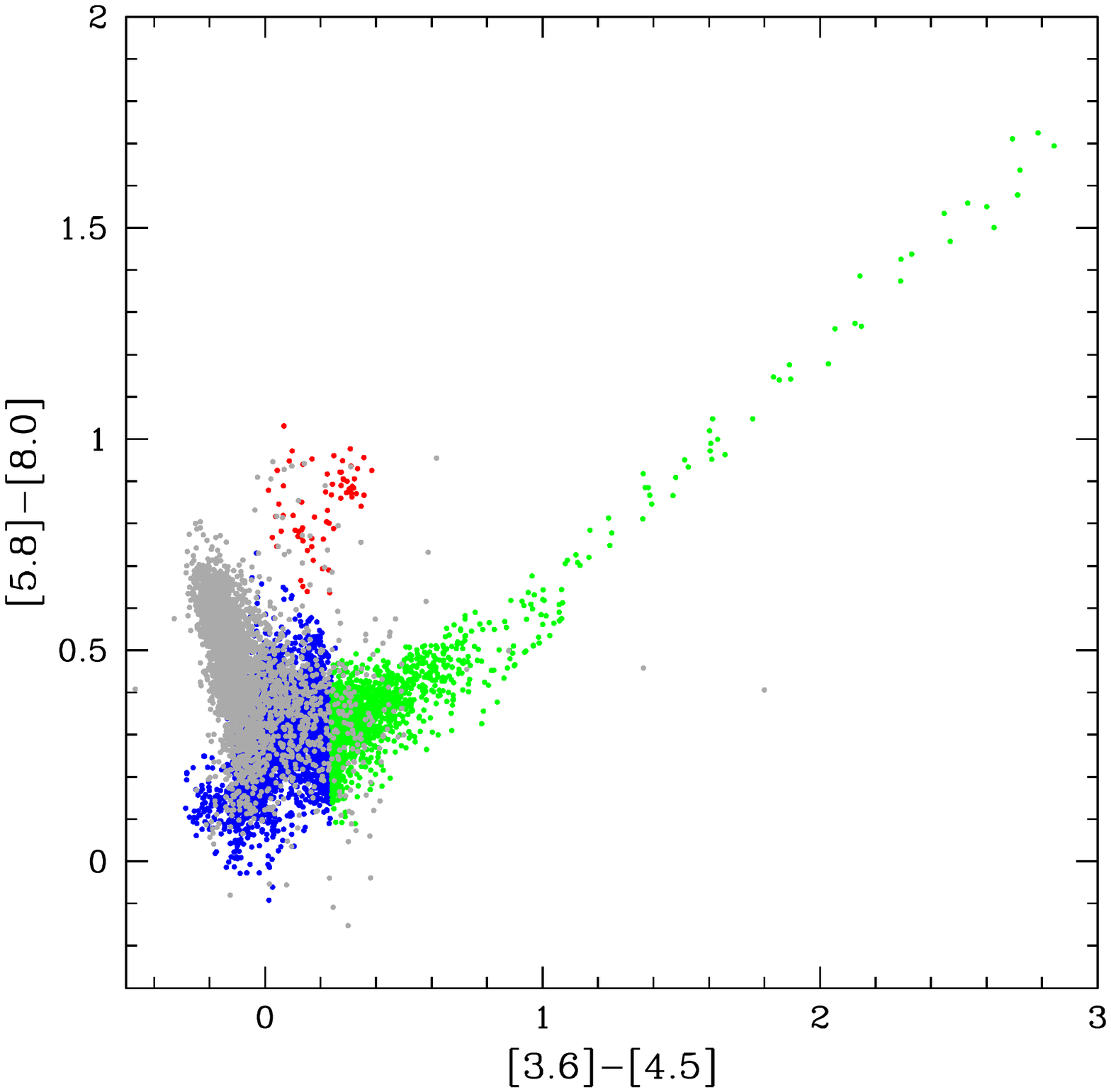}}
\end{minipage}
\begin{minipage}{0.33\textwidth}
\resizebox{1.\hsize}{!}{\includegraphics{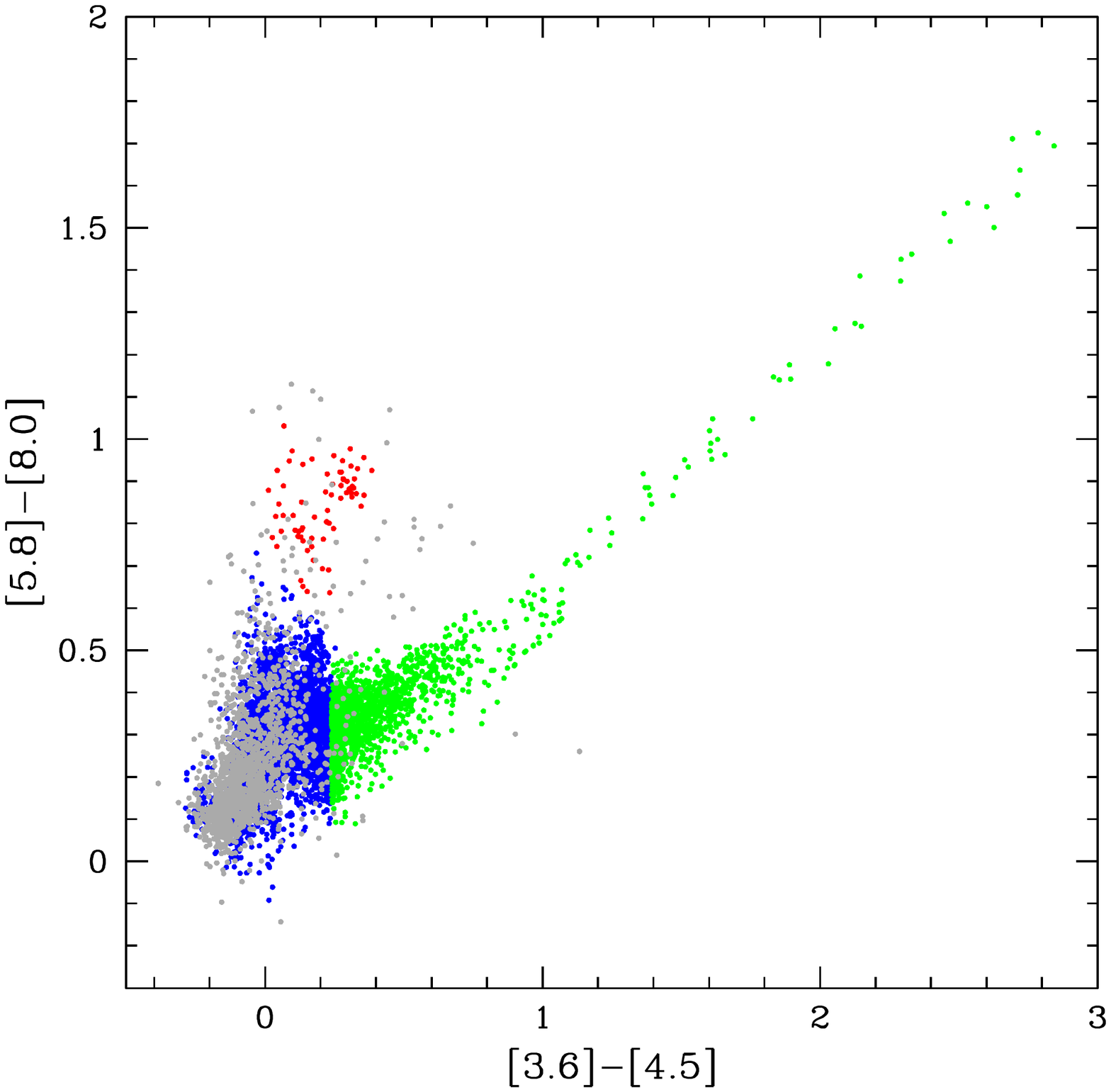}}
\end{minipage}
\begin{minipage}{0.33\textwidth}
\resizebox{1.\hsize}{!}{\includegraphics{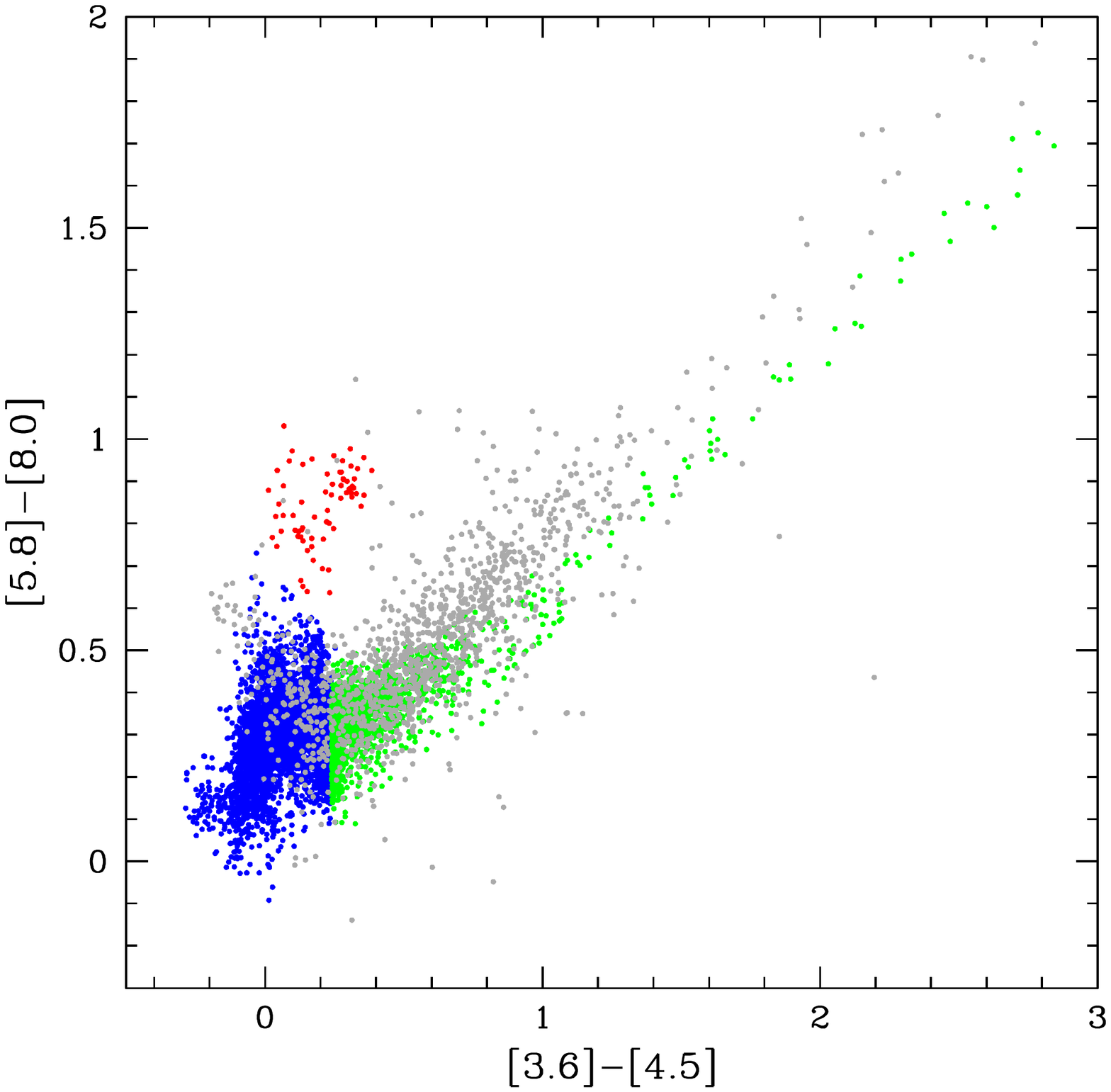}}
\end{minipage}
\vskip-30pt
\caption{Comparison between the position (grey dots) of the AGBs classified by 
\citet{riebel10} as C--rich (left panel) and O--rich (middle) candidates
and extreme stars (right) in the colour--colour ($[3.6]-[4.5],[5.8]-[8.0]$) diagram, 
compared to the locii defined by OCS (green points), HBBS (red points), FS and CMS stars. 
In this plane FS are indistinguishable from CMS, thus we show the FS+CMS groups 
together (blue points).
}
\label{fclass}
\end{figure*}

\subsection{The scarcely dust obscured objects}
The last group of objects considered are those in the Spitzer catalogue populating 
region III in the CCD1 (see left panels of Fig. \ref{ftracceccd} and \ref{synpop}); 
because this region is also populated by FS stars, the latter have been subtracted in the 
identification of this sample. Clearly the statistical analysis will be restricted to the 
sources with $[24]<9.5$, in agreement with the definition of the AGB sample used in this
work, specified in Section \ref{newsample}.

This subsample accounts for $\sim 65 \%$ of the entire population (see Table 2); it is 
made up of both carbon and oxygen--rich stars (CMS), with a modest degree of obscuration. 
Our models indicate that the optical depth 
$\tau_{10}$ is generally below 0.01, reaching $\tau_{10}=0.1$ in a limited number of cases. 
In the color--magnitude diagrams, shown in the bottom panels of Fig. \ref{fcms},  
they are within the regions with $8 < [8.0] < 10$ and $7.5 < [24] < 9.5$\footnote{The 
upper limits given above depend on our choice to limit this investigation to the stars 
observed at $[24] < 9.5$}. The agreement between the observations and the predictions is 
once more satisfactorily in these planes.

According to our analysis, the majority of CMS are low--metallicity models whose
initial mass is in the range $1-2M_{\odot}$. This is represented in the left panel of 
Fig. \ref{mucchio}, showing the mass and metallicity distribution of CMS. 
$\sim 90\%$ of this sample are carbon stars, either with a small C/O ratio, or in the phases
immediately following the extinction of thermal pulses. The remaining
$\sim 10\%$ are oxygen--rich stars, mainly low-mass objects in the early 
AGB phases, when they are still oxygen--rich; a limited number of more massive 
sources in the phases before the activation of HBB is also expected. 

The relative fraction of stars in the observed sample used here, belonging to the
CMS group, is found to be in nice agreement with the expected percentage of CMS, as
indicated in Table 2. This indicates that the duration of the various evolutionary phases
obtained via the theoretical models is consistent with the observational picture.

Compared with the observations of spectroscopically--confirmed stars, we see in
Fig. \ref{synpop} that our models of CMS explain the IR colours in the CCD1 and CMD24
diagrams of the O--rich stars by \citet{sloan08} and of part of the C--rich stars by
\citet{woods11}.

In the CMD2 the position of the points from our simulation
nicely fit the observed distribution, as shown in the top--right panel of 
Fig. \ref{fcms}. A few of CMS populate the region $1 < [8.0]-[24] < 2$, mainly occupied
by FS (see top--right panel of Fig. \ref{ffs}). Concerning the CCD1, we note 
that the observed distribution of colours of CMS is 
not fully reproduced by the models: in particular, a group of stars in the region  
around $[3.6]-[4.5] \sim -0.2$, $[5.8]-[8.0] \sim 0.6$ are out of the range covered by 
the theoretical tracks (see top--left panel of Fig. \ref{fcms}).

We identify most of these stars as carbon--rich ($C/O > 3$) objects not heavily 
obscured, in the phases immediately following the thermal pulse, before dust is produced 
in great quantities: as a consequence, the synthetic spectra are essentially determined 
by the spectrum of the central star, barely modified by the optically thin envelope.
No influence of the dust formation scheme is expected here. We note that the same
problem in reproducing the position of these stars in the CCD1 was already found by
\citet{srinivasan11} (see their Fig. 7, and the discussion in section 4.2.5). Note that
a few stars on the O--rich sample by \citet{woods11} fall in this region of the CCD1.
Both theoretical tracks presented in Fig. \ref{ftracceccd} and \ref{ftraccecmd} and the 
spectroscopically confirmed objects in Fig. \ref{synpop} show that the distribution of 
M-stars and C--rich objects in the CMS sample overlap in the CCD1 and CMD24 planes, 
thus preventing the possibility to disentangle between the two populations. 
Conversely, as shown in the middle and right panel of Fig. \ref{mucchio}, in the CCD2 and
CMD80 plane the two groups are fairly separated, the oxygen--rich stars occupying the 
bluer region, C--rich models defining a redder, parallel sequence. Indeed the tracks of 
the two groups of models bifurcate in this plane, as shown in the right panels of 
Fig. \ref{ftraccecmd}. The reason for this is in the strong depression of the $3.6\mu$m 
flux that characterises the C--rich models, that makes the $[3.6]-[8.0]$ colours redder. 
Spectroscopic analysis of these stars could further confirm this interpretation.

In Table 3 we present a summary of our interpretation. For each of the four groups in 
which we divide the LMC AGBs we give information on the metallicity, epoch of formation, 
initial masses of the precursors, surface chemical composition and dust present in the 
circumstellar envelope. We also show (second column) how stars in each group would be 
sampled according to the original classification by \citet{riebel10}; we discuss this 
point in the next section.

\section{The interpretation of observed C--rich, O--rich candidates
and Extreme AGBs in the LMC}
As explained in section \ref{histclass}, \citet{riebel10} proposed a classification of AGB stars based on their 
Spitzer colour and magnitudes. The final sample of stars, published in \citet{riebel12},
is divided among carbon stars (C), oxygen--rich objects (O) and 'extreme' stars. The
criterion followed to distinguish among C--rich and O--rich objects is based on the
scheme by \citet{cioni06}, and is shown in Fig. 1 in \citet{riebel12}. Extreme AGB 
candidates were selected based on the $J-[3.6]$ colour.

We introduced slightly tighter sample selection 
than \citet{riebel10} for quality control purpose. We further define a new 
classification scheme, based on theoretically predicted tracks and LMC 
populations/star-forming history.

We compare now the results from our interpretation with the photometric classification by \citet{riebel10, riebel12}, to check whether our 
models are in agreement with the division in C-- and oxygen--rich stars given by these authors, 
and to characterize the sample of objects classified as "extreme" by \citet{riebel10}.

Table 3 (column 2) illustrates the composition of each of the four AGB groups in terms of the classification by \citet{riebel10}.
 For each population we also give information 
concerning the formation epoch, the degree of obscuration and the properties of the dust 
in the wind. Fig. \ref{fclass} shows the 
comparison between the position of the AGBs classified by \citet{riebel10} as C--rich 
(left panel), O--rich (middle) candidates and extreme stars (right) in the CCD, and the locii
defined by OCS (green points), HBBS (red points), FS and CMS stars. In this plane FS are 
indistinguishable from CMS, thus we show the FS+CMS groups together (blue points).

The stars classified as extreme are the $\sim 20\%$ of the sample examined here (see right 
panel of Fig. \ref{fclass}); this group is mainly composed by 
dust obscured carbon stars (OCS, $83\%$), with smaller contributions from CMS ($\sim 15 \%$)
and oxygen--rich HBBS ($\sim 2\%$). {\it This reinforces the idea that most ($\sim 95 \% - 97\%$) 
of the extreme AGBs are C--star candidates, surrounded by optically thick envelopes.}

C--stars, shown in the left panel of Fig. \ref{fclass}, are located on the left side of 
the diagram, in the region $[3.6]-[4.5] < 0.4$. In this group we find $\sim 55\%$ of the stars
of our sample. As expected, the vast majority of C--stars fall within the CMS+FS group,
shown in blue in Fig. \ref{fclass}. On the evolutionary side, these stars correspond
either to C--stars with a C/O ratio slightly above unity, or to objects with a much
larger C/O, in the phases following the thermal pulse. A few of C--stars ($\sim 3\%$)
were identified in this investigation as OCS, and correspond to the gray points in the
zone $0.2 < [3.6]-[4.5] < 0.4$ overlapped to the green dots in the left panel of
Fig. \ref{fclass}; this clearly depends on the assumed threshold in the $[3.6]-[4.5]$
colour used to define the OCS group. A few sources, that according to our interpretation
are HBBS, are classified as C--stars (31 objects, $\sim 1\%$ of the C--stars sample).  
We believe that this is due to the criterion, described in section \ref{class}, 
used to separate the C--star from the oxygen--rich star sample, based on the position of the individual
objects in the colour--magnitude diagram (J-Ks, Ks), and on the assumption that 
oxygen--rich stars populate an almost vertical strip, bluer than the C--stars sequence 
\citep[see Fig. 1 in][]{cioni06}. This definition holds as far as the stars are not 
obscured, but neglects the migration to the red (hence, to the zone occupied by C--stars) 
of the oxygen--rich star tracks once the envelope becomes optically thick 
\citep[see Fig. 1 in][]{riebel12}.

Finally, we focus on the O--rich candidates, shown in the middle panel of Fig. \ref{fclass}.
The $95\%$ of these stars belong to the CMS group, and also to the FS stars populating the
finger in the ([8.0]-[24], [24]) plane. A small fraction, below $2\%$, is part of the
HBBS, indicated in red in Fig. \ref{fclass}. A few of O--stars are classified as OCS,
and correspond to the grey dots overlapped to the green points in the middle panel
of Fig. \ref{fclass}: however, these points are very close to the assumed boundary 
separating OCS from the (FS+CMS) sample, with a difference in magnitude $\sim 0.1$ mag,
well below the photometric error on $[3.6]-[4.5]$.

\section{The dust production rate from AGBs in the LMC}
The determination of the dust production rate (DPR) by AGB in the LMC is currently 
a major issue, believed to provide important information on the role of AGB stars 
in the dust enrichment of the ISM.

Different modalities to find out the DPR have been adopted so far. The works by 
\citet{srinivasan09} and \citet{boyer12} are based on the assumed correlation between 
the rate at which dust is ejected from the star+envelope system, $\dot{M}_d$, 
and the 8.0 $\mu$m excess; \citet{matsuura09,matsuura13} use a relation between the 
gas mass loss rate, $\dot{M}$, and the IR colors; the approach by \citet{riebel12} 
relies on the fitting of the individual SEDs, in turn providing  $\dot{M}_d$.

Here we find the DPR based on the results of our simulation, giving the mass loss rate 
experienced by the stars populating the synthetic diagrams. Note that we do not need 
any a priori assumption of the gas/dust ratio, as the individual  $\dot{M}_d$'s are found 
by degree of condensation of the various key--species (carbon for C--stars, silicon for 
oxygen--rich stars), which is a result of our modelling \citep{ventura14}. 

To determine the dust production rate for each individual object coming from our
simulation in a given moment of the AGB evolution, we proceed as in \citet{fg06}.
For what concerns oxygen--rich stars, the dust production rate at a given evolutionary phase is 
given by the contribution of the dust produced under the form of silicates, corundum and iron:
$$
\dot{M}_d=\dot{M}_{sil}+\dot{M}_{cor}+\dot{M}_{Fe}
$$
with 
$$
\dot{M}_{sil}=\dot{M}X_{Si}{A_{Sil}/A_{Si}}frac(Si)
$$
$$
\dot{M}_{cor}=\dot{M}X_{Al}{A_{Al_2O_3}/A_{Al}}frac(Al)
$$
$$
\dot{M}_{Fe}=\dot{M}X_{Fe}frac(Fe)
$$

For what concerns C--stars, we consider the contributions from solid carbon, SiC and
iron:
$$
\dot{M}_d=\dot{M}_{C}+\dot{M}_{SiC}+\dot{M}_{Fe}
$$
with 
$$
\dot{M}_{C}=\dot{M}X_{C}frac(C)
$$
$$
\dot{M}_{SiC}=\dot{M}X_{Si}{A_{SiC}/A_{Si}}frac(Si)
$$
$$
\dot{M}_{Fe}=\dot{M}X_{Fe}frac(Fe)
$$

In the above expressions, the various $frac$'s indicate the fraction of the key--species
condensed into dust; $X_i$ represent the mass fractions of the key--species at the
surface of the stars; $A_i$ indicate the weight of the various species considered.
$\dot{M}_{sil}$, $\dot{M}_{cor}$, $\dot{M}_{Fe}$, $\dot{M}_{C}$, $\dot{M}_{SiC}$
indicate, respectively, the rate at which silicates, corundum, iron, solid carbon and
SiC particles are ejected from the envelope of the star.

In agreement with \citet{raffa14}, we find an overall DPR of 
$\sim4.5\times10^{-5}$ $M_{\odot}/yr$, with relative contribution from 
carbon and oxygen--rich stars of, respectively, 4$\times 10^{-5}$ $M_{\odot}/yr$ and 
5$\times 10^{-6}$ $M_{\odot}/yr$. Concerning the dust production from oxygen--rich stars, our
results indicate that it is almost entirely provided by HBBS, which represent only $\sim1\%$ 
of the whole AGB sample examined here.

Concerning the carbon component, half of the dust ejection rate is due to stars with 
$[3.6]-[4.5] > 1$, including only $Z > 4 \times 10^{-3}$ objects. 

Compared to the afore mentioned investigations, our DPR from C--rich stars is close to the 
results by \citet{matsuura09,matsuura13}, whereas for the O--rich component our findings 
are in better agreement with \citet{riebel12}.

\section{Conclusions}
We use models of stars of intermediate mass ($1M_{\odot} \leq M \leq 8M_{\odot}$), evolved
through the AGB phase, integrated with the description of the dust formation process in
the winds, to interpret Spitzer observations of AGBs in the LMC.

We find that the position of the individual sources in the various colour--colour and 
colour--magnitude diagrams obtained with the Spitzer bands is strongly connected with the 
mass and evolutionary phase experienced by the star; C--stars and oxygen--rich objects 
with dust in their envelopes occupy well separated zones in some of the observational diagrams. 
Metallicity is also found to play a role in this context.

Results from our modelling nicely reproduce the observations, in terms of the percentage
of stars found in the different regions of the observational planes. The spectroscopically
confirmed stars largely confirm our interpretation.

We find that the majority (97$\%$) of the most dust obscured objects, traditionally classified 
as "extreme", are carbon stars surrounded by carbonaceous particles. These stars, in the 
($[3.6]-[4.5]$, $[5.8]-[8.0]$) diagram, correspond to the observed diagonal strip at 
$[3.6]-[4.5] > 0.2$. Our models suggest that these stars are composed of 
multiple metallicities, starting from $Z=10^{-3}$ to $Z=8\times 10^{-3}$. 
Most of them are low--metallicity (Z $\sim 10^{-3}$) objects, 
with ages ranging from $4 \times 10^{8}$ yr to 3 Gyr; however, the reddest population, 
observed at $[3.6]-[4.5] > 2$ is entirely composed by higher--Z stars 
(Z $> 4 \times 10^{-3}$), with initial masses $M \sim 2.5-3 M_{\odot}$, in the 
very latest phases of the AGB evolution. They are expected to be surrounded by an internal 
layer hosting SiC particles of $\sim$ 0.08 $\mu$m size and a more external zone with 
0.15-- 0.2 $\mu$m sized carbon grains.

The remaining part of the extreme sample is composed by more massive objects, experiencing  
HBB, surrounded by alumina dust ($\sim 0.07 \mu$m) and silicates grains ($\sim 0.1 \mu$m); 
these sources are the descendants of stars with initial masses $M \sim 5-6 M_{\odot}$ formed 
$\sim 10^{8} $yr ago. Only metal--rich objects (Z $\sim8\times 10^{-3}$) are predicted to 
populate this sub-sample.

The overall dust production rate is $\sim4.5\times10^{-5}$ $M_{\odot}/yr$, with an 85$\%$ 
contribution from C--stars, the remaining 15$\%$ coming from M--stars. DPR from the latter 
sample is entirely given by stars experiencing HBB.

The results of this analysis, to be confirmed by further spectroscopic investigations, will
allow a considerable step forward in the characterization of the AGB population of the LMC,
opening the possibility of interpreting most of the stars observed, particularly those with the strongest dust 
thermal emission, in terms of mass of the
precursor, metallicity, surface chemistry, and the properties of dust present in their
winds. The same analysis can be extended to the Small Magellanic Cloud and, thanks to the expected 
performances of the future space missions, to other galaxies of the Local Group.

This study will also have an important feedback on the evolution properties of AGBs,
particularly towards a better understanding of the two most uncertain physical mechanisms
characterizing this evolutionary phase, namely the extent of the Third Dredge-Up and the
strength of the Hot Bottom Burning experienced by massive AGBs.

\section*{Acknowledgments}
The authors are indebted to the anonymous referee for the careful reading of the manuscript 
and for the detailed and relevant comments, that helped to increase the quality of this 
work. F.D. and P.V. are indebted to Simonetta Puccetti for the assistance in data selection and 
handling.
D.A.G.H. acknowledges support provided by the Spanish Ministry of Economy and 
Competitiveness under grants AYA–2011–27754 and AYA–2011–29060. P.V. was supported by 
PRIN MIUR 2011 "The Chemical and Dynamical Evolution of the Milk Way and Local Group 
Galaxies" (PI: F. Matteucci), prot. 2010LY5N2T. R.S. acknowledges funding from the 
European Research Council under the European Union’s Seventh Framework Programme 
(FP/2007- 2013)/ERC Grant Agreement n. 306476.

\end{document}